\documentclass[aps,prb,twocolumn,10pt,showkeys,superscriptaddress]{revtex4-2}
\usepackage[T1]{fontenc}
\usepackage[utf8]{inputenc}
\usepackage{microtype} 
\usepackage{newtxtext,newtxmath}
\usepackage{physics}   
\usepackage{graphicx}
\usepackage[dvipsnames]{xcolor}
\usepackage{booktabs}
\usepackage{multirow}
\usepackage{orcidlink}
\usepackage[normalem]{ulem}
\usepackage{cancel}
\usepackage{appendix}
\usepackage[dvipsnames]{xcolor}

\usepackage{hyperref}
\hypersetup{
  colorlinks=true,
  linkcolor=blue, citecolor=blue,
  filecolor=blue, urlcolor=blue,
  breaklinks=true
}

\newcommand{\weff}{\omega_{\text{eff}}}
\newcommand{\weffabs}{|\omega_{\text{eff}}|}

\begin{document}

\title{Single-Scale Magnetoelastic Landau Quantization: Thermodynamics, Quantum Oscillations, and Metrology}

\author{Denise Assafr\~{a}o\orcidlink{0000-0001-5465-0154}}
\email{denise.lima@ufes.br}
\affiliation{Departamento de F\'{\i}sica, Universidade Federal do Esp\'{\i}rito Santo, Vit\'oria - Esp\'{\i}rito Santo, 29075-910, Brazil}

\author{Faizuddin Ahmed\orcidlink{0000-0003-2196-9622}}
\email{faizuddinahmed15@gmail.com; fahmed@rgu.ac}
\affiliation{Department of Physics, The Assam Royal Global University, Guwahati, 781035, Assam, India}

\author{Edilberto O. Silva\orcidlink{0000-0002-0297-5747}}
\email{edilberto.silva@ufma.br}
\affiliation{Programa de P\'os-Gradua\c c\~ao em F\'{\i}sica \& Coordena\c c\~ao do Curso de F\'{\i}sica -- Bacharelado, Universidade Federal do Maranh\~{a}o, 65085-580 S\~{a}o Lu\'{\i}s, Maranh\~{a}o, Brazil}

\date{\today}

\begin{abstract}
We develop a unified, single-scale description of thermodynamics and quantum oscillations in electronic systems with a uniform areal density of screw dislocations under a uniform magnetic field. A single tunable gap, $\hbar|\omega_{\mathrm{eff}}|$ with $\omega_{\mathrm{eff}}=\omega_{c}+\omega_{cl}$, organizes all equilibrium observables obtained from a compact harmonic-oscillator partition function: free energy, internal energy, entropy, heat capacity, magnetization, magnetic susceptibility, and magnetocaloric responses collapse onto universal hyperbolic kernels in $x=\hbar|\omega_{\mathrm{eff}}|/(2k_{B}T)$. We identify a compensated-field regime where the transverse gap closes and the heat capacity reaches an equipartition plateau, providing a sharp signature of magnetoelastic interference. In transport and torque, the same scale rigidly shifts the Hall fan and compresses the $1/B$ period of de Haas-van Alphen and Shubnikov-de Haas oscillations when expressed in $1/B_{\mathrm{eff}}$, enabling a phase-unwarping protocol that metrologizes the dislocation density from a single field sweep. In mesoscopic samples, boundary corrections to the Landau degeneracy generate finite-size calorimetric oscillations that diagnose the effective magnetic length. Moderate disorder and weak interactions preserve the kernel structure while smoothing amplitudes. We outline an experimental roadmap combining on-chip calorimetry, torque magnetometry, and transport, and discuss device-level opportunities in caloritronics and strain engineering—magnetocaloric microcooling, magnetoelastic heat switching, and dilatometric transduction—where the single scale $\hbar|\omega_{\mathrm{eff}}|$ enables rational design and optimization.
\end{abstract}


\maketitle

\section{Introduction}

Topological defects provide a unifying language to connect elasticity, geometry, and electronic structure in crystalline media~\cite{Bilby1955,AoP.1992.216.1,Nelson2002,Bowick2009}. In the geometric theory of defects, disclinations and dislocations act as sources of curvature and torsion in a Riemann-Cartan background \cite{AoP.1992.216.1,LCR.2021.9.85}, enabling continuum field-theory descriptions of otherwise microscopic crystalline inhomogeneities~\cite{AoP.1992.216.1,Nelson2002,Bowick2009}. This perspective dovetails with electronic realizations of effective gauge fields in solids, notably in strained graphene and related two-dimensional materials, where lattice deformations mimic pseudomagnetic fields and reshape the low-energy spectrum~\cite{VozmedianoRMP2010,LevyScience2010}.

A particularly transparent setting emerges for a uniform areal density of screw dislocations \cite{PR.1959.116.1121,PRB.1996.53.1985,PRE.2004.69.011705,PRE.2010.82.031704,PRB.2005.72.045423,PC.1996.266.1,JAP.1959.30.89,JAP.2003.93.8918,PRE.2006.74.051710,ZPBCM.1992.86.163}. In this case, the transverse spectrum organizes into an equally spaced “elastic Landau” ladder, which maps the problem onto a harmonic-oscillator structure with a defect-controlled cyclotron-like frequency~\cite{JPCM.2008.20.125209,PLA.2012.376.2838}. When a uniform magnetic field is also present, the two contributions superpose into a single effective cyclotron scale,
$\omega_{\mathrm{eff}}=\omega_{c}+\omega_{cl}$, that governs spectral gaps and degeneracies.

The present work develops a single-scale thermodynamic framework for this elastic-magnetic ladder. Starting from the exact spectrum in the combined (torsion$+$field) background~\cite{JPCM.2008.20.125209}, we construct the canonical partition function at fixed longitudinal momentum and obtain closed forms for the Helmholtz free energy, internal energy, entropy, and heat capacity in terms of $|\omega_{\mathrm{eff}}|$. Two robust consequences follow. First, a Schottky-type crossover in $C_V(T)$ (see Refs.~\cite{PRB.1985.31.7452,PRL.2013.110.186405,PRB.2004.70.174429,PRB.1987.35.1397,PRE.2012.86.061104,PRL.1982.48.1208} for review) tracks the gap $\hbar|\omega_{\mathrm{eff}}|$, shifting with either field or dislocation density. Second, all temperature dependences in $C_V$, $(\partial M/\partial T)_B$, and related susceptibilities are governed by the same universal hyperbolic kernel, an experimentally favorable signature that collapses diverse datasets onto a single curve.

Magnetic responses inherit this organization. The equilibrium magnetization and (isothermal) susceptibility follow directly from free-energy derivatives and display controlled low-$T$ saturation and Curie-like high-$T$ trends set by $|\omega_{\mathrm{eff}}|$. On the caloric side, the adiabatic magnetocaloric coefficient and the magnetic Gr\"{u}neisen parameter, central figures of merit for solid-state cooling and scale diagnosis, reduce to compact expressions where the same kernel cancels or factors out, yielding strict linearity in $T$ (fixed-$k$ model) and a smooth plateau (after longitudinal $k$-integration)~\cite{Tokiwa2011,Kuechler2012,Zhu2003}. Including the free longitudinal motion restores the classical $3$D limit by adding the universal $+\tfrac12 k_B$ to $C_V$, clarifying the approach to Bohr-van Leeuwen at high temperature.

Quantum-oscillation physics provides an orthogonal, precision probe of the same scale. Within the Lifshitz-Kosevich (LK) framework~\cite{ShoenbergBook,LK1956,Dingle1952}, we show that torsion merely replaces $B\mapsto B_{\mathrm{eff}}$ in the dHvA/SdH phases, thereby \emph{compressing} the $1/B$ period and reshaping amplitudes through the $1/|\omega_{\mathrm{eff}}|$ prefactor and Dingle damping. This motivates a practical “phase-unwarping’’ protocol: by reparametrizing field sweeps in $1/(B+\delta B)$ and maximizing FFT sharpness, one extracts the offset $B_{\mathrm{tor}}$ that encodes the screw-dislocation density directly from a single trace. The same $B_{\mathrm{eff}}$ shift appears in Hall transport via the St\v{r}eda relation~\cite{Streda1982}, which rigidly translates the integer-quantized fan without altering level spacings in $1/B_{\mathrm{eff}}$. The correspondence between torque magnetometry, magnetocaloric observables, and transport thus closes a metrological loop tied to \emph{one} tunable energy scale.

Finite size and boundaries provide additional, mesoscopic diagnostics. Edge-state counting and perimeter corrections modify the Landau degeneracy in confined geometries, producing field-dependent oscillatory contributions to caloric quantities~\cite{PhysRevB.76.085308,PhysRevB.68.035326,AP1970,AP1971,PRL2003,PhysRevLett.102.066401,PMM.2023.124.1069,JCTE.2023.68.379}. Because $\ell_{\mathrm{eff}}\propto1/\sqrt{|\omega_{\mathrm{eff}}|}$, both the amplitude and period of these oscillations are torsion- and field-controllable, offering a complementary handle in patterned flakes, rings, and quantum dots.

Beyond noninteracting ladders, weak disorder and moderate interactions preserve the kernel structure while smoothing amplitudes, leaving intact the single-scale control of crossovers and oscillation periods~\cite{ShoenbergBook,Dingle1952}. Recent high-resolution studies combining torque magnetometry, on-chip calorimetry, and multi-axis field sweeps in metals, semimetals, and nodal-line systems, e.g., the ZrSiS family, demonstrate the maturity of these probes for quantitative metrology of small energy scales and subtle Fermi-surface features~\cite{ComprehensiveZrSiS2024}. Our results position defect engineering (torsion) on the same footing as magnetic field tuning: both feed the unique scale $\hbar|\omega_{\mathrm{eff}}|$ that organizes thermodynamics, quantum oscillations, and transport, enabling cross-validation and device-level design in caloritronics and straintronics~\cite{PhysRevLett.133.233801,PhysRevB.99.201301,PLA.2019.383.125974,PhysRevB.109.184109,PRB.2020.102.235163,Manesco2020,PhysRevB.109.184109,PhysRevB.99.201301}.

The paper is organized as follows. In Sec.~\ref{sec2a}, we build the canonical partition function from the elastic-Landau spectrum and obtain closed forms for $A$, $U$, $S$, and $C_V$ in terms of $\weff$. Section~\ref{sec3} analyzes their limiting behaviors and scaling. Section~\ref{sec4} discusses magnetic and torsional responses, the adiabatic magnetocaloric coefficient, longitudinal-momentum integration, finite-size corrections to degeneracy, and grand-canonical implications (dHvA and Hall transport). Section~\ref{sec:strain_engineering} develops strain engineering and defect control aspects, including torsional conjugates and susceptibilities, operating windows, and calibration protocols. Section~\ref{sec:metrology} presents the metrology suite and quantum-oscillation analysis, including phase unwarping and St\v{r}eda consistency. Conclusions are summarized in Sec.~\ref{sec5}. Technical derivations and additional materials are collected in Appendices~\ref{app:map-SF}-\ref{app:streda} and the numerical cookbook in Appendix~\ref{app:cookbook}.

\section{Thermodynamic Analysis}\label{sec2a}

\noindent\emph{Why a thermodynamic route?}
Spectral information alone does not tell how energy, entropy, and temperature respond when fields or defect densities are tuned.
A thermodynamic formulation is both \emph{timely} and \emph{actionable}
for at least three reasons:

\begin{itemize}
\item \emph{Caloritronics and solid-state cooling \cite{PRB.2022.105.144429,PRL.2012.108.075301,PRB.2018.97.235421,PRL.2012.108.075301,PRM.2021.5.114403,PRB.2016.94.245405,PRB.2020.101.205407,PRB.2016.93.224509,Energies.2023.16.5095,PT.2015.68.48,SSC.2010.150.500}.} On-chip
micro/nano-calorimetry now resolves Schottky-like signatures and
magnetocaloric coefficients in micron-scale devices. Quantities such as
$C_V(T)$, $\Gamma_B=(\partial T/\partial B)_S$, and the magnetic
Gr\"uneisen parameter are direct figures of merit for low-power cryogenic
cycles and heat management in 2D/vdW stacks.

\item \emph{Strain engineering and defect control \cite{NC.2021.12.3585,ASS.2017.425.696,PRB.2015.92.155417,PRL.2025.134.136704,PhysRevApplied.2016.6.034005,PRB.2015.92.155417,PRM.2021.5.124205,PRB.2024.110.214103,PRX.2023.13.041037,PRB.2024.109.054111,PRB.2020.102.085421}.} Modern growth,
transfer, and patterning techniques enable controlled dislocation
densities and torsional textures in semiconductors and van der Waals heterostructures. Treating the areal dislocation density (parametrized by $\Omega$) as a \emph{thermodynamic field} places defect engineering on the same footing as magnetic-field control: both feed a single tunable gap $\hbar\weffabs$ that organizes the entire response.

\item \emph{Metrology and quantum oscillations \cite{PRA.2023.107.022605,PRL.2014.112.010503,PRA.2024.109.042623,PRL.2018.120.213201,PRA.2025.111.042605,PRL.2020.124.223603,PRL.2014.113.073003,PRA.2016.93.023846,IJTP.2024.63.76}.} Torque magnetometry and SQUID readouts access $M(T,B)$, $\chi(T,B)$, and de Haas-van Alphen (dHvA) oscillations with high precision. Our framework predicts how torsion compresses the $1/B$ period and reshapes amplitudes, providing a route to \emph{quantify} $\Omega$ from standard magnetic measurements.
\end{itemize}

Below we construct the canonical partition function for the elastic-Landau ladder governed by the \emph{single} control scale $\weff=\omega_c+\omega_{cl}$. From it, we obtain closed forms for the Helmholtz free energy, internal energy, entropy, and heat capacity, all
expressed through the gap variable $x=\hbar\weffabs/(2k_B T)$. Two
consequences follow and will be used throughout the paper: (i) the universal thermal kernel $\sinh^{-2}x$ that links $C_V$, $(\partial M/\partial T)_B$, and normalized susceptibilities; and (ii) simple scaling laws in $T/\weffabs$ that make the roles of $B$ and $\Omega$
\emph{quantitatively} transparent. This provides the backbone for the experimental roadmap (Table~\ref{subsec:exp_signatures}) and for the
discussion of compensated fields, magnetocaloric efficiency, and torsion-induced dHvA shifts.

\subsection{Spectrum and mapping to an effective ladder}

Our starting point is the spectrum for a quantum particle in an elastic medium with a uniform density of screw dislocations under a uniform magnetic field, as derived in Ref.~\cite{JPCM.2008.20.125209}. The energy
eigenvalues can be written as
\begin{equation}
E_{nmk}=\hbar(\omega_{cl}+\omega_c)
\left(n+\frac{|m|}{2}+\frac{m}{2}+\frac{1}{2}\right)
+\frac{\hbar^2 k^2}{2\mu},
\label{eq:energy_spectrum}
\end{equation}
with $n=0,1,2,\dots$, $m\in\mathbb{Z}$, longitudinal wave number $k$ (along the dislocation axis $z$), and effective mass $\mu$. The elastic and magnetic cyclotron contributions are
\begin{equation}
\omega_{cl}=\frac{2\hbar k\,\Omega}{\mu},\qquad
\omega_c=\frac{eB}{\mu}.
\label{eq:omegas_def}
\end{equation}
Here $\Omega$ encodes the (uniform) screw-dislocation density and $B$ is the external field. It is convenient to introduce the \emph{effective cyclotron frequency}
\begin{equation}
\weff=\omega_c+\omega_{cl}.
\label{eq:weff_def}
\end{equation}

For fixed $k$ (held as a parameter), the transverse spectrum groups into equally spaced levels. One can relabel quantum numbers so that
\begin{equation}
E_{N',k}=\hbar\weffabs\Big(N'+\tfrac12\Big)+\frac{\hbar^2 k^2}{2\mu},
\qquad N'=0,1,2,\dots,
\label{eq:ladder_fixedk}
\end{equation}
which makes explicit the Landau-like ladder with spacing
$\hbar\weffabs$. Because thermodynamic sums depend on Boltzmann weights, only $\weffabs$ enters the thermal kernels; we keep track of the sign of $\weff$ separately when differentiating with respect to $B$ or $\Omega$ (e.g., for magnetization), via $\mathrm{sgn}(\weff)$~\footnote{All temperature-dependent kernels (e.g., $\coth x$, $\csch^2 x$) use $x=\hbar\weffabs/(2k_B T)$. The sign $\mathrm{sgn}(\weff)$ appears only through chain-rule factors such as
$\partial\weffabs/\partial B=(e/\mu)\,\mathrm{sgn}(\weff)$.}.

\subsection{Canonical partition function at fixed longitudinal momentum}

We work in the canonical ensemble at fixed $k$.
Let $\beta=1/(k_B T)$. Including the macroscopic degeneracy per Landau-like level, $g$, the partition function reads
\begin{align}
Z(T,V;k)
&= g\,e^{-\beta\hbar^2 k^2/2\mu}
\sum_{N'=0}^{\infty}e^{-\beta\hbar\weffabs(N'+1/2)} \notag\\
&= g\,e^{-\beta\hbar^2 k^2/2\mu}\,
\frac{e^{-\beta\hbar\weffabs/2}}{1-e^{-\beta\hbar\weffabs}} \notag\\
&= g\,e^{-\beta\hbar^2 k^2/2\mu}\,
\frac{1}{2\sinh\big(\beta\hbar\weffabs/2\big)}.
\label{eq:partition_function}
\end{align}

It is convenient to use the dimensionless gap variable $x\equiv \beta\hbar\weffabs/2=\hbar\weffabs/(2k_B T)$ so that
\begin{equation}
Z=g\,e^{-\beta\hbar^2 k^2/2\mu}\,\frac{1}{2\sinh x}.
\label{eq:Zx_form}
\end{equation}

\paragraph*{Degeneracy.}
In the thermodynamic limit for a two-dimensional transverse area $A$, the Landau-like degeneracy per level is
\begin{equation}
g=\frac{A}{2\pi \ell_{\mathrm{eff}}^{2}},
\qquad
\ell_{\mathrm{eff}}=\sqrt{\frac{\hbar}{\mu\,\weffabs}},
\label{eq:degeneracy}
\end{equation}
the magnetic length built from $\weffabs$. In finite/mesoscopic samples, boundary corrections deplete the bulk value by $O(\ell_{\mathrm{eff}}/R)$
(Sec.~\ref{subsec:finite_size}). For the fixed-$k$ thermodynamics, $g$ simply multiplies $Z$; if one keeps $g$ inside $A=-k_BT\ln Z$, the term $-k_BT\ln g$ alters $A$ and $S$ but not $U$ or $C_V$. When reporting intensive, per-channel results we set $g=1$ before taking derivatives; sample totals are obtained a posteriori by multiplying by $g(B,\Omega)$ (or by the finite-size, corrected $g_{\mathrm{corr}}$).

\paragraph*{Range of validity (fixed-$k$ approximation).}
Holding $k$ fixed isolates the transverse ladder. For bulk 3D samples, one should integrate over the free longitudinal motion; this adds a universal $+\tfrac12 k_B$ to the heat capacity per particle and restores the
classical limit at high $T$  (Sec.~\ref{subsec:k_integration_details}). All
results below using Eq.~\eqref{eq:partition_function} are therefore most reliable for $k_B T\lesssim \hbar\weffabs$ or for quasi-2D/mesoscopic
geometries where a single $k$ dominates.

\paragraph*{Convention.} Throughout the thermodynamic analysis, we report intensive, per-channel results by setting $g=1$ (no system size specified). Whenever a sample area is specified, for macroscopic totals or transport, we reinstate the Landau-like degeneracy $g(B,\Omega)=A/(2\pi\ell_{\mathrm{eff}}^{2})$. Thus, $U,C_V,S,M,\chi,\Pi_\Omega,\chi_\Omega$ are given per channel unless otherwise noted; sample totals follow by multiplying by $g(B,\Omega)$ (or $g_{\mathrm{corr}}$ in mesoscopic geometries).

\subsection{Useful dimensionless parameter and limiting forms}

With $x=\hbar\weffabs/(2k_B T)$:

\begin{itemize}
\item \textbf{Low-$T$ (large $x$):} The ladder is gapped,
$\sinh x\simeq \tfrac12 e^{x}$, and excited-state occupation is
exponentially suppressed. Thermodynamic quantities acquire activated forms $\propto e^{-2x}$.

\item \textbf{High-$T$ (small $x$):} $\sinh x\simeq x$, so
$Z\propto 1/x$ and the transverse sector approaches equipartition.
Upon $k$-integration (3D gas), $C_V$ tends to the classical
$\tfrac32 k_B$ per particle.

\item \textbf{Compensation line:} when $\weff=0$ the transverse gap closes. The kernels should be taken in the $x\to 0$ limit (finite and regular); this produces the compensated-field plateau discussed later.
\end{itemize}

\noindent
Equations~\eqref{eq:partition_function}-\eqref{eq:Zx_form} are the only
ingredients we need to obtain, by standard derivatives, the closed forms
for $A$, $U$, $S$, and $C_V$ in Sec.~\ref{sec3}. The sign of $\weff$ will
reappear solely in response functions that differentiate with respect to
$B$ or $\Omega$ (magnetization, susceptibilities), via chain-rule factors
containing $\mathrm{sgn}(\weff)$.

\section{Thermodynamic Properties}\label{sec3}

From the partition function, we can derive all other thermodynamic quantities \cite{helrich2008modern,pathria2021statistical}.

\paragraph*{Helmholtz Free Energy (A).} The Helmholtz free energy is given by $A = -k_B T \ln(Z)$. Using the properties of the logarithm and Eq. \ref{eq:partition_function}, we find:
\begin{align}
A(T, V, k) &= \frac{\hbar^2 k^2}{2\mu} + k_B T \ln\left[ 2 \sinh\left(\frac{\hbar\weffabs}{2k_B T}\right) \right]\notag\\& -k_B T \ln(g) 
\label{eq:helmholtz}
\end{align}

\paragraph*{Internal Energy (U).} The internal energy is calculated via $U = -\frac{\partial \ln(Z)}{\partial \beta}$.
Differentiating with respect to $\beta$ and using the identity $\coth(x) = \cosh(x)/\sinh(x)$, we obtain
\begin{equation}
U(T, V, k) = \frac{\hbar^2 k^2}{2\mu} + \frac{\hbar\weffabs}{2} \coth\left(\frac{\hbar\weffabs}{2k_B T}\right).
\label{eq:internal_energy}
\end{equation}

\paragraph*{Heat Capacity at Constant Volume ($C_V$).} The heat capacity is the derivative of the internal energy with respect to temperature, $C_V = \left(\frac{\partial U}{\partial T}\right)_V$. Using Eq. (\ref{eq:internal_energy}), the chain rule and the derivative $\frac{d}{dx}\coth(x) = -\text{csch}^2(x)$, we get
\begin{equation}
C_V(T, V, k) = k_B \left( \frac{\hbar\weffabs}{2k_B T} \right)^2 \text{csch}^2\left(\frac{\hbar\weffabs}{2k_B T}\right).
\label{eq:heat_capacity}
\end{equation}

\paragraph*{Entropy (S).}
Finally, the entropy is obtained from the thermodynamic relation $S = (U-A)/T$. By using Eqs. (\ref{eq:internal_energy}) and (\ref{eq:helmholtz}) and simplifying, we obtain
\begin{align}
S(T, V, k) &= k_B \ln(g) + \frac{\hbar\weffabs}{2T} \coth\left(\frac{\hbar\weffabs}{2k_B T}\right) \notag\\&- k_B \ln\left[ 2 \sinh\left(\frac{\hbar\weffabs}{2k_B T}\right) \right].
    \label{eq:entropy}
\end{align}
These findings highlight that the macroscopic thermodynamic behavior of the system is governed by the intricate interplay between the material's topological characteristics, represented by 
$\omega_{cl}$, and the external magnetic field, represented by 
$\omega_c$.

\section{Analysis of the results and discussion}\label{sec4}

In this section, unless stated otherwise, thermodynamic formulas and figures are reported per transverse channel (we set $g=1$) to highlight the universal kernels. Whenever we discuss macroscopic transport or sample-level observables (grand-canonical/dHvA, IQHE/St\v{r}eda, and finite-size corrections), we work with sample totals and keep the degeneracy explicit, using $g(B,\Omega)$, or the corrected $g_{\mathrm{corr}}$ when applicable (see Eqs.~\eqref{eq:degeneracy} and \eqref{eq:weff_def}.

\subsection{Magnetic Field Effects}

\subsubsection{Constant Magnetic Field Regime}

\begin{figure*}[tbhp]
\centering
\includegraphics[width=0.49\textwidth]{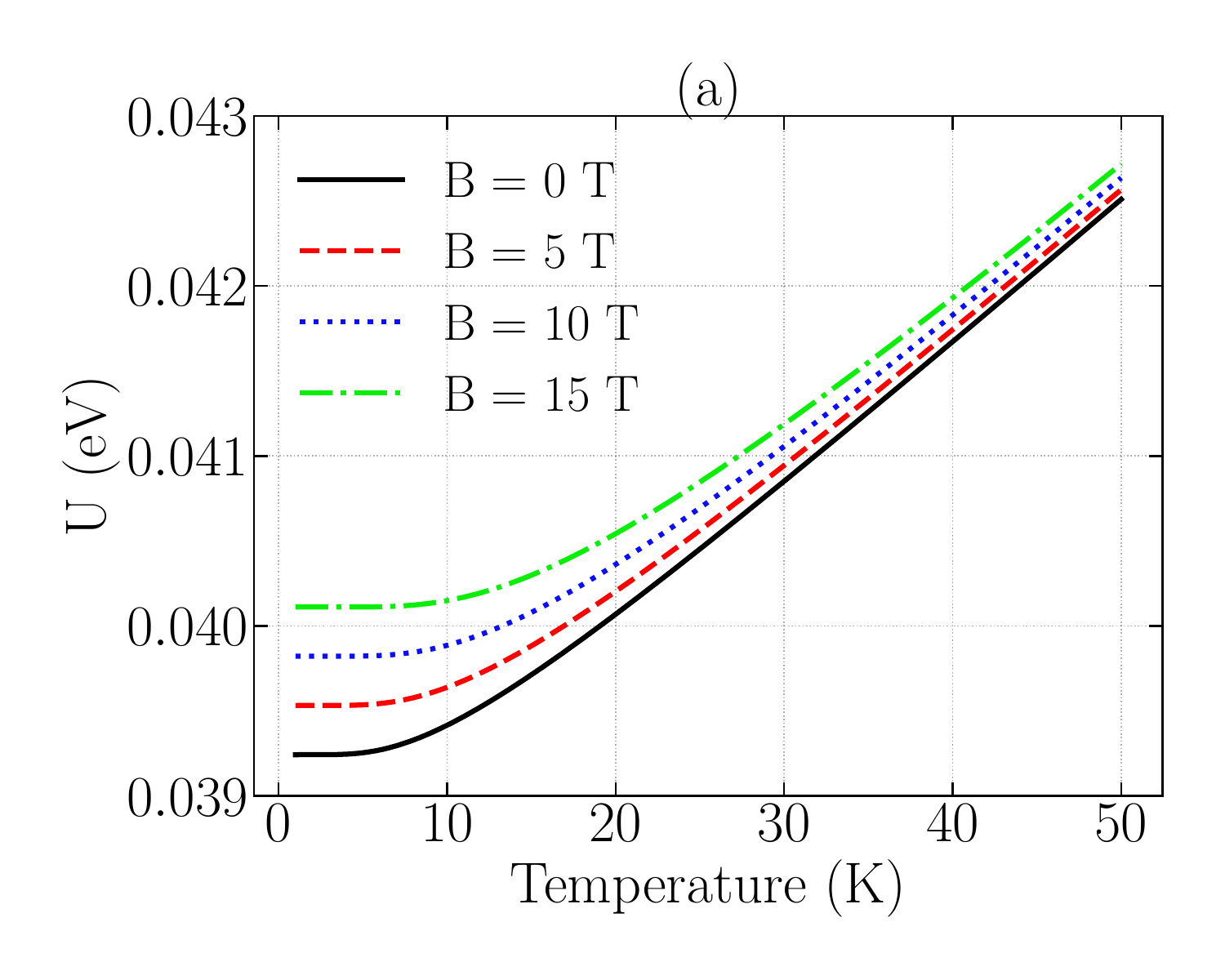}
\hfill
\includegraphics[width=0.49\textwidth]{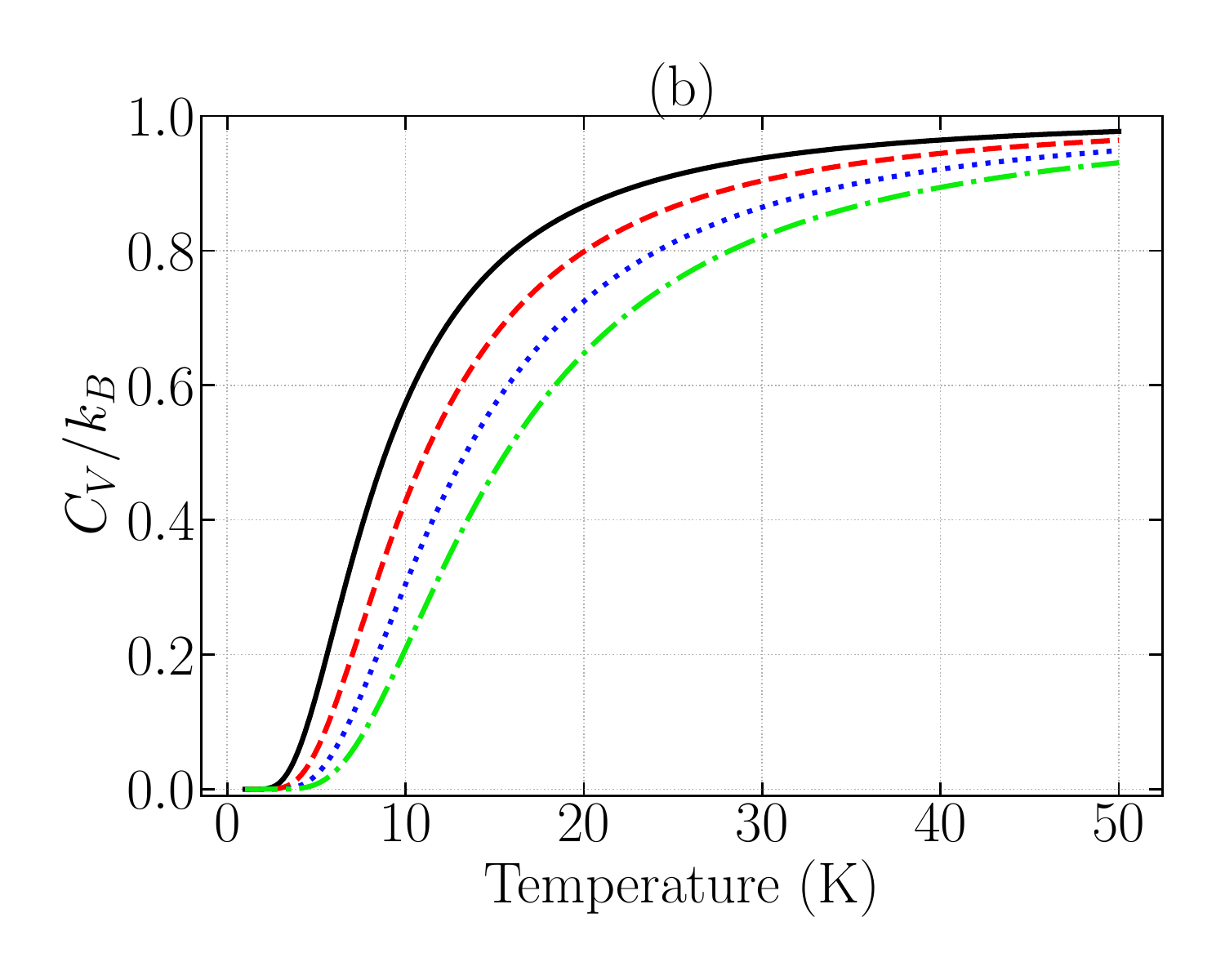}
\vfill
\includegraphics[width=0.49\textwidth]{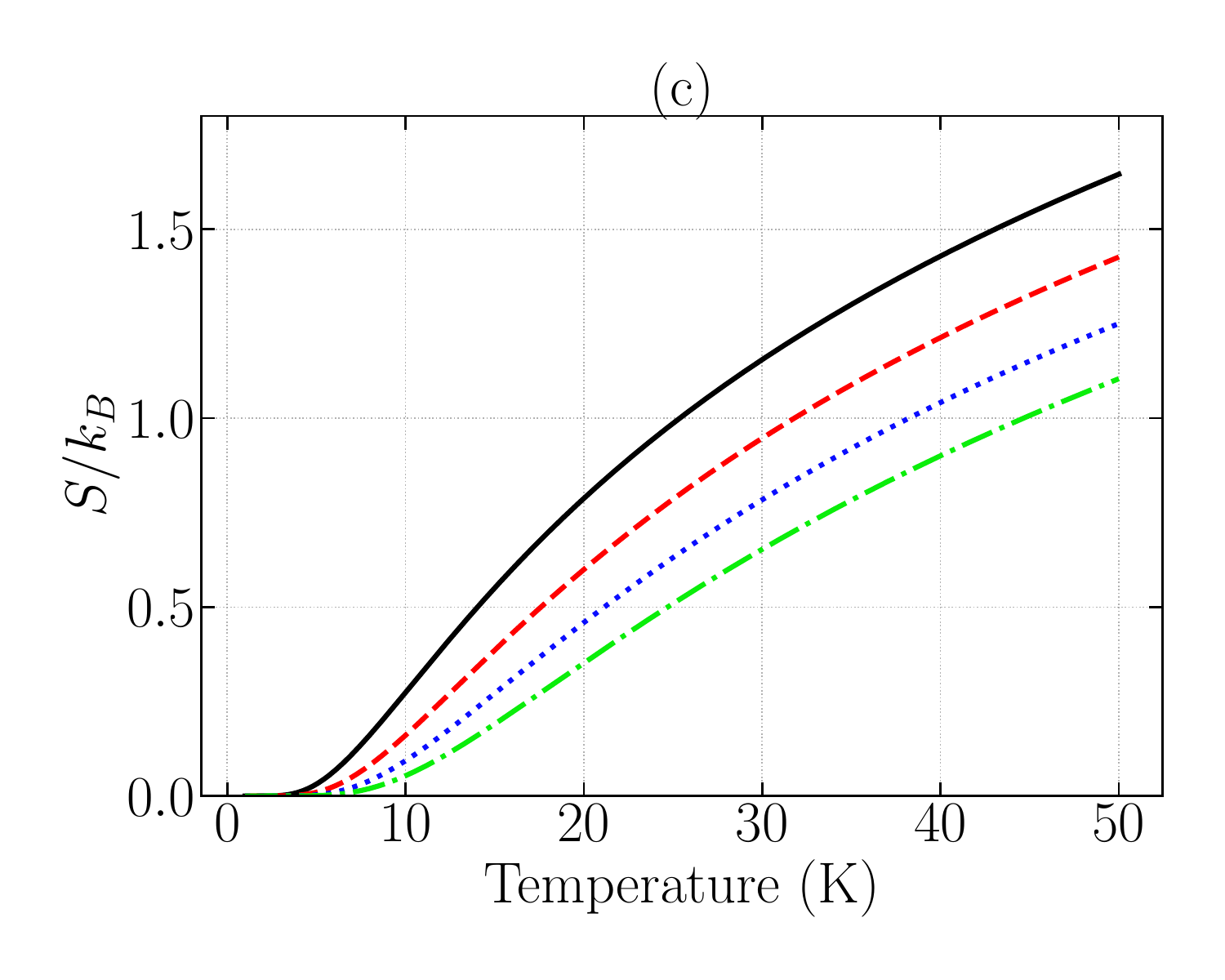}
\hfill
\includegraphics[width=0.49\textwidth]{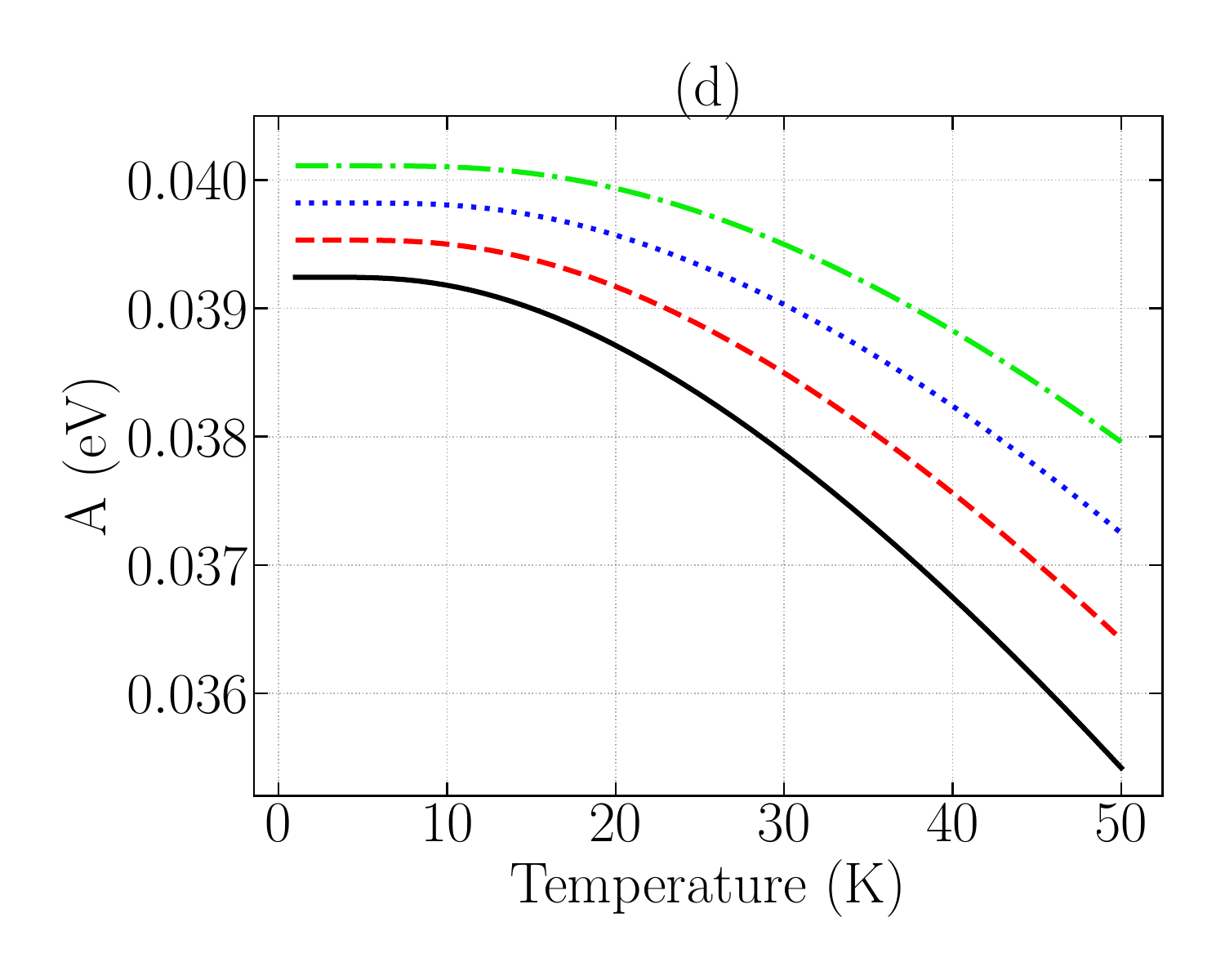}
\caption{\footnotesize Thermodynamic properties of the system as a function of temperature for a fixed torsion parameter ($\Omega = 1.5 \times 10^7\, \mathrm{m^{-1}}$) and different values of the external magnetic field, $B$. (a) The internal energy $U$ increases with $B$ due to the widening of the energy gap. (b) The heat capacity $C_V$ shows a Schottky-like peak that shifts to higher temperatures as $B$ increases, indicating a larger energy spacing. (c) The entropy $S$ rises more slowly for stronger magnetic fields, as fewer states become less thermally accessible. (d) The Helmholtz free energy $A$ is higher for larger $B$ at low temperatures, reflecting the dominance of the internal energy contribution.}
\label{fig:thermo_vs_B}
\end{figure*}

In this section, we analyze the thermodynamic behavior of the system under the influence of an external magnetic field, $B$, while keeping the torsion parameter fixed at $\Omega = 1.5 \times 10^7 \, \mathrm{m^{-1}}$. The results, plotted as a function of temperature, are shown in Fig.~\ref{fig:thermo_vs_B}.

\paragraph*{Internal Energy.} Figure~\ref{fig:thermo_vs_B}(a) displays the internal energy, $U$, of the system. At the zero-temperature limit ($T \to 0$), the internal energy converges to the ground-state energy, $E_0 = \hbar^2 k^2 / (2\mu) + \hbar\weffabs/2.$ As the magnetic field $B$ increases, the effective cyclotron frequency $\weffabs$ also increases, leading to a higher ground-state energy. This behavior is clearly observable as an upward shift of the curves for larger values of $B$. As the temperature rises, the thermal energy becomes sufficient to populate higher energy levels, leading to a monotonic increase in $U$. In the high-temperature regime ($k_B T \gg \hbar\weffabs$), all curves approach the classical equipartition limit, where the energy increases linearly with temperature but remains offset by their respective ground-state energies.

\paragraph*{Heat Capacity.} The heat capacity, $C_V$, plotted in Fig.~\ref{fig:thermo_vs_B}(b), provides deeper insight into the energy spectrum of the system. Each curve exhibits a prominent peak, a characteristic feature known as a Schottky anomaly. This anomaly arises in systems with a discrete energy spectrum, where the peak temperature, $T_{peak}$, corresponds to the thermal energy ($k_B T$) required to excite the system across its dominant energy gap, which in this case is $\hbar\weffabs$. As the magnetic field $B$ increases, $\weffabs$ becomes larger, thus widening the gap between the elastic Landau levels. Consequently, a higher thermal energy is needed to populate the first excited state, causing the peak of the heat capacity to shift toward higher temperatures. This behavior demonstrates that the magnetic field acts as an external tuning parameter for the effective energy quantization of the system. At both very low and very high temperatures, the heat capacity tends to zero and the classical limit ($k_B$), respectively, in agreement with the third law of thermodynamics and the equipartition theorem.

\paragraph*{Entropy.} The behavior of the entropy, $S$, is shown in Fig.~\ref{fig:thermo_vs_B}(c). At $T=0$, the entropy for all configurations vanishes (assuming a non-degenerate ground state, $g=1$), in accordance with the third law of thermodynamics. As the temperature increases, the entropy rises, reflecting the growing number of thermally accessible microstates. A key observation is that, for a given temperature, the entropy is lower for higher values of the magnetic field $B$. This behavior results directly from the larger energy gap $\hbar\weffabs$ at stronger fields. A larger gap implies that fewer excited states are available for a given amount of thermal energy, resulting in a less disordered system and, therefore, a smaller entropy. At sufficiently high temperatures, all curves eventually converge to the same saturation value, where the details of the energy level spacing become negligible.

\paragraph*{Helmholtz Free Energy.} Fig.~\ref{fig:thermo_vs_B}(d) displays the Helmholtz free energy, $A = U - TS$. The free energy represents the balance between the system's tendency to minimize its internal energy and to maximize its entropy. At low temperatures, the internal energy term ($U$) dominates; consequently, $A$ is higher for larger magnetic fields, following the same trend as the ground-state energy. As the temperature increases, the entropic contribution ($-TS$) becomes progressively more significant, leading to a decrease in the free energy for all curves. This reduction is more pronounced for weaker fields because, as shown in Fig.~\ref{fig:thermo_vs_B}(c), the entropy is higher, making the ($-TS$) term more influential. This behavior illustrates the fundamental thermodynamic principle that, at elevated temperatures, systems with a greater number of available states (i.e., smaller energy gaps) are thermodynamically favorable.

\begin{figure*}[tbhp]
    \centering
    \includegraphics[width=0.49\textwidth]{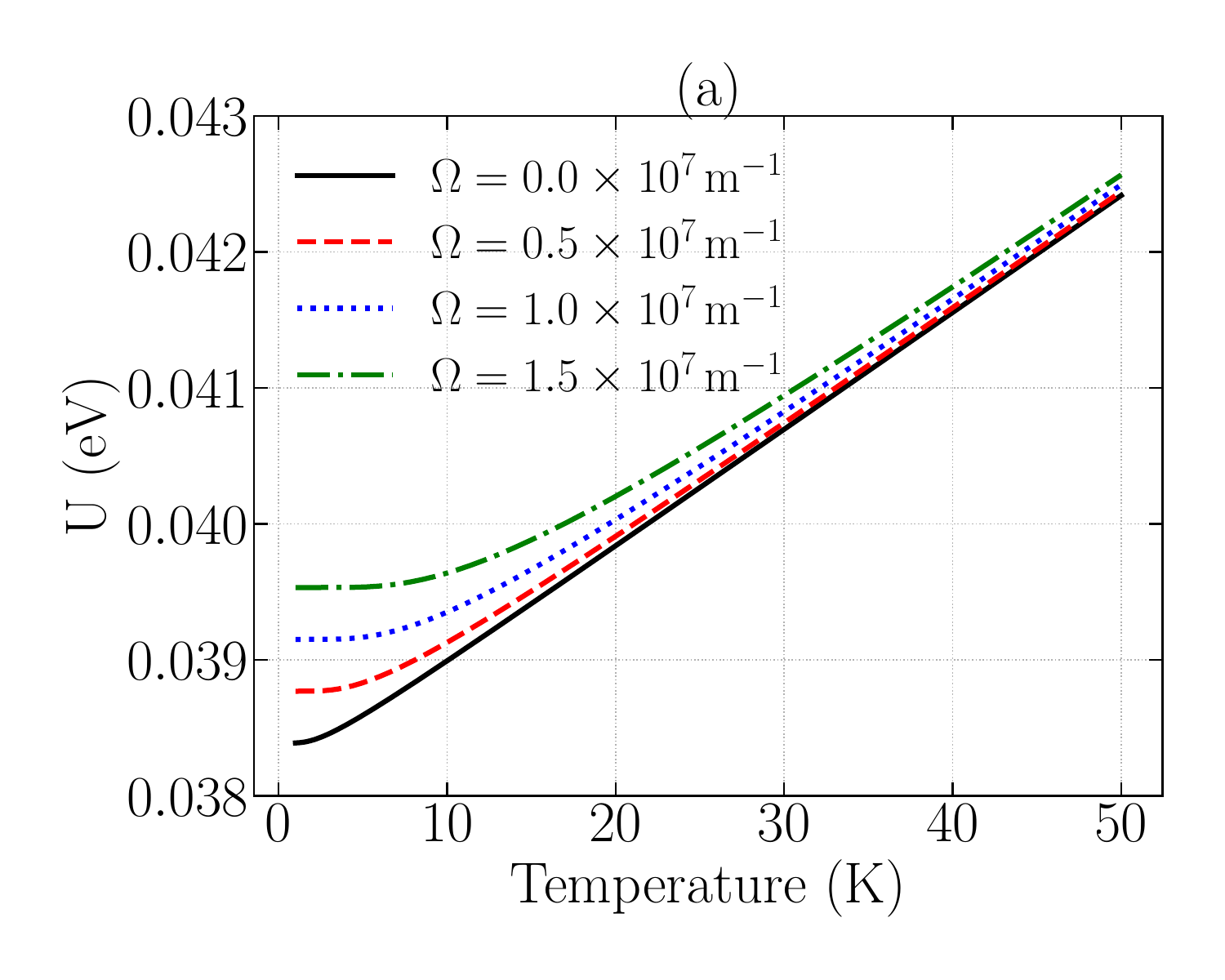}
    \hfill
    \includegraphics[width=0.49\textwidth]{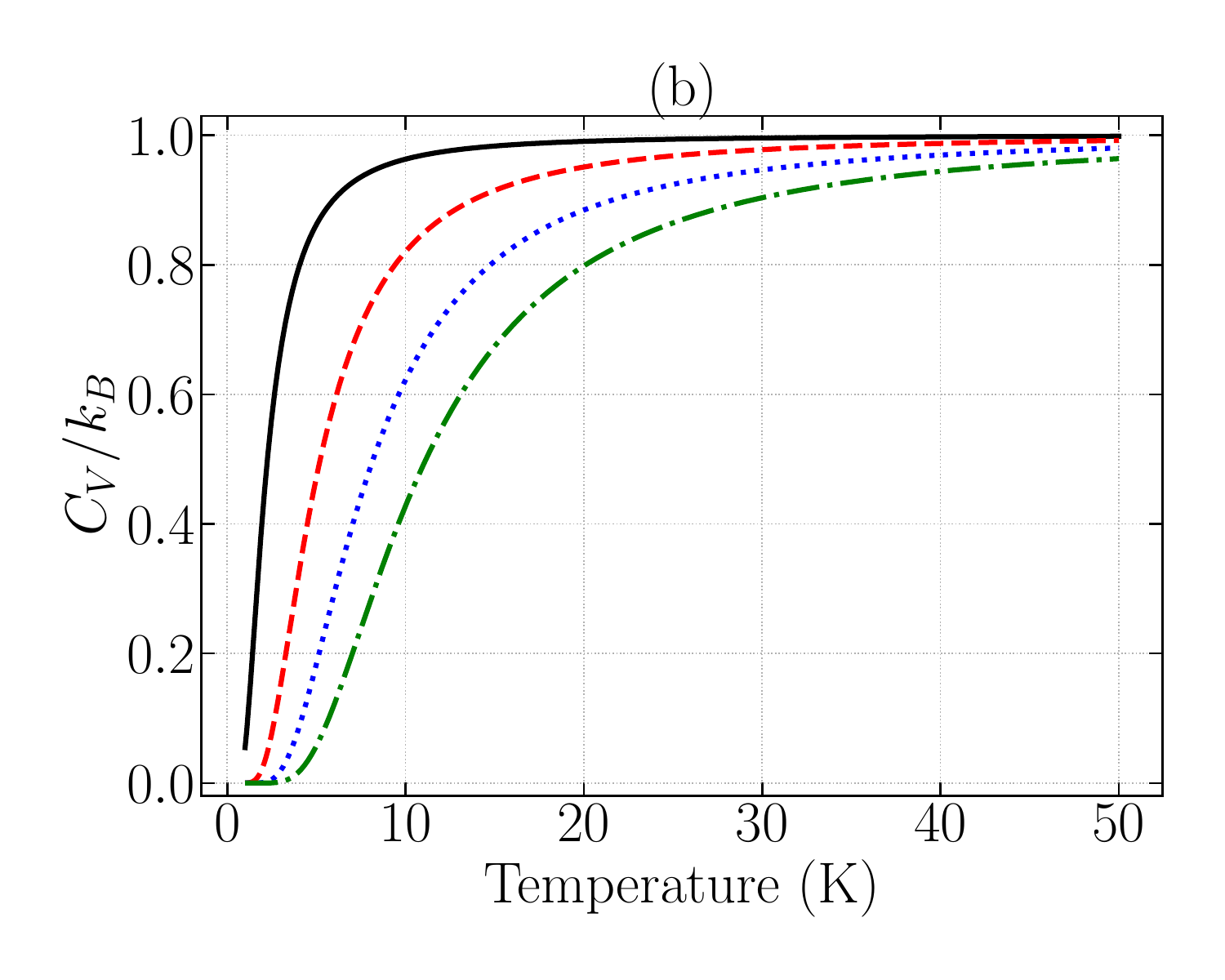}
    \vfill
    \includegraphics[width=0.49\textwidth]{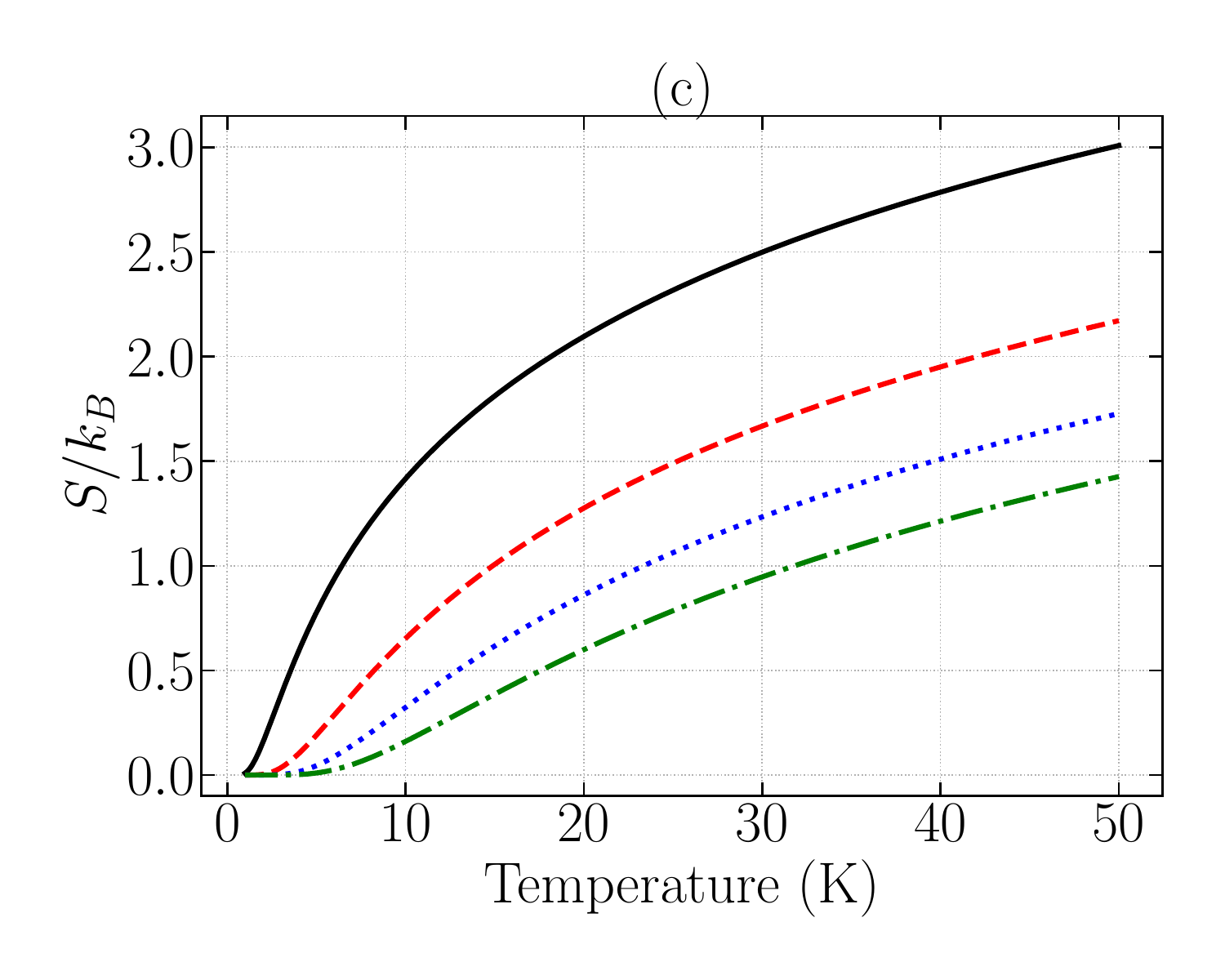}
    \hfill
    \includegraphics[width=0.49\textwidth]{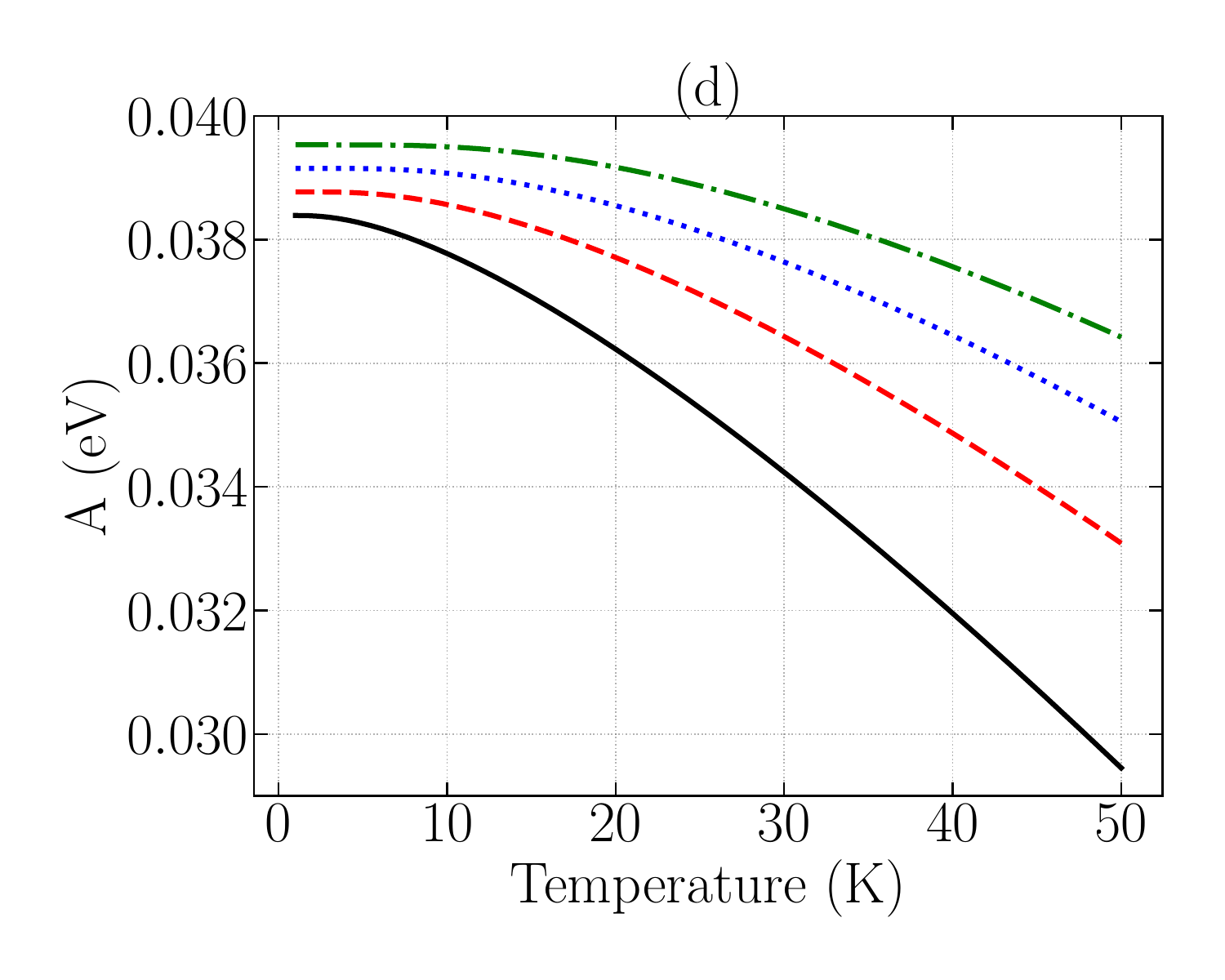}
    \caption{\footnotesize Thermodynamic properties of the system as a function of temperature for a fixed magnetic field ($B = 5.0 \, \mathrm{T}$) and different values of the torsion parameter, $\Omega$. (a) The internal energy $U$ increases with torsion density. (b) The Schottky-like peak of the heat capacity $C_V$ shifts to higher temperatures for larger $\Omega$, indicating an increase in the effective energy gap. (c) The entropy $S$ is lower for higher torsion densities at a given temperature. (d) The Helmholtz free energy $A$ increases with $\Omega$, as the internal energy contribution dominates in this temperature range.}
    \label{fig:thermo_vs_Omega}
\end{figure*}

\subsubsection{Torsion-induced Regime}
We now investigate the impact of the material's intrinsic torsion on its thermodynamic properties. In this analysis, the external magnetic field is fixed at $B = 5.0 \, \mathrm{T}$, while the torsion parameter, $\Omega$, which is proportional to the density of screw dislocations, is varied. The results are presented in Fig.~\ref{fig:thermo_vs_Omega}.

\paragraph*{Internal Energy.} Figure~\ref{fig:thermo_vs_Omega}(a) shows the internal energy $U$ as a function of temperature. The overall behavior is analogous to the case in which the magnetic field was varied. At any given temperature, the internal energy is higher for larger values of $\Omega$. This occurs because the torsion contributes to the effective cyclotron frequency $\weff$ through the elastic term $\omega_{cl} \propto \Omega$. A higher torsion density therefore results in a larger effective energy gap $\hbar\weffabs$ and, consequently, a higher ground-state energy. As the temperature increases, all configurations exhibit a monotonic rise in energy as higher "elastic Landau levels" become populated.

\paragraph*{Heat Capacity.} The effect of the torsion parameter is most clearly observed in the heat capacity, shown in Fig.~\ref{fig:thermo_vs_Omega}(b). Similar to the case of the magnetic field, increasing the torsion density causes the Schottky peak to shift toward higher temperatures. This reveals an important insight: the structural torsion of the material acts as an intrinsic "pseudo-magnetic field," effectively enhancing the quantization energy. The case $\Omega = 0$ (black solid line) corresponds to a system subject only to the standard magnetic field. As the density of topological defects increases (larger $\Omega$), the system becomes "stiffer" from a quantum mechanical standpoint, requiring greater thermal energy to excite higher levels. These results suggest that the thermodynamic response of a material can be engineered through controlled manipulation of defect density.

\paragraph*{Entropy.} The entropy curves shown in Fig.~\ref{fig:thermo_vs_Omega}(c) further corroborate this interpretation. For any given temperature, a higher value of $\Omega$ corresponds to a lower entropy. This occurs because the increased energy spacing between levels reduces the number of thermally accessible states for a fixed amount of thermal energy $k_B T$. The system with the highest torsion density is therefore the most "ordered" in this sense. All curves correctly start at $S=0$ at zero temperature and converge to the same classical limit at very high temperatures, where the specific details of the level spacing become negligible.

\paragraph*{Helmholtz Free Energy.} Figure~\ref{fig:thermo_vs_Omega}(d) displays the Helmholtz free energy. At low temperatures, the behavior of $A$ is dominated by the internal energy term $U$. Consequently, systems with higher torsion density, which have a larger ground-state energy, also exhibit a higher free energy. As the temperature increases, the entropic contribution  $-TS$ becomes significant, leading to a decrease in $A$. This reduction is less pronounced for larger values of $\Omega$ because, as seen in the entropy plot, these systems have lower entropy, thereby mitigating the influence of the $-TS$ term. This analysis highlights the competition between energy minimization and disorder maximization, with the torsion parameter playing a crucial role in setting the overall energy scale of the system.

\subsection{Magnetization and magnetic susceptibility}
\label{subsec:mag_results}

Orbital magnetization and the isothermal susceptibility follow here from the Helmholtz free energy of the elastic-Landau ladder, in the spirit of Landau’s diamagnetism \cite{Landau1930} and the modern Berry-phase formulation \cite{XiaoNiuRMP2010,Thonhauser2005,Ceresoli2006}. Within the single-scale structure set by $x=\hbar|\omega_{\mathrm{eff}}|/(2k_BT)$, temperature dependences factor through universal hyperbolic kernels, while the explicit field dependence of the Landau-like degeneracy $g\propto|\omega_{\mathrm{eff}}|$ adds a $\partial_B\ln g$ contribution to $M$. The low- and high-$T$ limits recover Landau saturation and the Bohr-van Leeuwen classicality, respectively.

Starting from $A(T)=k_{B}T\ln\bigl[2\sinh x\bigr]+\text{const}$ with
\[
x\equiv \frac{\beta\hbar|\omega_{\mathrm{eff}}|}{2},\qquad
\omega_{\mathrm{eff}}=\omega_{c}+\omega_{cl},
\]
the orbital magnetization and the (isothermal) susceptibility per channel are
\begin{align}
M_k(B,\Omega,T)
& \equiv -\Bigl(\frac{\partial A}{\partial B}\Bigr)_{T,\Omega}=-\,\mathrm{sgn}(\omega_{\mathrm{eff}})\,
   \frac{\hbar e}{2\mu}\,\coth x,
\label{eq:M_theory}
\\[4pt]
\chi_k(B,\Omega,T)
&\equiv\Bigl(\frac{\partial M_k}{\partial B}\Bigr)_{T,\Omega}
 = \Bigl(\frac{\hbar e}{2\mu}\Bigr)^{2}\beta\,\csch^{2}x .
\label{eq:chi_theory}
\end{align}
For $k_BT\ll\hbar|\omega_{\mathrm{eff}}|$ ($x\gg1$): $M_k\to-\mathrm{sgn}(\omega_{\mathrm{eff}})\,\hbar e/2\mu$ and $\chi_k\to0$.
For $k_BT\gg\hbar|\omega_{\mathrm{eff}}|$ ($x\ll1$): using $\coth x=1/x+x/3+\cdots$ and $\csch^{2}x=1/x^{2}-1/3+\cdots$,
\[
M_k \simeq -\,\mathrm{sgn}(\omega_{\mathrm{eff}})\Bigl[\,
\frac{e}{\mu}\frac{k_BT}{|\omega_{\mathrm{eff}}|}
+\frac{e}{\mu}\frac{\hbar^{2}|\omega_{\mathrm{eff}}|}{12k_BT}\Bigr],
\quad
\chi_k \simeq \Bigl(\frac{e}{\mu}\Bigr)^{2}\frac{k_BT}{|\omega_{\mathrm{eff}}|^{2}}.
\]
The linear-in-$T$ high-$T$ trends above are artifacts of the fixed-$k$, per-channel model. For sample totals one must either (i) keep the degeneracy $g(B,\Omega)$ inside the free energy, which adds the compensating term $-k_BT\,\partial_B\ln g$ to $M$, or (ii) include the longitudinal $k$-integration.
Both procedures remove the spurious growth and recover the Bohr-van Leeuwen
behavior ($M\to0$, $\chi\propto1/T$). When needed, totals follow from
$M_{\mathrm{tot}}=g(B,\Omega)\,M_k$ and
$\chi_{\mathrm{tot}}=g(B,\Omega)\,\chi_k$ after reinstating $g$.

\paragraph*{Low- and high-temperature limits.}
For $k_{B}T\ll\hbar\weffabs$ ($x\gg1$): $\coth x\to1$ and $\csch^{2}x\to0$, hence
\[
M\to -(\hbar e/2\mu)\,\mathrm{sgn}(\weff)+ \mathrm{sgn}(\weff)\,\frac{e}{\mu}\,\frac{k_{B}T}{\weffabs},
\qquad
\chi\to 0.
\]
In the opposite limit $k_{B}T\gg\hbar\weffabs$ ($x\ll1$), the terms $\propto T/\weffabs$
cancel between the two contributions, giving
\[
M \simeq -\,\mathrm{sgn}(\weff)\,
\frac{e}{\mu}\,\frac{\hbar^{2}\weffabs}{12\,k_{B}T},
\qquad
\chi \simeq -\Bigl(\frac{e}{\mu}\Bigr)^{2}\,
\frac{\hbar^{2}}{12\,k_{B}T},
\]
so that $M,\chi\to0$ as $T\to\infty$, in agreement with the Bohr-van Leeuwen theorem. Since $\weff=\omega_{c}+\omega_{cl}$ depends on both $B$ and $\Omega$, varying either shifts the crossover scale $k_{B}T\sim\hbar\weffabs$, providing a direct probe of magnetoelastic coupling.

Equations \eqref{eq:M_theory}-\eqref{eq:chi_theory} show that the elastic cyclotron contribution $\omega_{cl}$ (set by dislocation density) and the magnetic cyclotron frequency $\omega_{c}$ combine into a single scale $\hbar\weffabs$ that controls both the low-$T$ diamagnetic saturation of $M$ and the Schottky-type maximum in $\chi(T)$. These effects can be accessed via torque magnetometry in mechanically deformed semiconductor heterostructures.
\begin{figure}[tbhp]
  \centering
  \includegraphics[width=\linewidth]{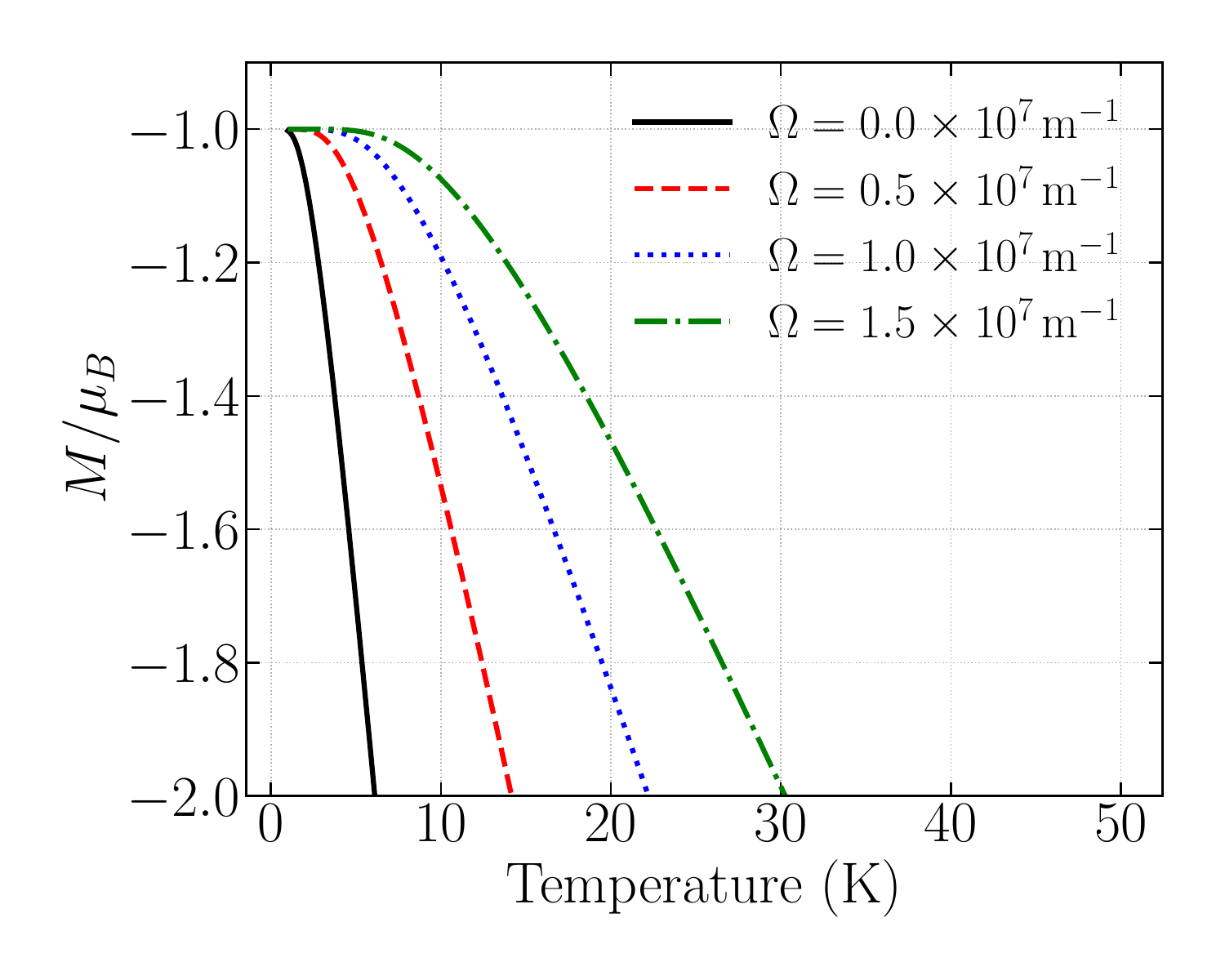}
  \caption{\footnotesize
  Magnetization per particle as a function of temperature for a fixed field $B = 5\,$T and four torsion densities
  $\Omega = 0$, $0.5$, $1.0$ and
  $1.5 \times 10^{7}\,$m$^{-1}$, shown in black, red, blue, green, respectively.
  The curves are in units of the Bohr magneton,
  $M/\mu_{B}$, and correspond to Eq.~\eqref{eq:M_theory}
  with $k = 1\times10^{9}\,$m$^{-1}$.\@ As $T\to0$ all curves converge to the universal
  Landau-diamagnetic value
  $M_{0}= -\hbar e/2\mu \simeq -\,\mu_{B}$.
  As $T$ increases and $k_{B}T\gg\hbar\weffabs$, the leading $T/\weffabs$ terms in Eq.~\eqref{eq:M_theory} cancel, leaving the next-order contribution
  $M\simeq -\,\mathrm{sgn}(\weff)\,(e/\mu)\,\hbar^{2}\weffabs/(12\,k_{B}T)$.
 Increasing $\Omega$ enhances $\weffabs$ and thus reduces the temperature-induced decay of $|M|$, meaning that torsion delays  the loss of diamagnetic saturation.}
  \label{fig:magnetisation_vs_Omega}
\end{figure}

The numerical traces follow Eq.~\eqref{eq:M_theory}. For $T\lesssim 5$\,K, the population is confined to the ground elastic-Landau level and $M$ saturates near $-\hbar e/2\mu$. Once $k_{B}T\sim\hbar\weffabs$, excited states contribute; however, the linear-in-$T$ pieces of the two contributions to $M$ cancel, and the high-$T$ behavior crosses over to $M\propto -\,\weffabs/T$, causing the curves to fan out more slowly than a naive $-T/\weffabs$ law. A higher dislocation density (larger $\Omega$) \emph{delays} the loss of diamagnetic saturation by enlarging~$\weffabs$.
\begin{figure}[tbhp]
  \centering
  \includegraphics[width=\linewidth]{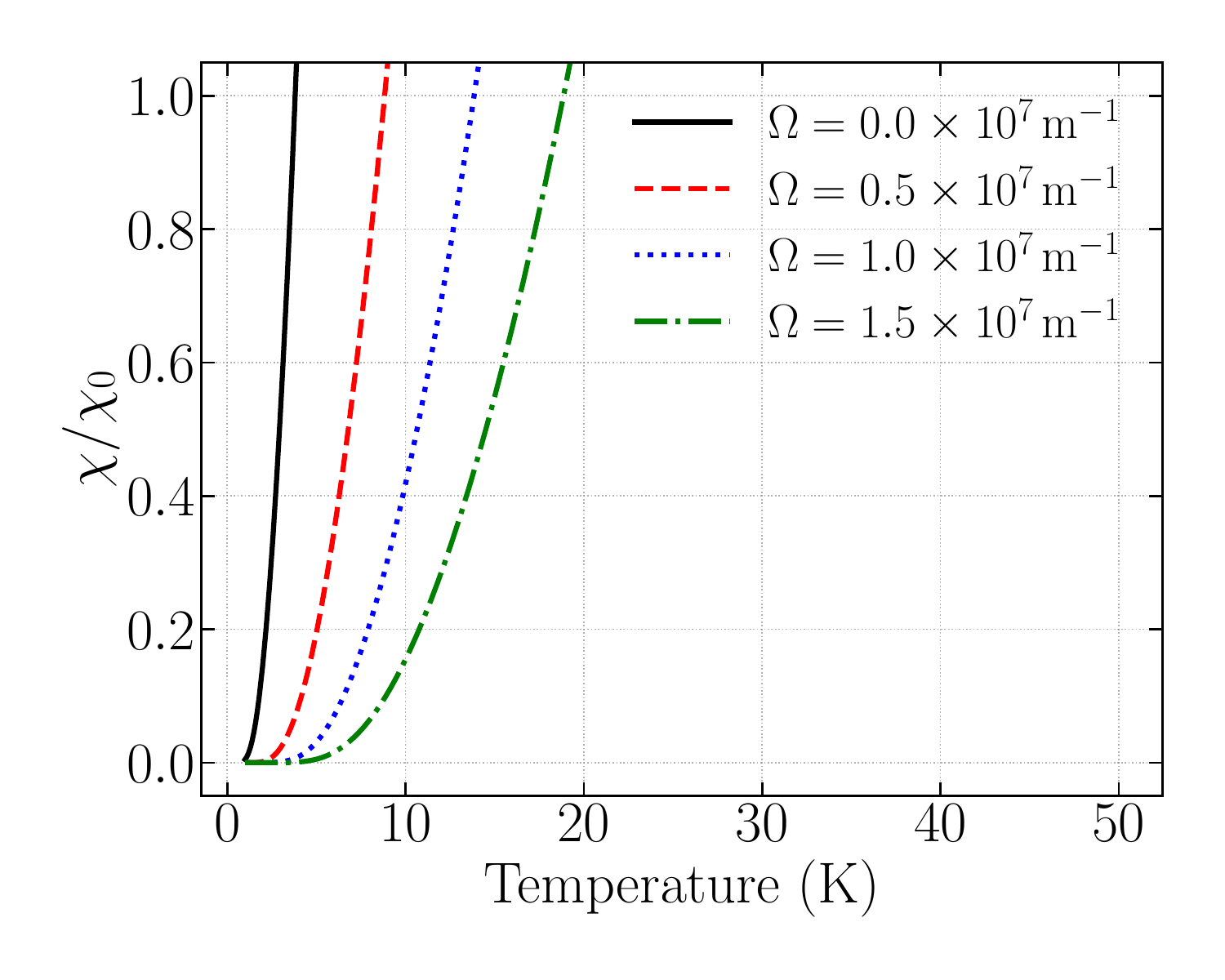}
  \caption{\footnotesize
  Normalized magnetic susceptibility
  $\chi/\chi_{0}$ as function of temperature, obtained from Eq.~\eqref{eq:chi_theory} using the same parameters of
  Fig.~\ref{fig:magnetisation_vs_Omega}. The normalization
  $\chi_{0}(T)= (\hbar e/2\mu)^{2}/(k_{B}T)$ removes the trivial Curie factor. With the corrected susceptibility, $\frac{\chi}{\chi_0}= \csch^{2}x \;-\; \frac{1}{x^{2}},\qquad
  x=\frac{\hbar\weffabs}{2k_{B}T},
  $ each curve starts from $\chi=0$ (ground-state rigidity) and, for $x \ll 1$ (high $T$), approaches the constant limit $\chi/\chi_0\to -\,1/3$, consistent with the diamagnetic high-$T$ behavior. Increasing $\Omega$ enlarges the effective gap, shifting the onset of strong susceptibility to higher $T$ and flattening the low-$T$ curvature.}
  \label{fig:susceptibility_vs_Omega}
\end{figure}

Plotting $\chi/\chi_{0}$ isolates the thermal kernel through the compact combination $\csch^{2}x-1/x^{2}$. At low $T$, the gap suppresses orbital excitations and $\chi$ vanishes rapidly; for $k_{B}T\gtrsim\hbar\weffabs$, $\chi$ follows the corrected Curie-normalized trend and tends to $-1/3$. Larger $\Omega$ increases $\weffabs$, shifting the crossover to higher temperature and producing broader, milder maxima, consistent with the flatter $M(T)$ trends in Fig.~\ref{fig:magnetisation_vs_Omega}.

\paragraph*{Range of validity.}
Equations \eqref{eq:M_theory}-\eqref{eq:chi_theory} presume a discrete transverse spectrum and fixed longitudinal momentum $k$. In the regime $k_{B}T\gg\hbar\weffabs$, the Bohr-van Leeuwen theorem requires that the orbital magnetization vanishes; the corrected expressions already capture the $M\to0$ trend. A more complete treatment that (i) integrates over the Maxwell distribution in $k$ and/or (ii) includes the $B$-dependent Landau degeneracy $g(B)$ ensures the expected scaling $M\propto B/T\to 0$ as $T\to\infty$. Consequently, the curves in Fig.~\ref{fig:magnetisation_vs_Omega} should be interpreted for $T\lesssim \hbar\weffabs/k_{B}$, beyond which the fixed-$k$ approximation ceases to be quantitatively reliable.

\subsection{Torsional Conjugate and Torsional Susceptibility}
\label{subsec:torsion_response}

Treating the areal density of screw dislocations $(\Omega)$ as a bona fide thermodynamic field invites the definition of a \emph{torsional conjugate},
$\Pi_{\Omega}\equiv-(\partial A/\partial\Omega)_{T,B}$, and its linear response, the \emph{torsional susceptibility} $\chi_{\Omega}=(\partial \Pi_{\Omega}/\partial\Omega)_{T,B}$.
This perspective is natural in the geometric theory of defects, where dislocations source spacetime \emph{torsion} in a Riemann-Cartan background and enter electronic Hamiltonians as elastic gauge fields \cite{AoP.1992.216.1}.
Closely related torsional responses are familiar in topological and quantum Hall fluids, in which coupling to background torsion underlies dissipationless (Hall) viscosity and other viscoelastic coefficients \cite{HughesLeighParrikar2011,BradlynRead2015,GromovAbanov2015}.
Here we evaluate $\Pi_{\Omega}$ and $\chi_{\Omega}$ for the elastic-Landau ladder, exposing the same single-scale kernel $x=\hbar|\omega_{\mathrm{eff}}|/(2k_{B}T)$ that organizes magnetization and heat capacity.
 
Here, we focus on the \emph{equilibrium} electronic free energy of the elastic-Landau ladder and define $\Pi_{\Omega}$ as the configurational work conjugate to $\Omega$. To our knowledge, explicit closed-form expressions for $\Pi_{\Omega}(T,B)$ and $\chi_{\Omega}(T,B)$ in the Landau-like spectrum induced by a uniform screw-dislocation density have not been tabulated previously; our results therefore provide a compact thermodynamic characterization that complements prior spectral studies of dislocation-induced Landau quantization \cite{JPCM.2008.20.125209,PLA.2012.376.2838}. In our single-scale framework, all temperature dependences again factor through $x=\hbar|\omega_{\mathrm{eff}}|/(2k_BT)$, so that $\Pi_{\Omega}$ and $\chi_{\Omega}$ share the same universal hyperbolic kernel structure encountered in the magnetic sector, offering experimentally accessible signatures (e.g., via dilatometry or strain-tuning under field \cite{ESLtd.2007.108,APL.2020.117.233502}) of magnetoelastic coupling in defect-rich media.

We treat the torsion parameter $\Omega$ (proportional to the screw-dislocation density) as a thermodynamic field. The variable conjugate to it is defined from the Helmholtz free energy $A(T,B,\Omega)$ as
\begin{equation}
\Pi_{\Omega}\equiv
  -\Bigl(\tfrac{\partial A}{\partial\Omega}\Bigr)_{T,B}.
\label{eq:Pi_def}
\end{equation}
Starting from the fixed-$k$ free energy,
\[
A(T)=k_{B}T\ln\bigl[2\sinh x\bigr]+\text{const},\qquad
x=\frac{\beta\hbar\weffabs}{2},
\]
one finds by the chain rule
\begin{equation}
\frac{\partial A}{\partial \Omega}
= k_{B}T\,\coth x\,\frac{\partial x}{\partial \Omega}, 
\end{equation}
with
\begin{align}
\frac{\partial x}{\partial\Omega}
&= \frac{\beta\hbar}{2}\,\frac{\partial |\omega_{\mathrm{eff}}|}{\partial\Omega}
= \beta\,\frac{\hbar^{2}k}{\mu}\,\mathrm{sgn}(\omega_{\mathrm{eff}}),
\end{align}
where we used $\weff=\omega_{c}+\omega_{cl}$,
$\partial\omega_{c}/\partial\Omega=0$ and
$\partial\omega_{cl}/\partial\Omega=2\hbar k/\mu$.
Inserting this result into \eqref{eq:Pi_def} yields
\begin{equation}
\Pi_{\Omega}(T)
= -\,\mathrm{sgn}(\weff)\,\frac{\hbar^{2}k}{\mu}\,
  \coth\Bigl(\tfrac{\beta\hbar\weffabs}{2}\Bigr),
\label{eq:Pi_Omega}
\end{equation}
which is Eq.~\eqref{eq:Pi_Omega} in compact form.

\paragraph*{Torsional susceptibility.}
The torsional susceptibility is
\begin{equation}
\chi_{\Omega}(T)\equiv
\Bigl(\tfrac{\partial \Pi_{\Omega}}{\partial\Omega}\Bigr)_{T,B}.
\label{eq:chiO_def}
\end{equation}
Away from the compensation point ($\weff\neq 0$), $\mathrm{sgn}(\weff)$
is constant and the derivative acts only on $x=\beta\hbar\weffabs/2$:
\begin{align}
\chi_{\Omega}(T)
&= -\,\mathrm{sgn}(\weff)\,\frac{\hbar^{2}k}{\mu}\;
\frac{d}{d\Omega}\left[\coth x\right] \notag\\
&= \frac{\hbar^{2}k}{\mu}\;\csch^{2}x\;
\frac{dx}{d\Omega}
= \Bigl(\frac{\hbar^{2}k}{\mu}\Bigr)^{2}\,
\beta\,\csch^{2}x ,
\label{eq:chi_Omega}
\end{align}
recovering Eq.~\eqref{eq:chi_Omega}. As expected for a response function, the final result is independent of the sign of~$\weff$ (since $\mathrm{sgn}^{2}=1$).

\paragraph*{Limiting behaviors.}

From \eqref{eq:Pi_Omega}, the $T\to0$ limit is $\Pi_{\Omega}^{0}=-\,\mathrm{sgn}(\weff)\,\hbar^{2}k/\mu$, while for $k_{B}T\gg\hbar\weffabs$,
$\coth x\simeq 1/x+O(x)$ and $\Pi_{\Omega}\simeq -\,(\hbar^{2}k/\mu)\,\mathrm{sgn}(\weff)\,(2k_{B}T/\hbar\weffabs) \propto T/\weffabs
$. Similarly, Eq.~\eqref{eq:chi_Omega} shows that $\chi_{\Omega}(T)\to 0$ exponentially at low $T$ and follows a Curie-like law $\propto 1/T$ at high $T$, with the universal thermal kernel $\csch^{2}x$.

\begin{figure}[ht!]
\includegraphics[width=\linewidth]{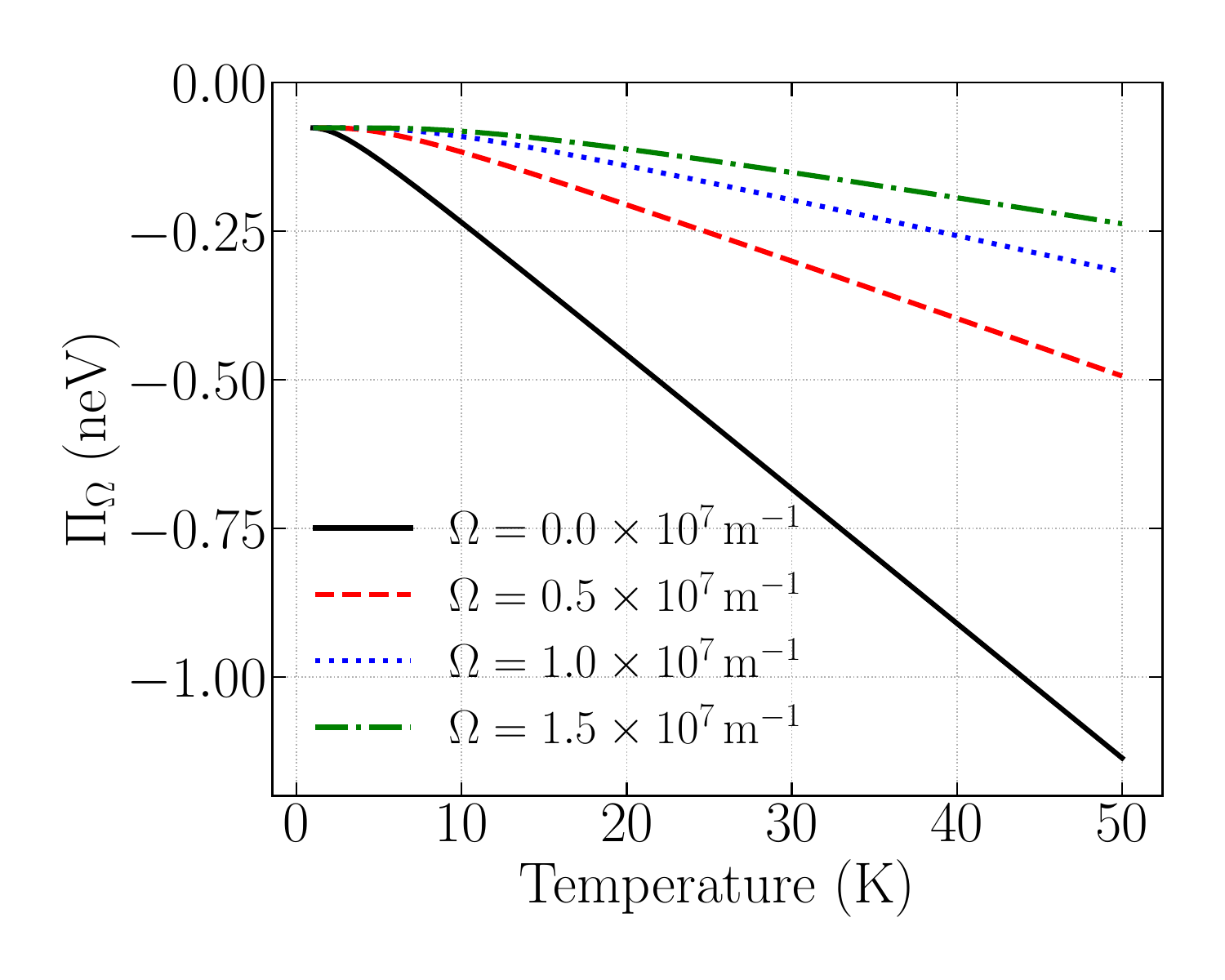}
\caption{\footnotesize
Torsional conjugate per particle, $\Pi_{\Omega}(T)$, for a fixed magnetic field ($B = 5\,\mathrm{T}$)
and four torsion densities
($\Omega = 0$, $0.5$, $1.0$, $1.5 \times 10^{7}\,\mathrm{m^{-1}}$) shown in black, red, blue, and green, respectively.
At $T \to 0$, all curves converge to
$\Pi_{\Omega}^{0}=-\,\hbar^{2}k/\mu$, according to
Eq.~\eqref{eq:Pi_Omega} with $\mathrm{sgn}(\weff)=+1$.
For $k_{B}T>\hbar\weffabs$, the $\coth$ term gives
$|\Pi_{\Omega}|\propto T/\weffabs$,
so larger $\Omega$ (corresponding to larger $\weffabs$) delays the onset of the increase.}
\label{fig:Pi_vs_Omega}
\end{figure}
\begin{figure}[tbhp]
\includegraphics[width=\linewidth]{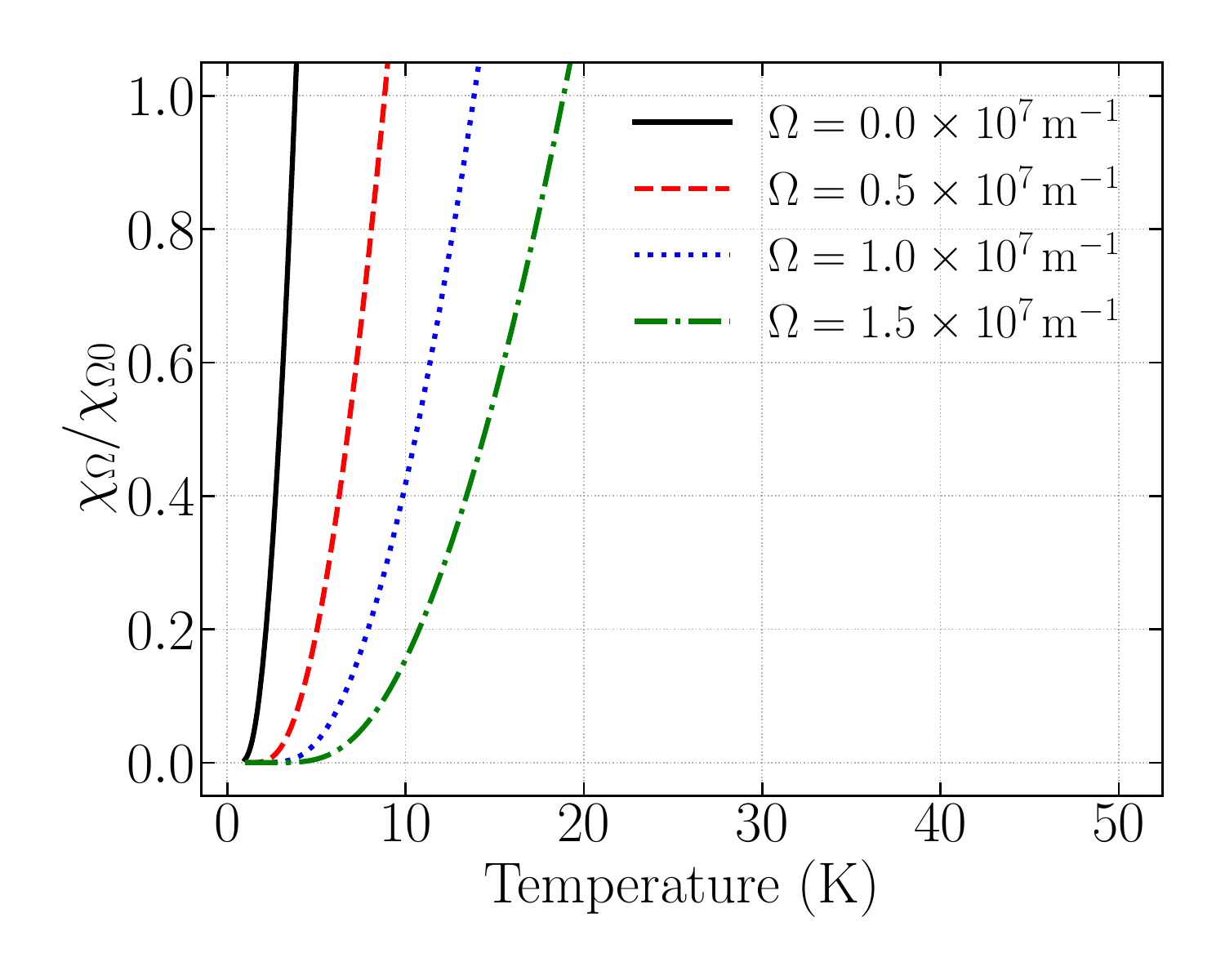}
\caption{\footnotesize
Normalized torsional susceptibility $\chi_{\Omega}/\chi_{\Omega 0}$,
with $\chi_{\Omega 0}(T)=(\hbar^{2}k/\mu)^{2}/(k_{B}T)$, calculated using the same parameters as in Fig.~\ref{fig:Pi_vs_Omega}.
The normalization removes the trivial Curie factor, leaving the universal kernel $\chi_{\Omega}/\chi_{\Omega 0}=1/\sinh^{2}x$,
$x=\hbar\weffabs/2k_{B}T$.
Increasing $\Omega$ enhances $\weffabs$ and shifts the midpoint of the
curve to higher $T$, mirroring the behavior of the magnetic susceptibility.}
\label{fig:chiOmega_vs_Omega}
\end{figure}

Figure~\ref{fig:Pi_vs_Omega} illustrates Eq.~\eqref{eq:Pi_Omega}. Because the $T\to 0$ limit depends only on $k$ and $\mu$, all traces share the same baseline $\Pi_{\Omega}^{0}$. As temperature grows and $x=\beta\hbar\weffabs/2$ decreases, the linearized $\coth x$ drives the increase of $|\Pi_{\Omega}(T)|$ with a slope $\propto 1/\weffabs$; hence, samples with higher dislocation density (larger $\Omega$) are \emph{stiffer} in the sense that $\Pi_{\Omega}(T)$ rises more slowly.

Equation~\eqref{eq:chi_Omega} exhibits the same thermal kernel as the magnetic susceptibility, $\csch^{2}x$, differing only by its prefactor. Normalizing by $\chi_{\Omega 0}(T)$ isolates this kernel in Fig.~\ref{fig:chiOmega_vs_Omega}: the curves start from zero (frozen lattice), reach one-half at $k_{B}T\simeq 0.88\,\hbar\weffabs$,
and decay as $1/T$ for $k_{B}T\gg\hbar\weffabs$. Increasing $\Omega$ shifts the crossover to higher temperatures, thereby reducing sensitivity
to defect-density fluctuations in the experimentally relevant regime.

\paragraph*{Remarks on the compensation point.}
At $\weff=0$ the derivative $\partial\mathrm{sgn}(\weff)/\partial\Omega$
is singular; Eqs.~\eqref{eq:Pi_Omega}-\eqref{eq:chi_Omega} apply for
$\weff\neq 0$, i.e.\ away from the compensated-field line.
In practice, disorder and level broadening smear this singularity,
and the trends in Figs.~\ref{fig:Pi_vs_Omega}-\ref{fig:chiOmega_vs_Omega}
remain as shown.

Together, Eqs.~\eqref{eq:Pi_Omega}-\eqref{eq:chi_Omega} and
Figs.~\ref{fig:Pi_vs_Omega}-\ref{fig:chiOmega_vs_Omega} provide a
thermodynamic portrait of magnetoelastic coupling:
$\Pi_{\Omega}$ measures the \emph{work} required to change the
dislocation density, whereas $\chi_{\Omega}$ quantifies the
sensitivity of that work to thermal agitation.
Both can, in principle, be accessed via high-resolution dilatometry under controlled defect injection, offering a direct probe of the
pseudomagnetic role of torsion in crystalline media.

\subsection{Magnetocaloric Effect}\label{subsec:magnetocaloric}

Framed by the Maxwell identity
$\bigl(\partial S/\partial B\bigr)_T=-\bigl(\partial M/\partial T\bigr)_B$ and the magnetic Gr{\"u}neisen definition
$\Gamma_B\equiv -\tfrac{1}{T}(\partial T/\partial B)_S=\tfrac{1}{C_V}(\partial M/\partial T)_B$, the \emph{magnetocaloric effect} (MCE) provides a direct thermodynamic probe of field-entropy coupling.
This route is standard in the MCE literature \cite{TishinSpichkinBook2003,JMMM.1999.200.44,GSCHNEIDNER2008945,JPD.2005.38.R381}
and underpins modern high-resolution protocols for adiabatic field sweeps and alternating-field thermometry in correlated and low-dimensional systems \cite{TokiwaGegenwart2011}.
Within our elastic-Landau framework, the same single scale $x=\hbar|\omega_{\mathrm{eff}}|/(2k_{B}T)$ that governs $C_V$ and $M$ controls $\Gamma_B(T)$:
the fixed-$k$ baseline exhibits an exact kernel cancellation, while the $k$-integrated (3D) case adds a universal $+\tfrac12 k_B$ to $C_V$, yielding a smooth crossover to the asymptotic plateau.

In our single-scale framework, all temperature dependences factor through the gap variable $x=\hbar|\omega_{\mathrm{eff}}|/(2k_BT)$, so that for the fixed-$k$ spectrum the thermal kernel cancels between
$(\partial M/\partial T)_B$ and $C_V$, yielding a strictly linear law
$\Gamma_B\propto 1/\omega_{\mathrm{eff}}$. Including the free longitudinal motion restores the bulk 3D limit by adding the universal $+\tfrac{1}{2}k_B$ to $C_V$, which breaks the cancellation and produces a smooth crossover to a constant high-$T$ plateau. Beyond our specific model, $\Gamma_B$ is widely used as a sensitive probe of field-tuned energy scales and quantum criticality, where it acquires universal scaling and may even diverge or change sign \cite{ZhuGarstRoschSi2003}.

Under adiabatic conditions ($dS=0$), the temperature change induced by a small field variation is \cite{ECMP.2005.236,IJAR.2023.31.5,PhysRevApplied.2023.20.014053}
\begin{equation}
\Bigl(\tfrac{\partial T}{\partial B}\Bigr)_{S}
= -\,\frac{T}{C_{V}}\,
   \Bigl(\tfrac{\partial M}{\partial T}\Bigr)_{B},
\label{eq:MCE} % keep legacy label for cross-refs
\end{equation}
which follows from the Maxwell identity
$\bigl(\partial S/\partial B\bigr)_{T}
 = -\bigl(\partial M/\partial T\bigr)_{B}$.
For a single harmonic mode with free energy  $A(T)=k_{B}T\ln[2\sinh x]+\text{const}$ with
\[
x=\frac{\beta\hbar|\omega_{\mathrm{eff}}|}{2}
=\frac{\hbar|\omega_{\mathrm{eff}}|}{2k_{B}T},
\]
and using the magnetization from Eq.~\eqref{eq:M_theory},
\begin{equation}
M(B,\Omega,T)=
-\,\mathrm{sgn}(\omega_{\mathrm{eff}})\,\frac{\hbar e}{2\mu}\,\coth x
\;,
\end{equation}
the temperature derivative at fixed $B$ is
\begin{align}
&\Bigl(\tfrac{\partial M}{\partial T}\Bigr)_{B}
= -\,\mathrm{sgn}(\omega_{\mathrm{eff}})
    \frac{\hbar e}{2\mu}\,
    \frac{d}{dT}\bigl[\coth x\bigr]
    \;%+\;\mathrm{sgn}(\omega_{\mathrm{eff}})\,\frac{e}{\mu}\,
    %\frac{k_{B}}{|\omega_{\mathrm{eff}}|}
    \notag\\
&= -\,\mathrm{sgn}(\omega_{\mathrm{eff}})
    \frac{\hbar e}{2\mu}\,
    \bigl[-\csch^{2}x\bigr]\,
    \frac{dx}{dT}
    \;%+\;\mathrm{sgn}(\omega_{\mathrm{eff}})\,\frac{e}{\mu}\,
    %\frac{k_{B}}{|\omega_{\mathrm{eff}}|}
    \notag\\
&= -\,\mathrm{sgn}(\omega_{\mathrm{eff}})
    \frac{\hbar e}{2\mu}\,
    \frac{\hbar|\omega_{\mathrm{eff}}|}{2k_{B}T^{2}}\,
    \csch^{2}x
    \;%+\;\mathrm{sgn}(\omega_{\mathrm{eff}})\,\frac{e}{\mu}\,
    %\frac{k_{B}}{|\omega_{\mathrm{eff}}|}.
\label{eq:dMdT}
\end{align}
\begin{figure}[tbhp]
\centering
\includegraphics[width=\linewidth]{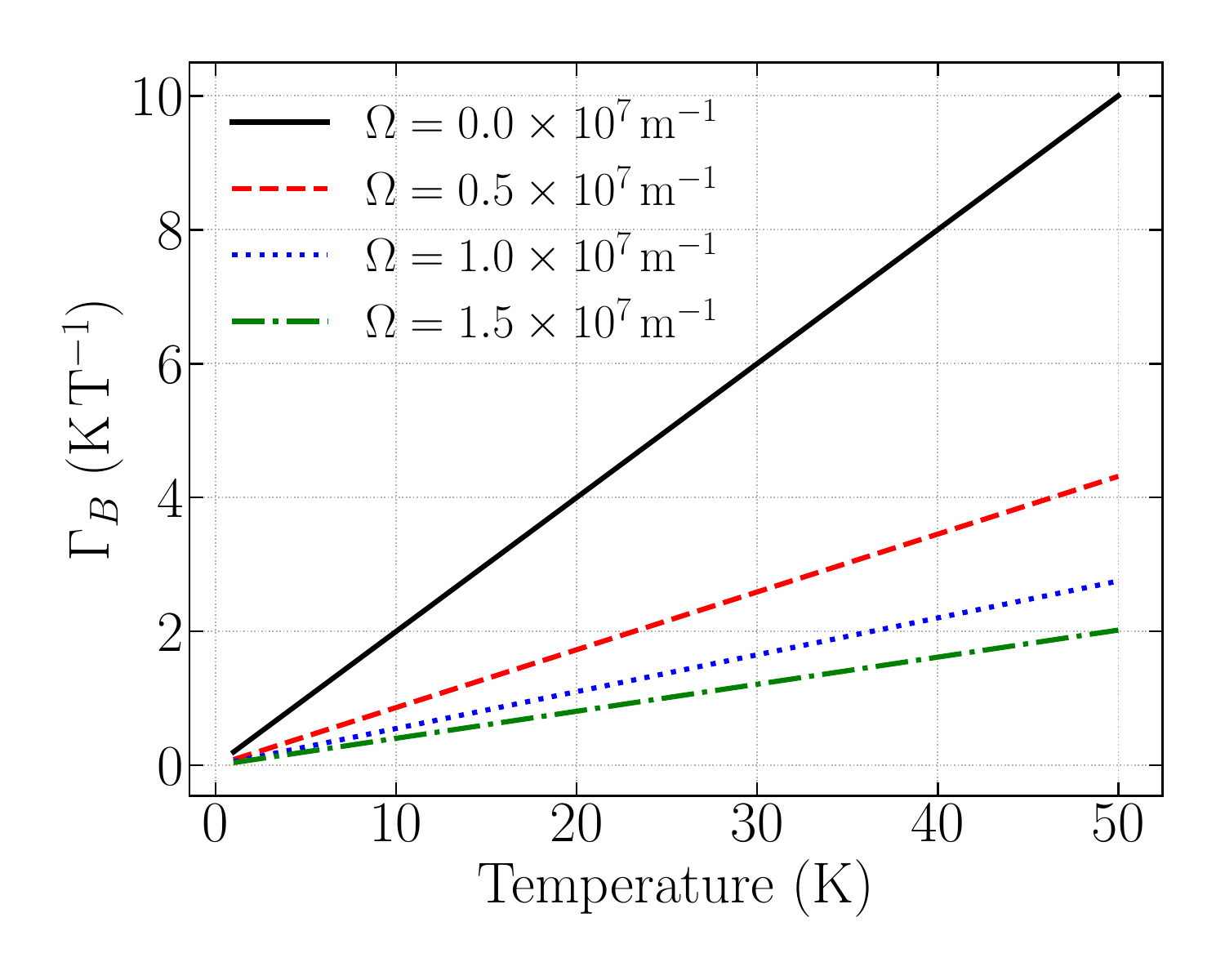}
\caption{\footnotesize
Adiabatic magnetocaloric coefficient
$\Gamma_{B}(T)=\bigl(\partial T/\partial B\bigr)_{S}$ for a fixed
magnetic field ($B = 5\,\mathrm{T}$) and four torsion densities
($\Omega = 0$, $0.5$, $1.0$, and
$1.5 \times 10^{7}\,\mathrm{m^{-1}}$) shown in black, red, blue, and green, respectively.
In the fixed-$k$ baseline (holding $g$ constant), the thermal kernel cancels between
$(\partial M/\partial T)_{B}$ and $C_{V}$, so
Eq.~\eqref{eq:MCE} reduces to the exact linear law
$\Gamma_{B}(T)=\dfrac{e}{\mu}\,\dfrac{T}{\omega_{\mathrm{eff}}}$
[cf.\ Eq.~\eqref{eq:MCE_linearT}], yielding straight lines whose slope
decreases as $|\omega_{\mathrm{eff}}|$ increases (larger $\Omega$).
Including the longitudinal degree of freedom replaces $C_{V}$ by
$C_{V}^{3\mathrm{D}}$ producing the smooth crossover described by
Eq.~\eqref{eq:MCE_3D}, with the high-$T$ asymptote reduced by a
factor $2/3$ [Eq.~\eqref{eq:MCE_3D_highT}].}
\label{fig:magnetocaloric_vs_Omega}
\end{figure}

Inserting Eq.~\eqref{eq:dMdT} into Eq.~\eqref{eq:MCE} gives the compact, signed form

\begin{align}
\Bigl(\tfrac{\partial T}{\partial B}\Bigr)_{S}
=& \mathrm{sgn}(\omega_{\mathrm{eff}})
   \frac{\hbar e}{2\mu}\,
   \frac{\hbar|\omega_{\mathrm{eff}}|}{2k_{B}}
   \frac{1}{T\,C_{V}}\,
   \csch^{2}x \notag\\&%-   
   %\mathrm{sgn}(\omega_{\mathrm{eff}})\,\frac{e}{\mu}\,
   %\frac{T}{|\omega_{\mathrm{eff}}|}\,
   %\frac{k_{B}}{C_{V}}
   \notag\\[4pt]
%&= \xcancel{\mathrm{sgn}(\omega_{\mathrm{eff}})
%   \frac{\hbar e}{2\mu k_{B}}\,
%   \frac{x\,\csch^{2}x}{C_{V}/k_{B}}.
%   \notag}\\%\\&-
  % \mathrm{sgn}(\omega_{\mathrm{eff}})\,\frac{e}{\mu}\,
  % \frac{T}{|\omega_{\mathrm{eff}}|}\,
   %\frac{1}{C_{V}/k_{B}},
   &= \mathrm{sgn}(\omega_{\mathrm{eff}})
   \frac{\hbar e}{2\mu }\,
   \frac{x\,\csch^{2}x}{C_{V}}.
  % \notag
\label{eq:MCE_xform}
\end{align}

% \rev{Equivalently, with $\beta=1/k_{B}T$ one may write}
% \rev{\begin{align}
% (\partial T/\partial B)_{S}
%  = \mathrm{sgn}(\omega_{\mathrm{eff}})
%    \dfrac{\hbar^{2}e}{4\mu}&\,
%    \dfrac{\beta\,\csch^{2}x}{C_{V}}
%   \notag\\%&-
% %   \mathrm{sgn}(\omega_{\mathrm{eff}})\,\dfrac{e}{\mu}\,
% %   \dfrac{T}{|\omega_{\mathrm{eff}}|}\,
% %   \dfrac{1}{C_{V}}.  
% \end{align}}

\paragraph*{Fixed-$k$ spectrum.}
In the fixed-$k$ baseline (degeneracy $g$ held constant), we have $C_{V}/k_{B}=x^{2}\csch^{2}x$. Using the Eq.~\eqref{eq:MCE_xform}, the thermal kernel cancels, yielding
\begin{align}
\Bigl(\tfrac{\partial T}{\partial B}\Bigr)_{S}^{\mathrm{fixed}\,k}
&= \mathrm{sgn}(\omega_{\mathrm{eff}})
   \frac{\hbar e}{2\mu k_{B}}\,
   \frac{x}{x^{2}}
 = \mathrm{sgn}(\omega_{\mathrm{eff}})\,
   \frac{\hbar e}{2\mu k_{B}}\,
   \frac{1}{x} \notag\\
&= \mathrm{sgn}(\omega_{\mathrm{eff}})\,
   \frac{e}{\mu}\,
   \frac{T}{|\omega_{\mathrm{eff}}|}
 = \frac{e}{\mu}\,\frac{T}{\omega_{\mathrm{eff}}}.
\label{eq:MCE_linearT}
\end{align}
If, instead, one keeps the physical $B$-dependence of the Landau-like degeneracy ($g\propto|\omega_{\mathrm{eff}}|$), the fixed-$k$ result acquires a multiplicative correction:
\[
\Bigl(\tfrac{\partial T}{\partial B}\Bigr)_{S}^{\mathrm{fixed}\,k,\;g(B)}
=
\frac{e}{\mu}\,\frac{T}{\omega_{\mathrm{eff}}}\,
\Biggl[\,1-\frac{1}{x^{2}\csch^{2}x}\,\Biggr],
\]
which smoothly suppresses the linear law as $T$ increases and will be regularized once the longitudinal motion is included.

\paragraph*{3D gas (including longitudinal motion).}
For a bulk system, integrating the free longitudinal motion adds a universal $+\tfrac12 k_{B}$ to the heat capacity (Sec.~\ref{subsec:k_integration_details}),
\[
C_{V}^{3\mathrm{D}}/k_{B}=x^{2}\csch^{2}x+\tfrac12 .
\]
Using this in Eq.~\eqref{eq:MCE_xform} (and adopting the $g$-constant baseline so that only the first term contributes) gives
\begin{equation}
\Bigl(\tfrac{\partial T}{\partial B}\Bigr)_{S}^{3\mathrm{D}}
=
\mathrm{sgn}(\omega_{\mathrm{eff}})
\frac{\hbar e}{2\mu k_{B}}\,
\frac{x\,\csch^{2}x}{x^{2}\csch^{2}x+\tfrac12}.
\label{eq:MCE_3D}
\end{equation}
Its limits are transparent:
(i) \emph{Low $T$ (large $x$)}: $x^{2}\csch^{2}x\to0$ so $(\partial T/\partial B)_{S}^{3\mathrm{D}}\to0$ exponentially.
(ii) \emph{High $T$ (small $x$)}: using $x^{2}\csch^{2}x = 1 - x^{2}/3 + \cdots$,
\begin{align}
\Bigl(\tfrac{\partial T}{\partial B}\Bigr)_{S}^{3\mathrm{D}}
&\xrightarrow[T\to\infty]{}
\mathrm{sgn}(\omega_{\mathrm{eff}})
\frac{\hbar e}{2\mu k_{B}}\,
\frac{x(1/x^{2})}{1+\tfrac12}\notag\\&
= \mathrm{sgn}(\omega_{\mathrm{eff}})
\frac{\hbar e}{3\mu k_{B}}\,
\frac{1}{x} \notag\\
&= \mathrm{sgn}(\omega_{\mathrm{eff}})\,
\frac{2}{3}\,\frac{e}{\mu}\,
\frac{T}{|\omega_{\mathrm{eff}}|}
= \frac{2}{3}\,\frac{e}{\mu}\,\frac{T}{\omega_{\mathrm{eff}}}.
\label{eq:MCE_3D_highT}
\end{align}
Hence, the longitudinal degree of freedom reduces the linear-in-$T$ slope by a factor $2/3$ at high temperature and smooths the crossover.

Because the same thermal kernel cancels between $(\partial M/\partial T)_{B}$ and $C_{V}$ in the fixed-$k$ baseline, Eqs.~\eqref{eq:MCE}-\eqref{eq:MCE_linearT} imply a strictly linear law with magnitude inversely proportional to the cyclotron gap $|\omega_{\mathrm{eff}}|$: a larger screw-dislocation density (larger $\Omega$) suppresses the adiabatic temperature change. Using $k=10^{9}\,\mathrm{m^{-1}}$ and $T=40\,\mathrm{K}$, one finds $(\partial T/\partial B)_{S}\approx 8\,\mathrm{K\,T^{-1}}$ for $\Omega=0$, decreasing to $\approx 1.6\,\mathrm{K\,T^{-1}}$ for $\Omega=1.5\times10^{7}\,\mathrm{m^{-1}}$. For a bulk 3D gas, the longitudinal contribution shifts the high-$T$ slope to the $2/3$ value in Eq.~\eqref{eq:MCE_3D_highT}.

\subsection{Integration over the longitudinal momentum \texorpdfstring{$k$}{k}}
\label{subsec:k_integration_details}

Up to this point, we have kept the longitudinal momentum $k$ \emph{fixed},
emphasizing the Landau-type quantization in the transverse plane.
For a real three-dimensional gas, however, the motion along the $z$-axis is free, so the
canonical partition function must be integrated over the continuum
$k\in(-\infty,\infty)$.

\paragraph*{Two standard ingredients of the derivation.}
(i) \emph{Separation of degrees of freedom.} The spectrum factorizes as
\begin{align}
E_{N',k}= \hbar\weffabs\Big(N'+\tfrac12\Big) + \frac{\hbar^2 k^2}{2\mu},    
\end{align}
so the fixed-$k$ partition function reads
\begin{align}
Z(T,k)&= g\,\sum_{N'=0}^{\infty}
e^{-\beta\hbar\weffabs(N'+1/2)}\,e^{-\beta\hbar^2 k^2/2\mu}
\notag\\&= g\,\frac{e^{-\beta\hbar\weffabs/2}}{1-e^{-\beta\hbar\weffabs}}\;
e^{-\beta\hbar^2 k^2/2\mu}.    
\end{align}
Using the identity $e^{-x}/(1-e^{-2x})=1/[2\sinh(x)]$ with $x=\beta\hbar\weffabs/2$, we obtain the compact form
\begin{equation}
Z(T,k)=
g\,\frac{1}{2\sinh(\beta\hbar\weffabs/2)}\;
e^{-\beta\hbar^2 k^2/2\mu}.
\label{eq:ZTk_compact}
\end{equation}
(ii) \emph{From sum to integral in $k$ (1D density of states).}
Imposing periodic boundary conditions in a box of length $L$ along $z$ gives
$k_n=2\pi n/L$ ($n\in\mathbb{Z}$), and in the thermodynamic limit
\begin{equation}
\sum_{n} f(k_n)\;\longrightarrow\; \frac{L}{2\pi}\int_{-\infty}^{\infty} f(k)\,dk.  
\end{equation}

\paragraph*{Carrying out the $k$-integration.}
Combining Eq. \eqref{eq:ZTk_compact} with the 1D density of states, we arrive at
\begin{align}
Z_{3\mathrm{D}}(T)
 &= \frac{L}{2\pi}
    \int_{-\infty}^{\infty}dk\;
    Z(T,k) \notag\\
 &= \frac{gL}{2\pi}
    \frac{1}{2\sinh(\beta\hbar\weffabs/2)}
    \int_{-\infty}^{\infty}
      dk\,e^{-\beta\hbar^{2}k^{2}/2\mu}.
\label{eq:Z3D_step1}
\end{align}
The remaining Gaussian integral equals
$
\sqrt{2\pi\mu/(\beta\hbar^{2})}
$,
which suggests introducing the thermal de Broglie wavelength
\begin{equation}
\lambda_{T}
   =\sqrt{\frac{2\pi\hbar^{2}\beta}{\mu}}
   =\sqrt{\frac{2\pi\hbar^{2}}{\mu k_{B}T}}.
\label{eq:lambdaT}
\end{equation}
Substituting the Gaussian result into Eq. \eqref{eq:Z3D_step1} and using
$\sqrt{2\pi\mu/(\beta\hbar^{2})}=1/\lambda_{T}$ cancels the factor
$2\pi$, yielding the compact three-dimensional partition function
\begin{equation}
Z_{3\mathrm{D}}(T)=
   \frac{gL}{\lambda_{T}}\,
   \frac{1}{2\sinh\bigl(\beta\hbar\weffabs/2\bigr)}.
\label{eq:Z_3D}
\end{equation}

\paragraph*{Thermodynamic consequences.}
Taking the derivative $U_{3\mathrm{D}}=-\partial_{\beta}\ln Z_{3\mathrm{D}}$ introduces an additional term $\tfrac12 k_{B}T$ that is absent when $k$ is fixed,
\begin{equation}
U_{3\mathrm{D}}(T)=U(T)+\frac12\,k_{B}T,
\label{eq:U3D}
\end{equation}
where $U(T)$ is the fixed-$k$ internal energy given in Eq.~\eqref{eq:internal_energy}.
Consequently,
\begin{equation}
C_{V}^{3\mathrm{D}}(T)=C_{V}(T)+\frac12\,k_{B},
\label{eq:Cv3D}
\end{equation}
reflecting the classical equipartition contribution from the free
quadratic degree of freedom along the $z$-direction.

With fixed $k$, the model behaves as a ``zero-dimensional'' system with a fully discrete spectrum in the transverse plane. Integrating over $k$ restores the continuum of longitudinal states, making the gas effectively one-dimensional and adding a universal $+\tfrac12 k_{B}$ contribution to the heat capacity per particle, in agreement with Eqs.~\eqref{eq:U3D}-\eqref{eq:Cv3D}.
At even higher temperatures, when
$k_{B}T\gg\hbar\weffabs$, the transverse ladder also overlaps and the classical three-dimensional limit $C_{V}\to\tfrac32 k_{B}$ is recovered, consistent with the Bohr-van Leeuwen theorem \cite{RMP.2015.27.1550019,PRA.2009.79.042107,PRE.2023.107.014125,PRR.2025.7.L022068,PRR.2023..5.043223,EPL.2009.87.17002}.
Including the $k$-integration is therefore essential for quantitative
comparison with bulk measurements and clarifies how the system
interpolates between strictly two-dimensional Landau physics and a
three-dimensional ideal gas.

\begin{figure}[tbhp]
\centering
\includegraphics[width=\linewidth]{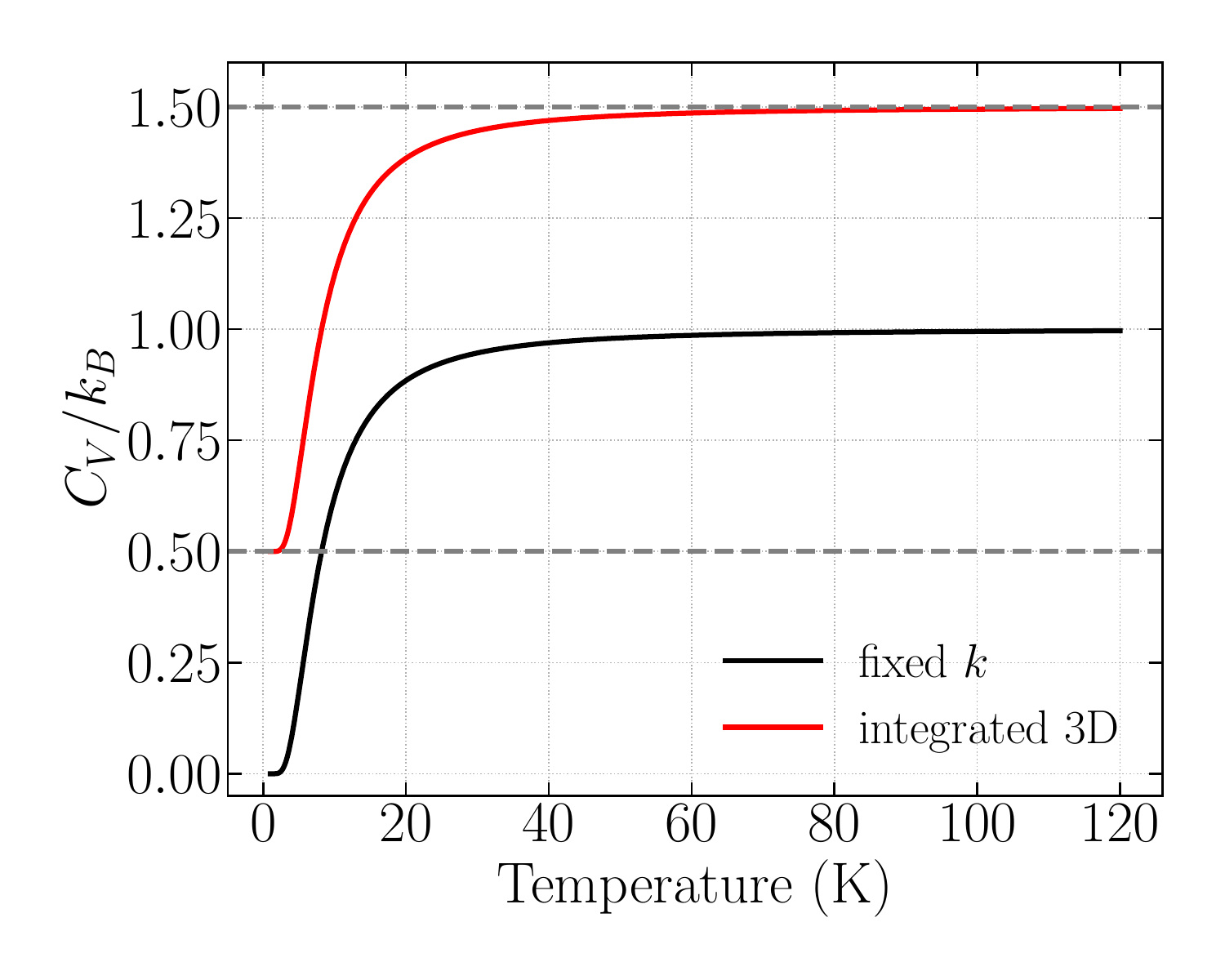}
\caption{\footnotesize
Heat capacity per particle, with and without longitudinal integration. Both curves correspond to the representative parameters $B = 5$\,T, $\Omega = 1.0\times 10^{7}\,\text{m}^{-1}$ and $k = 10^{9}\,\text{m}^{-1}$. The black line shows the fixed-$k$ result $C_{V}(T)$ from Eq.\,\eqref{eq:heat_capacity}, while the red line includes the longitudinal contribution, $C_{V}^{3\mathrm{D}} = C_{V} + \tfrac12 k_{B}$ [Eq.\,\eqref{eq:Cv3D}]. Horizontal dashed lines indicate the offset $\tfrac12 k_{B}$ and the classical limit $\tfrac32 k_{B}$, reached when both longitudinal and transverse continua are fully thermally populated.}
\label{fig:heatcap_k_integration}
\end{figure}

The comparison in Fig.~\ref{fig:heatcap_k_integration}
illustrates at a glance the effect of the longitudinal $k$-integration. The fixed-$k$ curve rises from zero and saturates at
$C_{V}=k_{B}$ once $k_{B}T\gtrsim\hbar\weffabs$, reflecting the two discrete (transverse) degrees of freedom. Including the continuum of longitudinal states shifts the entire heat-capacity curve upward by exactly $\tfrac12 k_{B}$, the equipartition term associated with free motion along the $z$-direction, and thus drives the high-temperature limit toward the classical 3D value $C_{V}\to\tfrac32 k_{B}$. Because the additional $\tfrac12 k_{B}$ contribution is parameter-independent, the offset is universal and should appear in any bulk measurement. Its experimental observation would confirm that the elastic Landau levels indeed coexist with a free longitudinal band. 
Curves labeled “transverse/fixed-$k$” exclude the longitudinal continuum and should be compared to quasi-2D or mesoscopic devices, whereas “3D” curves include the universal offset $+\tfrac12 k_B$ and approach $C_V\to \tfrac32 k_B$ at high $T$.

\subsection{Grand-canonical ensemble and de Haas-van Alphen oscillations}
\label{subsec:grand_canonical}

In this subsection we adopt the \emph{grand-canonical} description and keep the Landau-like degeneracy $g(B,\Omega)$ explicit to analyze quantum oscillations. Our approach follows the Lifshitz-Kosevich route, Poisson resummation \cite{PRX.2018.8.011027,PRB.2010.82.045123} of the Landau-level sum, leading to the thermal damping $R_T=X/\sinh X$ and the disorder (Dingle) factor $R_D=e^{-\pi/(\,|\omega_{\mathrm{eff}}|\,\tau_D)}$, as developed for metals and quasi-2D systems and extended to include level broadening and interlayer warping \cite{LTP.2014.40.270,PRB.2025.111.075131,WassermanSpringford1996,Dingle1952,Champel2001,JPC.1985.18.L783,PhysRevLett.76.1308}. We also connect the oscillatory thermodynamics to transport via the St\v{r}eda relation, which links field derivatives of the grand potential to the Hall conductivity \cite{Streda1982}. Within this framework, torsion enters through the effective-field substitution $B\mapsto B_{\mathrm{eff}}$ (hence phases and damping depend on $|\omega_{\mathrm{eff}}|$), while overall amplitudes carry the explicit sample degeneracy $g(B,\Omega)$.

Here, we work with \emph{sample totals} and therefore keep the Landau-like degeneracy
$g(B,\Omega)$ explicit, with
$g(B,\Omega)=A/(2\pi \ell_{\mathrm{eff}}^{2})$,
$\ell_{\mathrm{eff}}=\sqrt{\hbar/(\mu|\omega_{\mathrm{eff}}|)}$,
and $\omega_{\mathrm{eff}}=\omega_c+\omega_{cl}$. 
Consider non-interacting fermions occupying the elastic-Landau ladder
$
E_{N'}=\hbar\weffabs\bigl(N'+\tfrac12\bigr)
$,
with a macroscopic degeneracy per level $g$ (proportional to the sample area).
In the grand canonical ensemble, the grand potential is
\begin{equation}
\Omega_G(T,\mu_F,B,\Omega)
 = -k_BT\,g\sum_{N'=0}^{\infty}
 \ln\Bigl[1+e^{-\beta\bigl(E_{N'}-\mu_F\bigr)}\Bigr].
\label{eq:grand_potential_full}
\end{equation}
As in the usual Landau problem, the $1/B$-periodic oscillations emerge
once the sum over levels is recast via Poisson summation.

Introduce the discrete density of states (DoS)
$
\rho(E)=g\sum_{N'}\delta\bigl(E-\hbar\weffabs(N'+\tfrac12)\bigr)
$
and rewrite
\begin{equation}
\Omega_G=-k_BT\int_{-\infty}^{\infty} \rho(E)\,
\ln\bigl[1+e^{-\beta(E-\mu_F)}\bigr]\,dE.
\end{equation}

Applying the Poisson summation formula to the series in $N'$ with spacing
$\hbar\weffabs$ (Maslov phase $+1/2$) yields
\begin{align}
\rho(E)
&= \bar{\rho} \;+\;
2\,\bar{\rho}\sum_{r=1}^{\infty}
\cos\Bigl[\tfrac{2\pi r}{\hbar\weffabs}\,E\Bigr],
\label{eq:Poisson_rho}
\end{align}
where $\bar{\rho}= g/(\hbar\weffabs)$ is the smooth (non-oscillatory) part.
Only the second term generates the periodic structure in $1/\weffabs$ (hence in $1/B$).
Substituting Eq. \eqref{eq:Poisson_rho} into $\Omega_G$ and integrating by parts,
we split $\Omega_G=\Omega_{\mathrm{sm}}+\Omega_{\mathrm{osc}}$, with
\begin{equation}
\Omega_{\mathrm{osc}}
= -2\,\bar{\rho}\sum_{r=1}^{\infty}
\int_{-\infty}^{\infty} dE\,\,
\ln\bigl[1+e^{-\beta(E-\mu_F)}\bigr]\,
\cos\Bigl(\tfrac{2\pi r E}{\hbar\weffabs}\Bigr).
\label{eq:Omega_osc_start}
\end{equation}
Differentiating \eqref{eq:Omega_osc_start} with respect to $\mu_F$ and integrating back,
the relevant integral is the cosine transform of the Fermi kernel.
The standard Lifshitz-Kosevich (LK) procedure yields the thermal damping factor
\begin{equation}
R_T(r,T)=
\frac{X_r}{\sinh X_r},
\qquad
X_r=\frac{2\pi^2 r k_B T}{\hbar\weffabs}.
\label{eq:RT}
\end{equation}
Additionally, disorder (Dingle width $\Gamma_D=\hbar/2\tau_D$) produces the exponential damping
\begin{equation}
R_D(r,\tau_D)=
\exp\Bigl(-\frac{\pi r}{\weffabs\,\tau_D}\Bigr),
\label{eq:RD}
\end{equation}
written here in terms of the Dingle time $\tau_D$.

Combining these ingredients, the oscillatory part of the grand potential
assumes the compact form
\begin{align}
\Omega_{\mathrm{osc}}
= -\frac{g\,\mu_F^2}{\hbar\weffabs}
\sum_{r=1}^{\infty}\frac{1}{\pi^2 r^2}\,&
\cos\Bigl(\frac{2\pi r\,\mu_F}{\hbar\weffabs}\Bigr)\notag\\ & \times
R_T(r,T)\,R_D(r,\tau_D),
\label{eq:Omega_osc_final}
\end{align}
where constants of integration are absorbed into the smooth background
$\Omega_{\mathrm{sm}}$, irrelevant for the oscillatory piece.

\paragraph*{Oscillatory magnetization.}
The magnetization follows from $M=-(\partial\Omega_G/\partial B)_{T,\mu_F}$.
Differentiating \eqref{eq:Omega_osc_final} and using
$\partial\weffabs/\partial B = (e/\mu)\,\mathrm{sgn}(\weff)$,
we obtain
\begin{align}
M_{\mathrm{osc}}(B)
&= -\frac{\partial\Omega_{\mathrm{osc}}}{\partial B}
\notag\\[2pt]
&= -\,\mathrm{sgn}(\weff)
\,\frac{g\,\mu_F}{\hbar\weffabs}\,
\sum_{r=1}^{\infty}\frac{1}{\pi r}\,
\sin\Bigl(\frac{2\pi r\,\mu_F}{\hbar\weffabs}\Bigr)\notag\\& \times
R_T(r,T)\,R_D(r,\tau_D).
\label{eq:Mosc_full}
\end{align}
Here, the $\mathrm{sgn}(\weff)$ dependence appears only in the overall prefactor
via $\partial\weffabs/\partial B$, while the phase and denominators involve
$\weffabs$. This keeps the correct periodicity in $1/B$ even if $\weff$ can cross zero
(near the compensation condition).

\paragraph*{Practical limits and magnetization oscillations (dHvA)} \cite{PRB.2023.108.L201103,PRB.2002.65.205405,AdP.2019.531.1900254,PE.2021.132.114760,AdP.2022.535.2200371}.
(i) In the low-temperature, high-field regime
$k_BT\ll \hbar\weffabs$ (i.e., $X_r\ll1$),
$R_T\to 1$.
(ii) When higher harmonics are suppressed by temperature and/or disorder,
one may retain only $r=1$.
Under these conditions, \eqref{eq:Mosc_full} reduces to
\begin{equation}
M_{\mathrm{osc}}(B)\;\simeq\;
-\,\mathrm{sgn}(\weff)\,
\frac{g\,\mu_F}{\hbar\weffabs}\;
\sin\Bigl(\frac{2\pi \mu_F}{\hbar\weffabs}\Bigr)\;
e^{-\pi/(\weffabs\tau_D)}.
\label{eq:Mosc_simple}
\end{equation}
If, in addition, one adopts the usual convention $B>0$ and $\weff>0$,
so that $\mathrm{sgn}(\weff)=1$, this reproduces precisely the
expression used in the main text:
\begin{equation}
M_{\mathrm{osc}}(B)=
-\frac{g\,\mu_{F}}{\hbar\weffabs}
\sin\Bigl(\frac{2\pi \mu_{F}}{\hbar\weffabs}\Bigr)
e^{-\pi/(\weffabs\tau_{D})}.\label{eq:dHvA}
\end{equation}

\paragraph*{Period in $1/B$ and the role of torsion.}
The phase of the $r$-th harmonic is
$\varphi_r=2\pi r\,\mu_F/(\hbar\weffabs)$.
Since $\weff=\omega_c+\omega_{cl}
= eB/\mu + 2\hbar k\Omega/\mu$,
the inverse gap $1/\weffabs$ is approximately linear in $1/B$ when $|\omega_c|\gg|\omega_{cl}|$,
but the \emph{period is compressed} as $\Omega$ grows (or as $k$ increases), through the sum
$\omega_c+\omega_{cl}$ in the denominator. In addition, the amplitude is reduced by two
mechanisms: the prefactor $1/\weffabs$ and the Dingle damping
$\exp[-\pi/(\weffabs\tau_D)]$.

\paragraph*{Remarks on signs.}
\emph{(a)} Using $\weffabs$ in the phase and denominators guarantees the correct periodicity
even near the compensation point ($\weff=0$).
\emph{(b)} The only remaining sign sensitivity is global and originates from
$\partial\weffabs/\partial B=(e/\mu)\,\mathrm{sgn}(\weff)$ in the step
$M=-\partial\Omega/\partial B$.
If one fixes the convention $\weff>0$ (choose the field orientation so that the effective cyclotron frequency is positive), the $\mathrm{sgn}(\weff)$ factor in Eqs.
\eqref{eq:Mosc_full}-\eqref{eq:Mosc_simple} can be dropped.

\begin{figure}[tbhp]
  \centering
  \includegraphics[width=\linewidth]{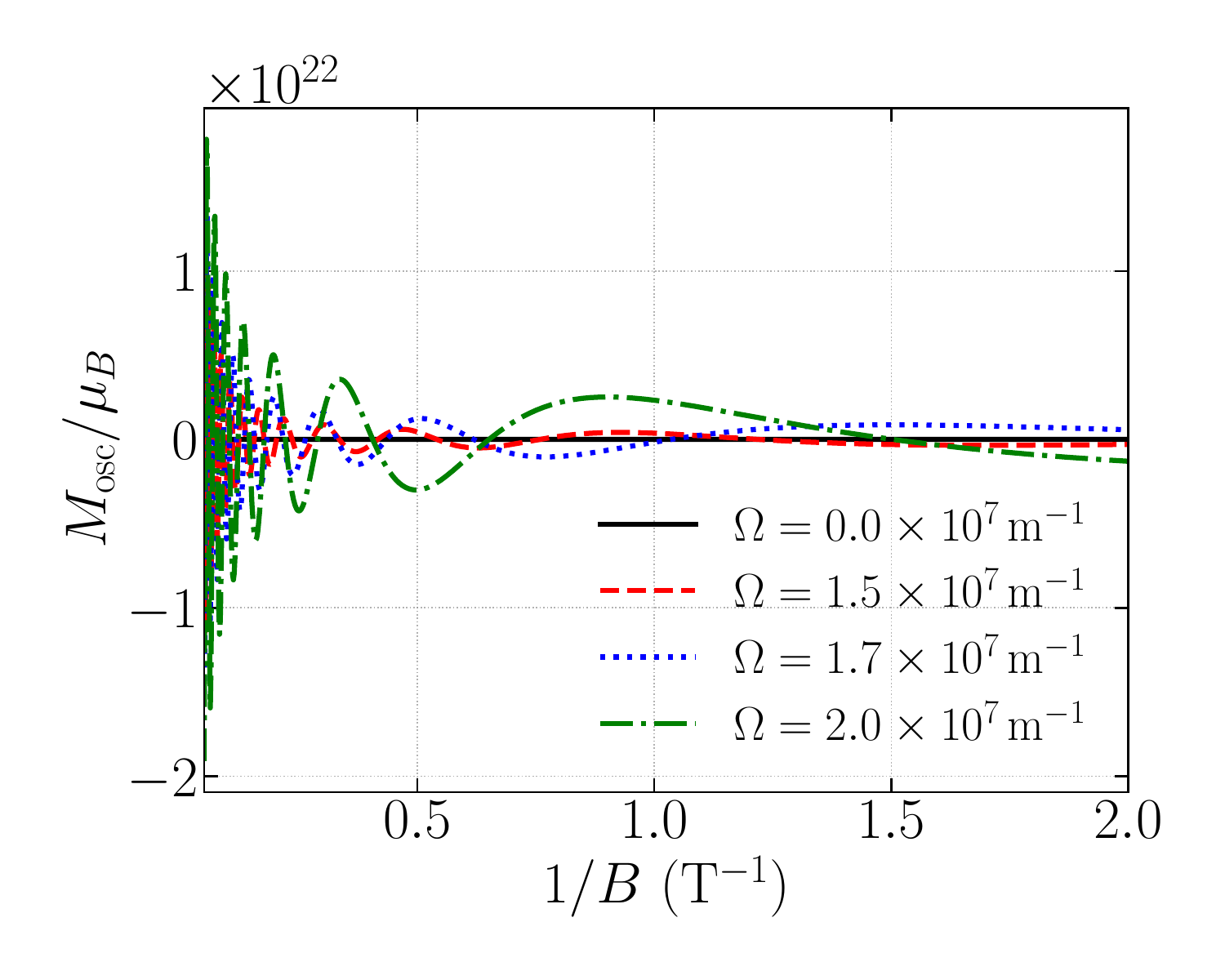}
  \caption{\footnotesize
  de Haas-van Alphen oscillations of the orbital magnetization.
  The curves show the normalized oscillatory component
$M_{\mathrm{osc}}/(10^{22}\mu_{B})$ as a function of the inverse magnetic
  field $1/B$ for four torsion densities
  ($\Omega = 0$, $1.5$, $1.7$, and
  $2.0 \times 10^{7}\,\mathrm{m^{-1}}$) shown in black, red, blue, and green, respectively.
  Parameters:
  longitudinal wave number $k=10^{9}\,\mathrm{m^{-1}}$,
  Fermi energy $\mu_{F}=0.05$ eV, and
  Dingle time $\tau_{D}=0.10$ ps.
  The analytical expression follows Eq.\,\eqref{eq:dHvA};
  the $10^{22}$ factor simply rescales the ordinate so that all traces
  fit within the same panel.
  Increasing $\Omega$ enhances the effective cyclotron frequency
$\omega_{\mathrm{eff}}=\omega_{c}+2\hbar k\Omega/\mu$, thereby shortening the oscillation period in $1/B$ and reducing the
  amplitude through both the $1/\omega_{\mathrm{eff}}$ prefactor and
  the Dingle damping factor
  $\exp[-\pi/(\weffabs\tau_{D})]$.}
  \label{fig:dHvA}
\end{figure}

Figure~\ref{fig:dHvA} clearly illustrates how screw-dislocation density
modifies the quantum magnetic oscillations.
Because the phase of the sine term in Eq.\,\eqref{eq:dHvA} is
$2\pi\mu_{F}/\hbar\weffabs$, a larger
$\Omega$, and hence a larger $\weffabs$, compresses the period in $1/B$.
At the same time, the amplitude decreases as
$1/\weffabs$ and is further suppressed by the Dingle factor, reflecting that torsion increases the Landau-like level
spacing and thereby reduces the density of states at the Fermi energy. Detecting this systematic shift provides a practical means to measure the screw dislocation density via high-precision de Haas-van Alphen experiments: the slope of the colored traces in $1/B$ space yields $\weff$, while their envelope reveals the interplay between topological strain and disorder broadening.

\begin{figure}[tbhp]
\centering
\includegraphics[width=0.45\textwidth]{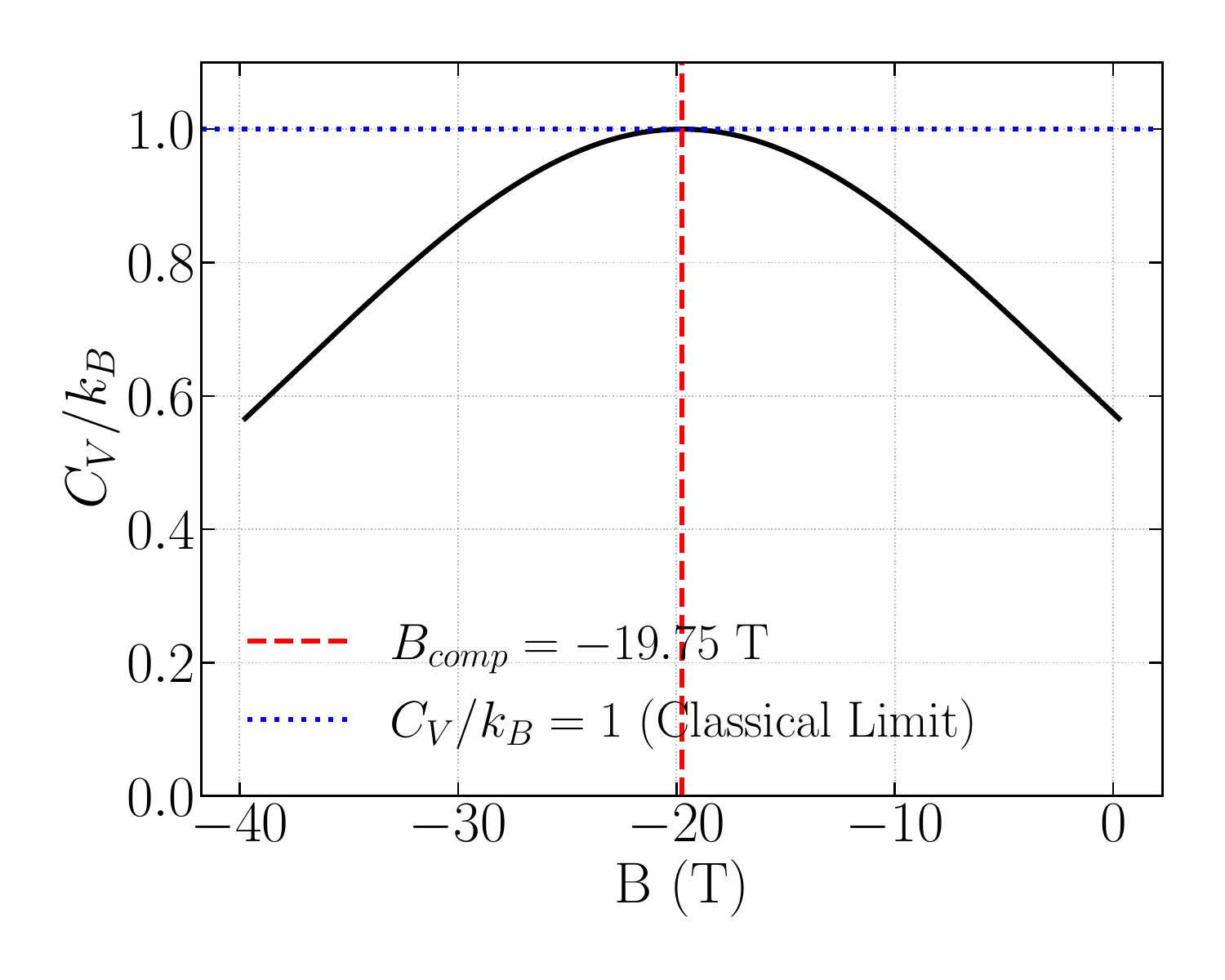}
\caption{\footnotesize Heat capacity $C_V$ as a function of the magnetic field $B$ at a fixed temperature of $T=10\,$K and for a torsion parameter $\Omega = 1.5 \times 10^7 \, \mathrm{m^{-1}}$. The plot displays a distinct peak at the compensation field $B_{\mathrm{comp}} \approx -19.75\,$T (red dashed line), where the heat capacity reaches the classical limit $C_V = k_B$ (blue dotted line). This feature illustrates the crossover from a quantum-gapped system to quasi-free-particle behavior.}
\label{fig:compensated_field}
\end{figure}

\subsection{Topological transport: integer quantum Hall plateaus with torsion}
\label{subsec:iqhe_torsion}

Again, we work with \emph{sample totals} and therefore keep the Landau-like degeneracy
$g(B,\Omega)$ explicit (cf. Eqs.~\eqref{eq:degeneracy} and \eqref{eq:weff_def} for
$\ell_{\mathrm{eff}}$ and $\omega_{\mathrm{eff}}$).
In disordered 2D electron systems, the integer quantum Hall effect (IQHE) emerges from localized bulk states and delocalized critical energies at Landau-level centers. Within our framework, torsion shifts the transverse ladder via the effective cyclotron
frequency $\omega_{\mathrm{eff}}$ (Sec.~\ref{sec2a}), which can be recast as a \emph{synthetic} magnetic field (Eq.~\eqref{eq:Beff_mapping}).
\begin{equation}
\omega_{\mathrm{eff}}=\frac{e}{\mu}\,B_{\mathrm{eff}},\qquad
B_{\mathrm{eff}}\equiv B + B_{\mathrm{tor}},\qquad
B_{\mathrm{tor}}=\frac{2\hbar k\,\Omega}{e}.
\label{eq:Beff_mapping}
\end{equation}
Thus, at fixed density, torsion rigidly shifts all quantum-Hall landmarks by $B_{\mathrm{tor}}$, compressing the $1/B$ periods (as already seen in dHvA) and displacing plateau transitions.

\paragraph*{St\v{r}eda relation from the grand potential \cite{PRB.2015.91.035423,PRB.2006.73.073304,JPA.2015.48.365002}.}
Starting from the grand potential in Eq.~\eqref{eq:grand_potential_full}
(or its broadened DOS version, Sec.~\ref{subsec:disorder_interactions}),
the St\v{r}eda formula $\sigma_{xy}=e(\partial n/\partial B)_{\mu,T}$
can be written as
\begin{align}
\sigma_{xy}(T,\mu_F,B,\Omega)
&= -\,e\,\frac{\partial^2 \Omega_G}{\partial \mu_F\,\partial B}\Bigg|_{T}\notag\\
&= -\,e\,\frac{\partial}{\partial B}
\int dE\,\rho(E)\,f(E-\mu_F),
\label{eq:Streda_from_Omega}
\end{align}
where $f$ is the Fermi function and $\rho(E)$ the effective DOS.
At $T\to0$, for a set of broadened Landau bands carrying Chern number $C_{n}=1$, this reduces to the standard quantization
\begin{equation}
\sigma_{xy}(0,\mu_F,B,\Omega)
= \frac{e^2}{h}\,\nu,\qquad
\nu=\sum_{n=0}^{\infty}\Theta\big(\mu_F-E_n(B_{\mathrm{eff}})\big),
\label{eq:sigma_xy_step}
\end{equation}
with $E_n=\hbar|\omega_{\mathrm{eff}}|(n+\tfrac12)$ and $\nu$ the filling factor.
Finite $T$ and disorder enter through a thermal/Dingle smoothing of the steps via
$\rho(E)$ and $f$.
\begin{figure}[tbhp]
  \centering
  \includegraphics[width=\linewidth]{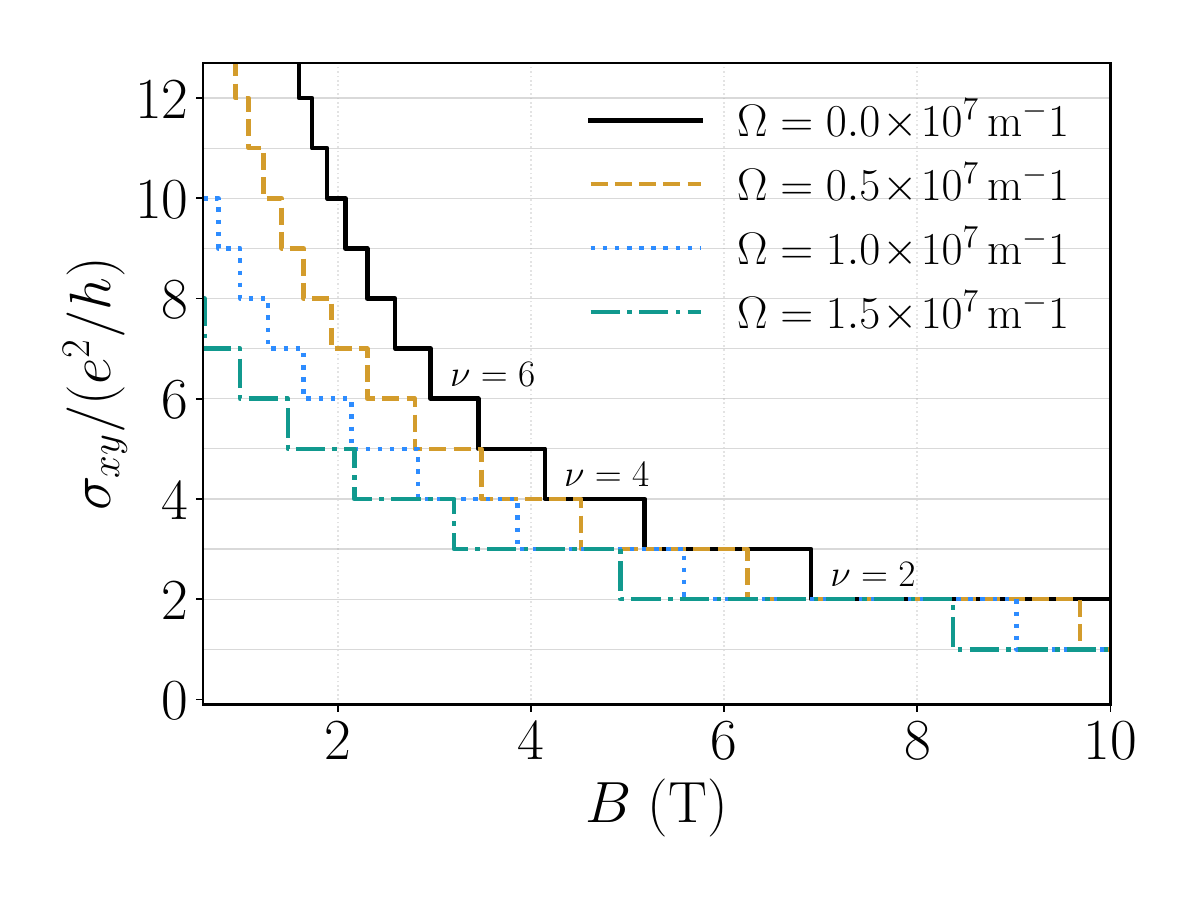}
  \caption{\footnotesize
  Integer-quantized Hall conductivity as a function of magnetic field. The curves show $\sigma_{xy}(B)$ in units of $e^{2}/h$ for a 2D electron gas with longitudinal wave number $k=10^{8}\,\mathrm{m^{-1}}$ and areal density   $n_{2D}=5\times10^{15}\,\mathrm{m^{-2}}$, for four torsion densities ($\Omega=0$, $0.5$, $1.0$, and $1.5\times10^{7}\,\mathrm{m^{-1}}$) shown in black, orange, blue, and green, respectively. The step-like structure reflects the integer filling factor $\nu(B,\Omega)=2\pi\hbar n_{2D}/\bigl(\mu\,|\omega_{\mathrm{eff}}|\bigr)$, with the effective cyclotron frequency $\,\omega_{\mathrm{eff}}=\omega_{c}+\omega_{cl}
  = eB/\mu + 2\hbar k\Omega/\mu$. Increasing $\Omega$ enhances $|\omega_{\mathrm{eff}}|$ and thereby compressing $\nu(B)$: the plateau-to-plateau transitions shift to higher $B$, and the overall $\sigma_{xy}$ decreases at fixed field. Thin horizontal guide lines mark integer values of $\nu$, and annotations highlight a few plateaus ($\nu=6,4,2$) for the $\Omega=0$ trace. The curves correspond to the $T\to 0$ (disorder-free) limit; weak disorder would merely round the steps without altering their systematic shift with~$\Omega$.}
  \label{fig:sigma_xy_plateaus}
\end{figure}

\paragraph*{Plateau transitions and critical fields.}
Plateau transitions (localization-delocalization) occur when $\mu_F$ crosses the
center of a broadened elastic-Landau level,
$\mu_F=E_n(B_{\mathrm{eff}})$.
Solving for $B$ gives an \emph{exactly linear} torsion shift of the critical fields:
\begin{equation}
B_c^{(n)}(\Omega)
= \frac{\mu}{e}\,\frac{2\mu_F}{\hbar(2n+1)}\;-\;B_{\mathrm{tor}}
= \frac{\mu}{e}\,\frac{2\mu_F}{\hbar(2n+1)}\;-\;\frac{2\hbar k\,\Omega}{e}.
\label{eq:Bc_linear_shift}
\end{equation}
Hence, increasing $\Omega$ translates the entire plateau fan rigidly with
slope $-2\hbar k/e$, while preserving the spacings in $1/B_{\mathrm{eff}}$.

Figure~\ref{fig:sigma_xy_plateaus} displays the sequence of integer plateaus in the Hall conductivity produced by the elastic-Landau spectrum. As $\Omega$ increases, the elastic contribution $\omega_{cl}\propto\Omega$ enlarges the effective cyclotron gap, so the filling factor $\nu(B,\Omega)$ decreases faster with $B$ and the plateau transitions shift rightward. This torsion-controlled compression of the $\nu$ ladder is the direct transport analog of the thermodynamic trends governed by the single scale $\hbar|\omega_{\mathrm{eff}}|$.

\begin{figure}[tbhp]
  \centering
  \includegraphics[width=\linewidth]{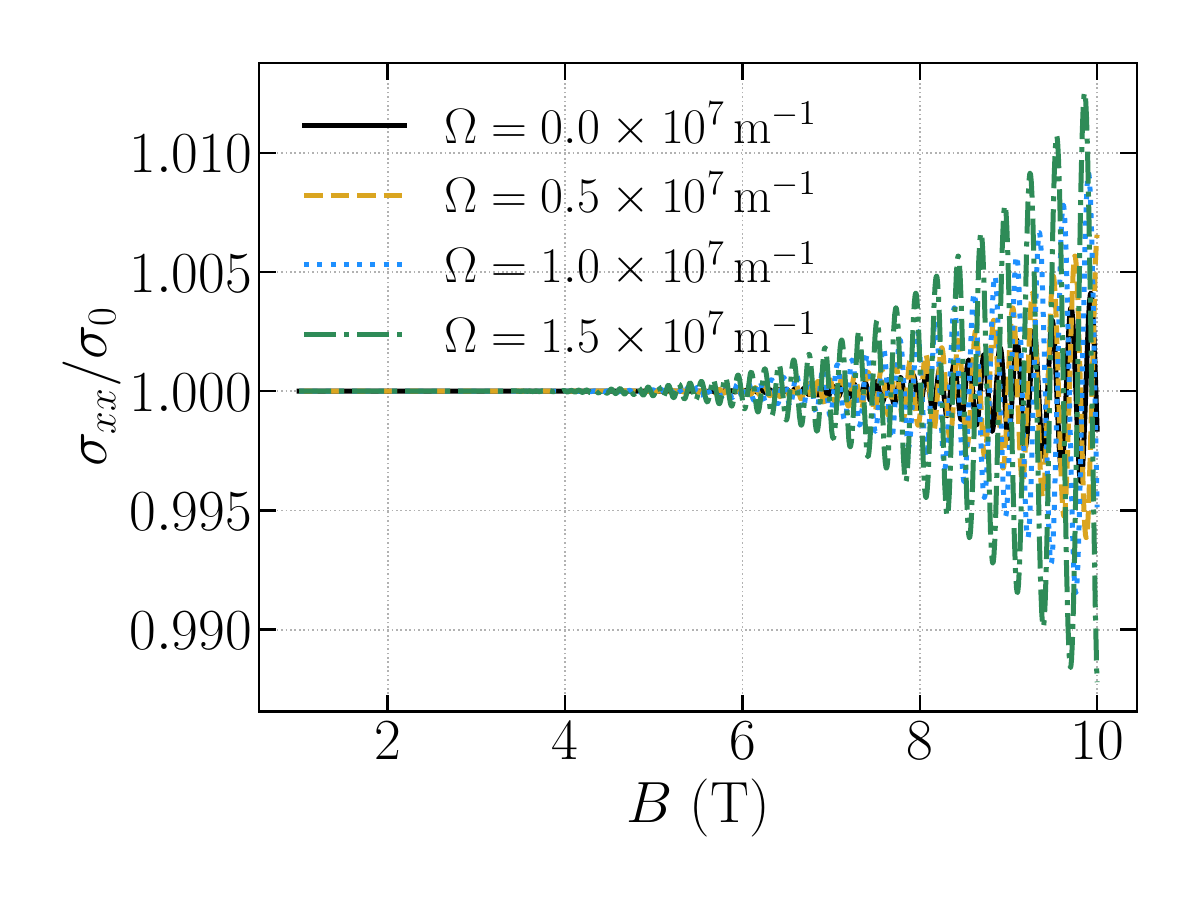}
  \caption{\footnotesize
  Shubnikov-de Haas longitudinal conductivity at $T=4.2\,$K for four
  torsion densities ($\Omega=0$, $0.5$, $1.0$, and
  $1.5\times10^{7}\,\mathrm{m^{-1}}$) shown in black, orange, blue, and green, respectively.
  The model incorporates the Lifshitz-Kosevich thermal factor
  $R_T=X/\sinh X$ with $X=2\pi^{2}k_B T/(\hbar|\omega_{\mathrm{eff}}|)$
  and the Dingle damping factor
  $R_D=\exp[-\pi/(|\omega_{\mathrm{eff}}|\tau_D)]$,
  where $\omega_{\mathrm{eff}}=\omega_c+\omega_{cl}
  = eB/\mu + 2\hbar k\Omega/\mu$,
  $k=10^{8}\,\mathrm{m^{-1}}$, $\mu_F=0.05\,$eV, and $\tau_D=1$\,ps.
  Increasing $\Omega$ enhances $|\omega_{\mathrm{eff}}|$, thereby  \emph{compressing the oscillation period in $1/B$} (i.e., increasing the oscillation frequency in
  field sweeps) and reducing the LK/Dingle damping, resulting in a
  slightly stronger high-$B$ envelope. The overall trend mirrors the
  torsion-shifted quantum-oscillation frequency discussed for the
  thermodynamic sector.}
  \label{fig:sxx_sdh_vs_B}
\end{figure}

\paragraph*{Longitudinal channel and scaling.}
The longitudinal conductivity exhibits peaks at $B_c^{(n)}$.
A minimal phenomenology that captures temperature and disorder is
\begin{align}
\sigma_{xx}(B)&\simeq \sum_{n}\,
\sigma_{xx}^{\mathrm{max}}\,
\exp\left[-\frac{(B-B_c^{(n)})^2}{2\,w^2(T,\Gamma)}\right],
\notag\\ &
w(T,\Gamma)\propto T^{\kappa}+\alpha\,\Gamma,
\label{eq:sigma_xx_gauss}
\end{align}
where $\Gamma=\hbar/2\tau_D$ is the Dingle width and $\kappa$ the
critical exponent controlling the thermal-dephasing broadening of the transition.
Equation~\eqref{eq:Bc_linear_shift} implies that all $\sigma_{xx}$ peaks shift
\emph{linearly} with $\Omega$, while keeping fixed separations in $1/B_{\mathrm{eff}}$.

Figure~\ref{fig:sxx_sdh_vs_B} shows SdH oscillations of the
longitudinal conductivity, normalized by a smooth background
$\sigma_{0}$. The amplitude grows with the field because both the LK factor
$R_T=X/\sinh X$ approaches unity and the Dingle factor
$R_D=\exp[-\pi/(|\omega_{\mathrm{eff}}|\tau_D)]$ becomes less
suppressive as $|\omega_{\mathrm{eff}}|$ increases.
Torsion enters through $\omega_{cl}\propto \Omega$ and therefore shifts
the oscillation frequency: larger $\Omega$ compresses the period in
$1/B$, in agreement with the torsion-induced shift of the de Haas-van Alphen frequency.

\begin{figure*}[tbhp]
\centering
\includegraphics[width=0.45\linewidth]{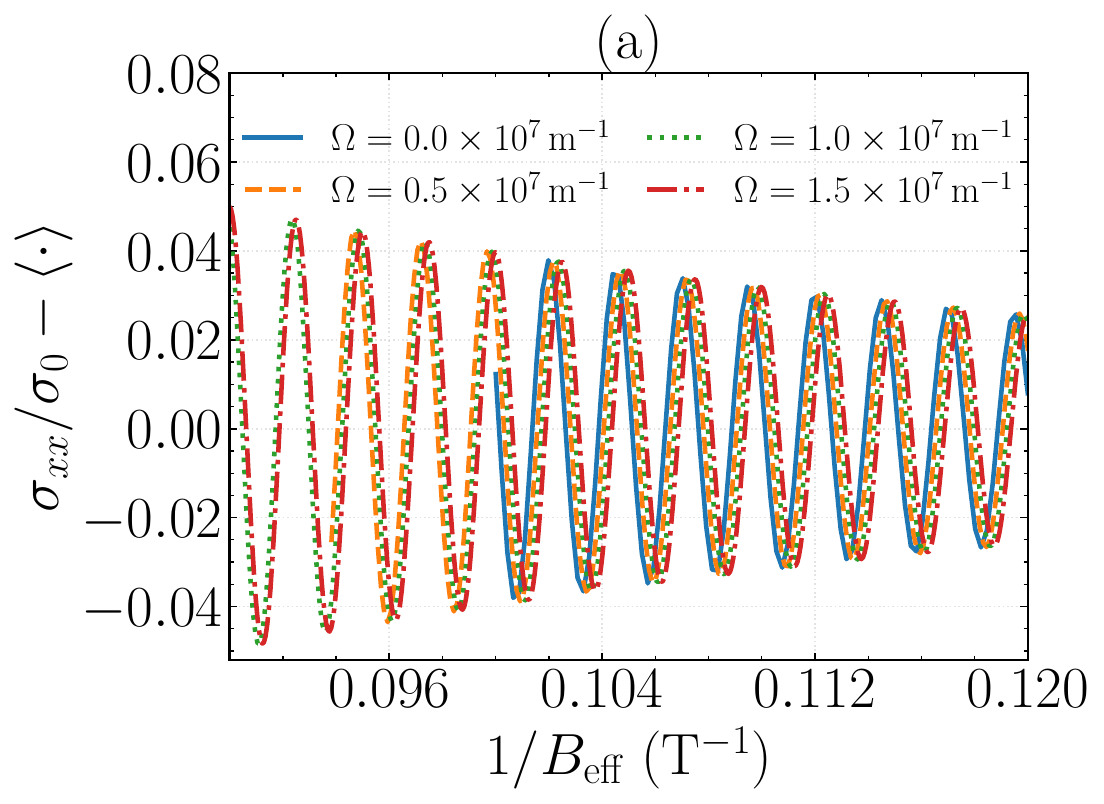}
\includegraphics[width=0.45\linewidth]{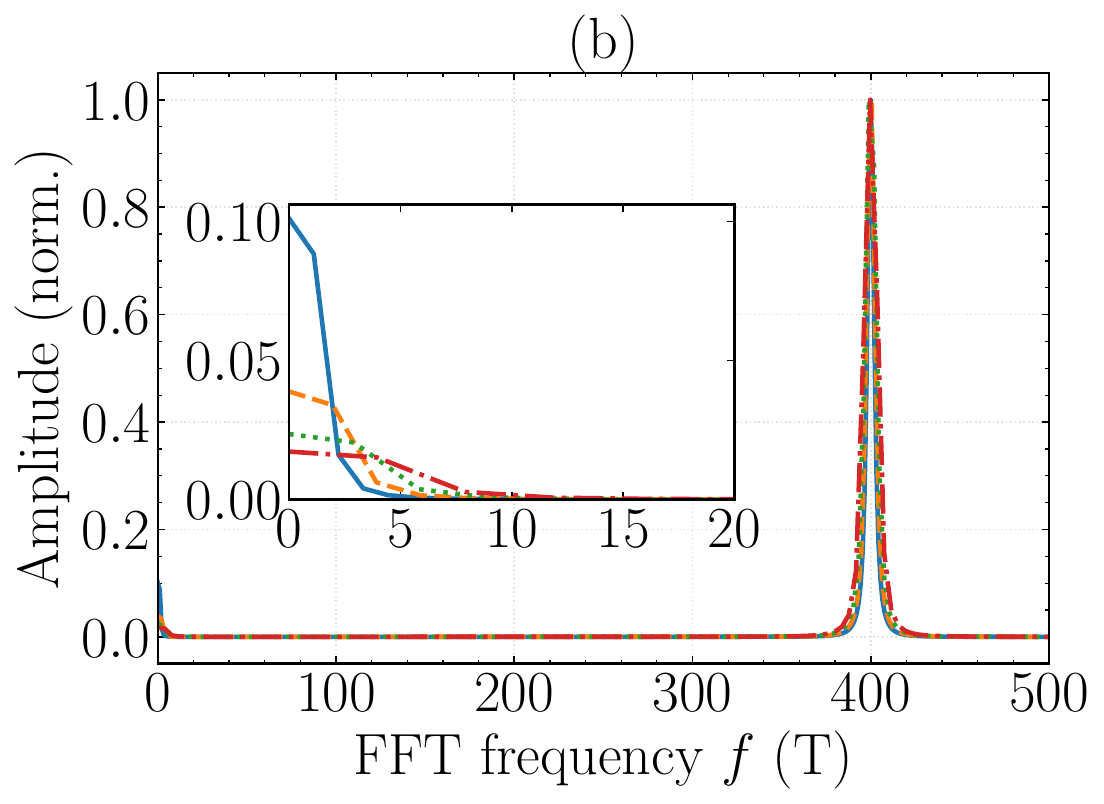}
\caption{\footnotesize
Shubnikov-de Haas oscillations with torsion. The effective magnetic field is
$B_{\rm eff}=B+B_{\mathrm{tor}}$ with $B_{\mathrm{tor}}=2\hbar k\Omega/e$ ($k\equiv k$).
Four torsion densities are shown,
$\Omega=\{0,0.5,1.0,1.5\}\times10^{7}\,\mathrm{m^{-1}}$, using distinct line styles, while the magnetic field is swept over
$B\in[1,10]\,\mathrm{T}$.
(a) Mean-subtracted longitudinal conductivity $\sigma_{xx}/\sigma_{0}$ plotted against $1/B_{\rm eff}$.
(b) Normalized FFT amplitude of $\sigma_{xx}(1/B_{\rm eff})$; the inset magnifies the
low-frequency window $f\in[0,40]\,\mathrm{T}$ showing a small common background.
Physical parameters: $T=4.2\,\mathrm{K}$, $k=10^{8}\,\mathrm{m^{-1}}$,
$m^\star=0.067\,m_e$ ($m^\star\equiv\mu$), $\tau_q=10^{-13}\,\mathrm{s}$, $\sigma_0=1$.}
\label{fig:transport_sxx_full}
\end{figure*}

\paragraph*{Real-space trace and spectral view} [Fig.~\ref{fig:transport_sxx_full}].
Figure~\ref{fig:transport_sxx_full}(a) shows the mean-subtracted longitudinal conductivity $\sigma_{xx}/\sigma_{0}$ plotted against the reciprocal effective field $x=1/B_{\mathrm{eff}}$ (with $B_{\mathrm{eff}}=B+B_{\mathrm{tor}}$). After mapping from $B$ to $x$, the uniform resampling renders the SdH series nearly periodic and exposes a systematic torsion-induced phase shift: increasing $\Omega$ advances the oscillations because it raises the effective cyclotron scale $|\omega_{\mathrm{eff}}|=|\omega_{c}+\omega_{cl}|$ with $\omega_{cl}=2\hbar k\Omega/m^{\star}$.

We then analyze the oscillatory content via the fast Fourier transform (FFT), i.e., the discrete Fourier transform evaluated with an $O(N\log N)$ algorithm [Fig.~\ref{fig:transport_sxx_full}(b)]. The horizontal axis is the FFT frequency $f$ (in T), conjugate to $x$ (in T$^{-1}$), so the SdH period satisfies $\Delta(1/B_{\mathrm{eff}})=1/f$. The dominant cyclotron peak shifts to higher $f$ as $\Omega$ increases, reflecting the compression of the oscillation period in $1/B$ caused by the enhancement of $|\omega_{\mathrm{eff}}|$. The inset magnifies the low-frequency window $f\in[0,40]$\,T and reveals a small, rapidly decaying background common to all traces; it arises from finite record length together with windowing/detrending (spectral leakage and mean/trend removal), carries negligible weight relative to the main peak, and does not affect its position. Taken together, panels (a) and (b) show that the magnetic field and torsion enter on the same footing through $B_{\mathrm{eff}}$ and $\omega_{\mathrm{eff}}$, tying the transport oscillation frequency to the same thermodynamic scale $\hbar|\omega_{\mathrm{eff}}|$ that governs $C_{V}(T)$, $M(T)$, and $\chi(T)$.

\paragraph*{Thermodynamic-transport dictionary.} Because $B_{\mathrm{eff}}$ simultaneously controls dHvA, $M(T)$, and $C_V(T)$, there is a direct connection between the thermodynamic traces and the plateau transitions \cite{Streda1982}:
\begin{equation}
\frac{\partial M_{\mathrm{osc}}}{\partial \mu_F}
= -\,\frac{\partial^2 \Omega_G}{\partial \mu_F\,\partial B}
= -\,\frac{\sigma_{xy}}{e},
\label{eq:Streda_M_relation}
\end{equation}
with the denotation $k\equiv k$ for the longitudinal wave number throughout.

Thus, \emph{slope breaks} (“sawtooth” features) in $M(\mu_F,B,\Omega)$
coincide with the plateau edges of $\sigma_{xy}$, and torsion shifts both by
$B_{\mathrm{tor}}$.

\subsection{Compensated Field Condition: \texorpdfstring{$\omega_{\mathrm{eff}}=0$}{omega\_eff=0}}
\label{subsec:compensated}

A particularly interesting physical regime occurs when the effective cyclotron frequency vanishes, i.e., $\weff = \omega_{cl} + \omega_c = 0$. This condition reveals the profound analogy between structural torsion and the magnetic field, as one can be used to precisely cancel the effect of the other. Such compensation occurs when the magnetic field is tuned to a specific negative value, $B_{\mathrm{comp}}$, given by
\begin{equation}
B_{\mathrm{comp}} = -\frac{2\hbar k\Omega}{e}.
\label{eq:B_comp}
\end{equation}
At this compensation point, the spacing between the elastic Landau levels collapses, and the energy spectrum given by Eq.~\eqref{eq:energy_spectrum} reduces to that of a free particle moving along the dislocation line:
\begin{equation}
E(k)=\frac{\hbar^{2}k^{2}}{2\mu}.
\label{eq:free_line_spectrum}
\end{equation}
This restoration of quasi-free motion has a distinct thermodynamic signature. Away from the compensation point, the system possesses a gapped spectrum and exhibits a Schottky-like heat capacity. At the compensation point, however, the gap closes. In the limit $\weff \to 0$, the heat capacity from Eq.~\eqref{eq:heat_capacity} approaches the classical equipartition value corresponding to two translational degrees of freedom, $C_V \to k_B$.

To illustrate this crossover, we plot the heat capacity as a function of the magnetic field $B$ for a fixed temperature, as shown in Fig.~\ref{fig:compensated_field}. The plot clearly demonstrates that for magnetic fields far from $B_{\mathrm{comp}}$, the heat capacity is suppressed, as the thermal energy is insufficient to overcome the large effective energy gap. As $B$ approaches $B_{\mathrm{comp}}$, the gap narrows, and $C_V$ increases, reaching its maximum value of $k_B$ precisely at the compensation point. This behavior reveals a remarkable crossover from a quantum-gapped regime to a classical-like plateau, tunable by the external magnetic field.
\begin{figure}[tbhp]
\centering
\includegraphics[width=0.45\textwidth]{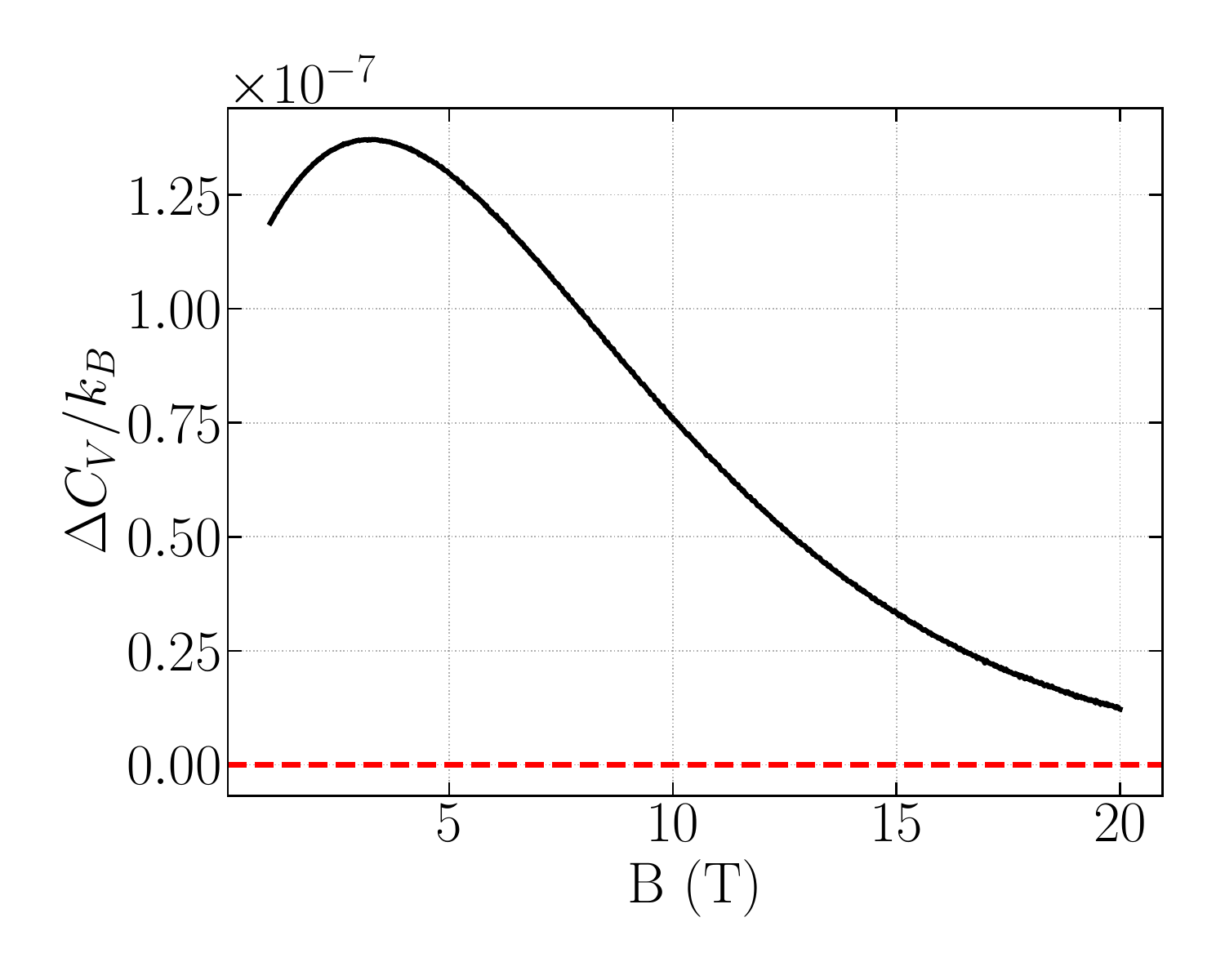}
\caption{\footnotesize Oscillatory correction to the heat capacity, $\Delta C_V = C_{V,\mathrm{corr}} - C_{V,\mathrm{bulk}}$, as a function of the magnetic field $B$ for a mesoscopic system of radius $R=50$\,nm at a fixed temperature of $T=4$\,K. The oscillations arise directly from the finite-size corrections to the degeneracy of the elastic Landau levels.}
\label{fig:finite_size}
\end{figure}

\subsection{Finite-Size Corrections}
\label{subsec:finite_size}

As for other macroscopic observables, results below are \emph{sample totals} with explicit degeneracy; in mesoscopic devices, we use the corrected $g_{\mathrm{corr}}$ (Eqs.~\eqref{eq:degeneracy} and \eqref{eq:finite_size_correction}).

Boundary effects in mesoscopic samples are well known to produce subleading, geometry-dependent corrections to spectral counting and
thermodynamic densities. In zero field, these manifest as the perimeter
term in the Weyl (or multiple-reflection) expansion of the density of
states~\cite{AP1970, AP1971, Baltes1976}. In the presence of a magnetic field, they admit a complementary edge-state interpretation in
terms of chiral boundary modes that carry part of the Landau
degeneracy~\cite{PRB1982}. Related finite-size and edge
phenomena have been extensively documented in quantum dots and rings subjected to magnetic fields~\cite{PhysRevB.76.085308,PhysRevB.68.035326,PRL2003,PhysRevLett.102.066401,PMM.2023.124.1069,JCTE.2023.68.379}.

The thermodynamic expressions derived so far assume the
thermodynamic limit. For a finite 2D sample of area $A$, the bulk Landau
degeneracy per level is
$g = A/(2\pi \ell_{\mathrm{eff}}^{2})$, with
$\ell_{\mathrm{eff}}=\sqrt{\hbar/(\mu\weffabs)}$. In a
bounded domain, the spectral counting acquires a perimeter correction
proportional to the boundary length $P$. For a connected region with an effective radius $R\sim\sqrt{A/\pi}$ this correction can be expressed, within a minimal model, as
\begin{equation}
g \longrightarrow g_{\mathrm{corr}}
= g\left(1-\frac{\ell_{\mathrm{eff}}}{R}\right),
\label{eq:finite_size_correction}
\end{equation}
which captures the leading $O(\ell_{\mathrm{eff}}/R)$ depletion of the
bulk degeneracy due to edge states \footnote{In a more general Weyl
setting one writes
$g_{\mathrm{corr}}\simeq A/(2\pi\ell_{\mathrm{eff}}^{2})
-\eta\,P/(2\pi\ell_{\mathrm{eff}})+\cdots$,
with a geometry/boundary-condition coefficient $\eta$; for a disk one
recovers a form equivalent to Eq.~\eqref{eq:finite_size_correction}.}.
Because $\ell_{\mathrm{eff}}$ depends on
$\weff$, the corrected
degeneracy $g_{\mathrm{corr}}$ inherits both \emph{field and torsion
dependence}. Consequently, derivatives of the free energy acquire additional terms that contribute to $U$, $S$, and $C_{V}$.

A salient manifestation of this effect is the emergence of oscillatory, field-dependent corrections in caloric observables, the finite-size
analog of de Haas-van Alphen oscillations. Physically, as $B$ is
swept, the ratio $\ell_{\mathrm{eff}}/R$ moves the edge-state ladder
relative to the bulk Landau bands, thereby modulating the level filling and the
degeneracy entering the partition function. Damping due to temperature
and disorder plays the familiar role described, in the context of magnetic oscillations, by Dingle-type factors~\cite{PRL2003,PhysRevLett.102.066401}.

To isolate these effects, we compute the heat-capacity correction
$\Delta C_{V}=C_{V,\mathrm{corr}}-C_{V,\mathrm{bulk}}$, where the
``corrected'' quantity is obtained by replacing $g \to g_{\mathrm{corr}}$
in the free energy and propagating the change to $C_{V}$. The result,
shown in Fig.~\ref{fig:finite_size}, displays a characteristic
nonmonotonic structure as a function of $B$: a thermally broadened
oscillation around zero, determined by the effective magnetic length and the sample size. Its amplitude scales, to leading order, with
$\ell_{\mathrm{eff}}/R$ and decreases with increasing temperature (level
smearing), while its period follows the expected $1/\weffabs$
trend characteristic of Landau quantization. These finite-size signatures  
therefore emerge as natural targets in mesoscopic platforms, such as quantum rings,
dots, and patterned van der Waals flakes \cite{PRM.2025.9.024001,PRB.2024.109.064416}, where boundary control is available, and the interplay of torsion and magnetic field can be tuned
\emph{in situ}.
\begin{figure*}[tbhp]
  \centering
  \includegraphics[width=0.65\linewidth]{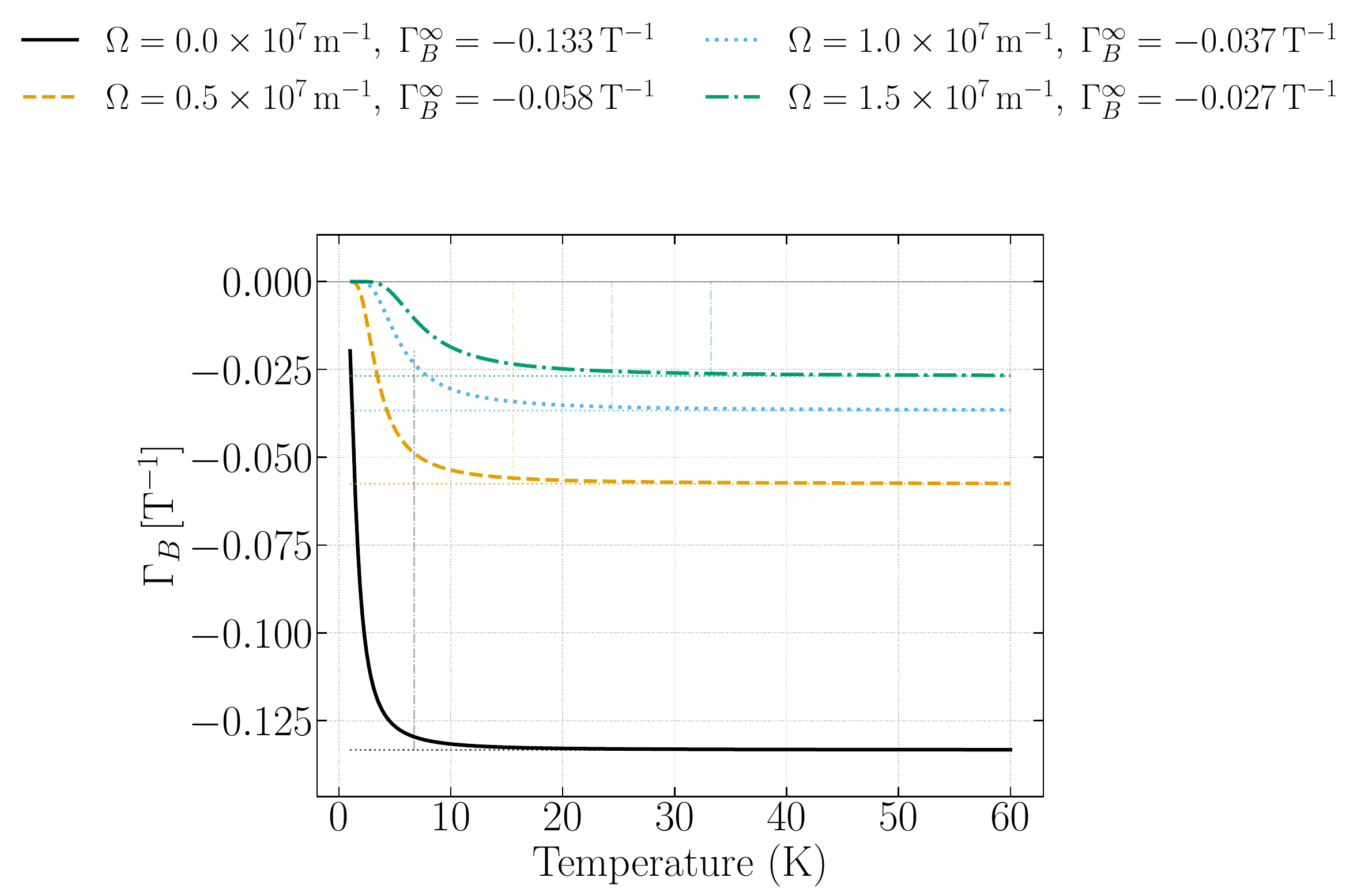}
  \caption{\footnotesize
  Magnetic Gr\"uneisen coefficient $\Gamma_{B}(T)$ at fixed magnetic field
  $B=5\,\mathrm{T}$ for four torsion densities
  ($\Omega=0$, $0.5$, $1.0$, and
  $1.5\times10^{7}\,\mathrm{m^{-1}}$) represented by solid, dashed, dotted, dash-dotted lines, respectively.
  The calculation includes the longitudinal contribution to the heat
  capacity, $C_{V}^{3\mathrm{D}}=C_{V}+\tfrac{1}{2}k_{B}$, which yields a
  smooth approach to the high-temperature asymptotes (horizontal dotted
  lines) given by
  $\Gamma_{B}^{\infty}=-(2/3)\,e/(\mu\omega_{\mathrm{eff}})$
  [cf.\ Eqs.~\eqref{eq:Gruneisen}-\eqref{eq:Gruneisen_explicit} and \eqref{eq:Gamma_infty}].
  Vertical dash-dotted markers indicate the crossover scale
  $T^{\star}=\hbar\omega_{\mathrm{eff}}/k_{B}$.
  Increasing $\Omega$ enhances $\omega_{\mathrm{eff}}$ and therefore (i) shifts the
  crossover to higher temperatures and (ii) reduces the magnitude of the
  asymptotic value, i.e., $|\Gamma_{B}^{\infty}|$ decreases with increasing~$\Omega$.}
  \label{fig:gruneisen_vs_Omega_kint}
\end{figure*}

\subsection{Magnetic Gr\"uneisen parameter}
\label{subsec:gruneisen}

Anchored in entropy scaling, the \emph{magnetic Gr\"uneisen parameter} has become a sensitive probe of low-energy scales and quantum criticality: as the dominant energy scale collapses near a field-tuned quantum critical point, $\Gamma_{B}$ is expected to diverge and may change sign \cite{Zhu2003}. Experimentally, high-resolution alternating-field and quasistatic protocols access $\Gamma_{B}$ either directly from adiabatic $\partial T/\partial B$ measurements or indirectly via $(\partial M/\partial T)/C_V$ \cite{Tokiwa2011,Kuechler2012}. Within our elastic-Landau framework, the same single gap $x=\hbar|\omega_{\mathrm{eff}}|/(2k_BT)$ that organizes $C_V$ and $M$ also controls $\Gamma_{B}$: in the fixed-$k$ baseline the thermal kernel cancels exactly, whereas including the free longitudinal motion (bulk 3D) adds a universal $+\tfrac12 k_{B}$ to $C_V$ and produces a smooth crossover to a constant high-$T$ plateau.

The magnetic Gr\"uneisen parameter quantifies the adiabatic temperature response to a field change and admits two equivalent forms,
\begin{equation}
\Gamma_{B}\equiv -\frac{1}{T}\Bigl(\frac{\partial T}{\partial B}\Bigr)_{S}
=\frac{1}{C_{V}}\Bigl(\frac{\partial M}{\partial T}\Bigr)_{B},
\label{eq:Gruneisen}
\end{equation}
where $M$ is the magnetization and $C_{V}$ the heat capacity at constant volume. In our model, all thermodynamics is controlled by the elastic-Landau gap $\hbar\weffabs$, with $\weff=\omega_{c}+\omega_{cl}$ and
$\omega_{c}=eB/\mu$, $\omega_{cl}=2\hbar k\Omega/\mu$.

From the Helmholtz free energy $A(T)=k_{B}T\ln\bigl[2\sinh x\bigr]+\text{const}$ with
\[
x=\frac{\beta\hbar\weffabs}{2}=\frac{\hbar\weffabs}{2k_{B}T},
\]
the magnetization follows as
\begin{align}
M(B,\Omega,T)
&= -\Bigl(\frac{\partial A}{\partial B}\Bigr)_{T}
= -\Bigl(\frac{\partial A}{\partial x}\Bigr)
  \Bigl(\frac{\partial x}{\partial\weffabs}\Bigr)
  \Bigl(\frac{\partial\weffabs}{\partial B}\Bigr) \notag\\
&= -\,\mathrm{sgn}(\weff)\,\frac{\hbar e}{2\mu}\,
   \coth x,
\label{eq:M_coth}
\end{align}
since $\partial A/\partial x=k_{B}T\,\coth x$, $\partial x/\partial\weffabs=\beta\hbar/2$, and
$\partial\weffabs/\partial B=(e/\mu)\,\mathrm{sgn}(\weff)$.
Differentiating \eqref{eq:M_coth} at fixed $B$ gives
\begin{equation}
\Bigl(\frac{\partial M}{\partial T}\Bigr)_{B}
= -\,\mathrm{sgn}(\weff)\,\frac{\hbar e}{2\mu}\,
\frac{\hbar\weffabs}{2k_{B}T^{2}}\,
\csch^{2}x.
\label{eq:dMdT_exact}
\end{equation}

\paragraph*{Working expression for $\Gamma_{B}$.}
Using $C_{V}(T)=k_{B}\,x^{2}\,\csch^{2}x$ for the (transverse) fixed-$k$ model,
\begin{equation}
\Gamma_B(T)=\frac{1}{C_V}\left(\frac{\partial M}{\partial T}\right)_B
= -\,\mathrm{sgn}(\omega_{\mathrm{eff}})\,\frac{\hbar e}{2\mu k_B}\,
\frac{x\,\csch^{2}x}{C_V/k_B}, \label{eq:Gruneisen_explicit}
\end{equation}

with $x=\hbar|\omega_{\mathrm{eff}}|/2k_B T$, which is the expression used throughout the analysis.

\paragraph*{Two regimes: fixed-$k$ vs.\ 3D ($k$-integrated).}
(i) \emph{Fixed-$k$ model.} Substituting $C_{V}/k_{B}=x^{2}\csch^{2}x$ into
\eqref{eq:Gruneisen_explicit} gives a \emph{temperature-independent} result,
\begin{equation}
\Gamma_{B}^{\mathrm{fixed}\,k}
= -\,\mathrm{sgn}(\weff)\,\frac{e}{\mu\,\weffabs}
= -\,\frac{e}{\mu\,\weff}.
\label{eq:Gamma_fixedk}
\end{equation}
The equality on the right uses $\mathrm{sgn}(\weff)/\weffabs=1/\weff$.

(ii) \emph{Three-dimensional gas (including longitudinal motion).}
For a bulk sample, integrating over the free $k$-motion adds a universal
$+\tfrac{1}{2}k_{B}$ to the heat capacity (Sec.~\ref{subsec:k_integration_details}),
so
\begin{equation}
C_{V}^{3\mathrm{D}}(T)=k_{B}\,x^{2}\csch^{2}x+\frac{1}{2}k_{B}.
\end{equation}
Replacing $C_{V}$ by $C_{V}^{3\mathrm{D}}$ in \eqref{eq:Gruneisen_explicit} yields a smooth $T$-dependence:
\[
\Gamma_{B}^{3\mathrm{D}}(T)
= -\,\mathrm{sgn}(\weff)\,\frac{\hbar e}{2\mu k_{B}}\,
\frac{\beta\,\csch^{2}x}{x^{2}\csch^{2}x+\tfrac12}.
\]
Its asymptotes follow directly from the small/large $x$ limits:

\emph{Low $T$ (large $x$):} since $x^{2}\csch^{2}x\to 0$,
$\Gamma_{B}^{3\mathrm{D}}\to 0$ exponentially.

\emph{High $T$ (small $x$):} using
$x^{2}\csch^{2}x=1-\tfrac{x^{2}}{3}+O(x^{4})$ and $\beta=1/(k_{B}T)$,
\begin{equation}
\Gamma_{B}^{3\mathrm{D}}(T\to\infty)
\longrightarrow
-\,\frac{2}{3}\,\mathrm{sgn}(\weff)\,\frac{e}{\mu\,\weffabs}
= -\,\frac{2}{3}\,\frac{e}{\mu\,\weff}.
\label{eq:Gamma_infty}
\end{equation}
Thus, the longitudinal degree of freedom reduces the fixed-$k$ value in Eq.
\eqref{eq:Gamma_fixedk} by a factor $2/3$ at high temperatures and restores the expected approach to a constant plateau.

\paragraph*{Remarks on signs and robustness.}
The kernels $\coth x$, $\csch^{2}x$ depend only on the \emph{gap} $x\propto\weffabs$. The $\mathrm{sgn}(\weff)$ factor enters \emph{only} through
$\partial\weffabs/\partial B=(e/\mu)\,\mathrm{sgn}(\weff)$ in the chain rule for $M=-\partial A/\partial B$, and therefore multiplies $\Gamma_{B}$ globally in Eqs.
\eqref{eq:Gruneisen_explicit}-\eqref{eq:Gamma_infty}. If a fixed field orientation is adopted such that $\weff>0$, one may set $\mathrm{sgn}(\weff)=+1$ everywhere without changing any trend or magnitude.

\paragraph*{Role as a scale probe.}
Operationally, $\Gamma_{B}$ measures the \emph{magnetocaloric efficiency} (cooling/heating upon sweeping $B$ under adiabatic conditions) and serves as a sensitive probe of the dominant low-energy scale. In systems near field-tuned quantum critical points, $\Gamma_{B}$ often diverges or changes sign, reflecting the collapse of the characteristic energy scale and the emergence of universal entropy scaling laws. In the present context, the existence of a \emph{single} tunable scale, $\hbar\weff$, makes $\Gamma_{B}$ a direct “thermometer” of magnetoelastic coupling: increasing the screw-dislocation density raises $\weff$, shifts the crossover in $\Gamma_{B}(T)$ to higher temperatures, and reduces the magnitude of the high-$T$ asymptotic plateau.

In the minimal fixed-$k$ model, the same thermal kernel $\sinh^{-2}x$ ($x=\hbar\weffabs/2k_{B}T$) appears in both $C_{V}$ and in $(\partial M/\partial T)_{B}$, leading to an exact cancellation in Eq. \eqref{eq:Gruneisen_explicit} and resulting in a strictly linear $T$-dependence. For a bulk three-dimensional gas, however, the longitudinal motion is free and must be integrated out (Sec.~\ref{subsec:k_integration_details}), which adds a universal $+\tfrac{1}{2}k_{B}$ contribution to the heat capacity, $C_{V}^{3\mathrm{D}}=C_{V}+\tfrac{1}{2}k_{B}$. This additional term breaks the cancellation and produces a smooth crossover of $\Gamma_{B}(T)$: a rapid low-$T$ decay followed by a monotonic approach to a constant high-$T$ limit. Within this $k$-integrated description, the asymptote value is
\begin{equation}
\Gamma_{B}^{\infty}=
   -\frac{2}{3}\,\frac{e}{\mu\,\omega_{\mathrm{eff}}},
\qquad
\omega_{\mathrm{eff}}=\omega_{c}+\omega_{cl},
\label{eq:Gamma_infty}
\end{equation}
while the crossover scale is $T^{\star}\sim\hbar\weffabs/k_{B}$.

Figure~\ref{fig:gruneisen_vs_Omega_kint} shows that all curves start
near zero and bend toward negative values once
$k_{B}T$ becomes comparable to the elastic-Landau gap,
$k_{B}T\sim\hbar\weffabs$.
The additional $+\tfrac{1}{2}k_{B}$ contribution in $C_{V}^{3\mathrm{D}}$
delays and smooths this drop, replacing the linear law of the fixed-$k$
model with a gradual approach to a constant negative plateau.
Because $\weff$ increases with dislocation density,
larger $\Omega$ values push the crossover to higher $T$ (vertical markers)
and reduce the magnitude of the magnetocaloric efficiency, as indicated by the less negative $\Gamma_{B}^{\infty}$ (horizontal
guides). When both longitudinal and transverse continua are thermally
populated, the four traces collapse onto their respective asymptotes,
reflecting that $\Gamma_{B}(T)$ is governed predominantly by the single
tunable scale $\weff$.
\begin{figure*}[t]
	\centering
	\includegraphics[scale=1.05,trim=2pt 2pt 2pt 2pt,clip]{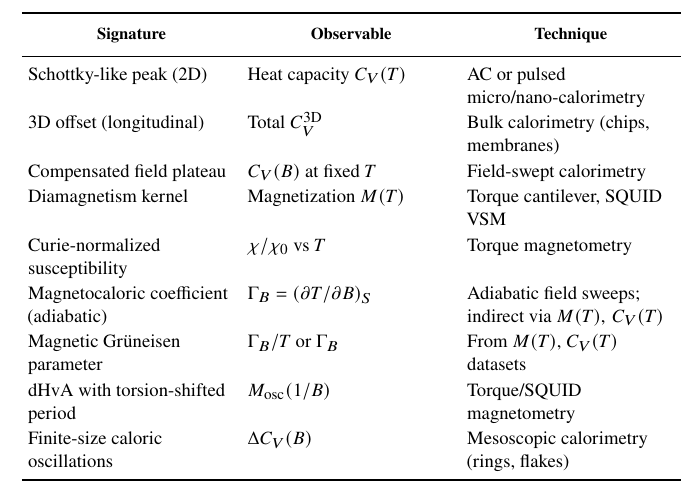}
	\caption{Experimental signatures, observables, and measurement techniques.}
	\label{fig:tableA}
\end{figure*}
\begin{figure*}[t]
	\centering
	\includegraphics[scale=1.0,trim=2pt 2pt 2pt 2pt,clip]{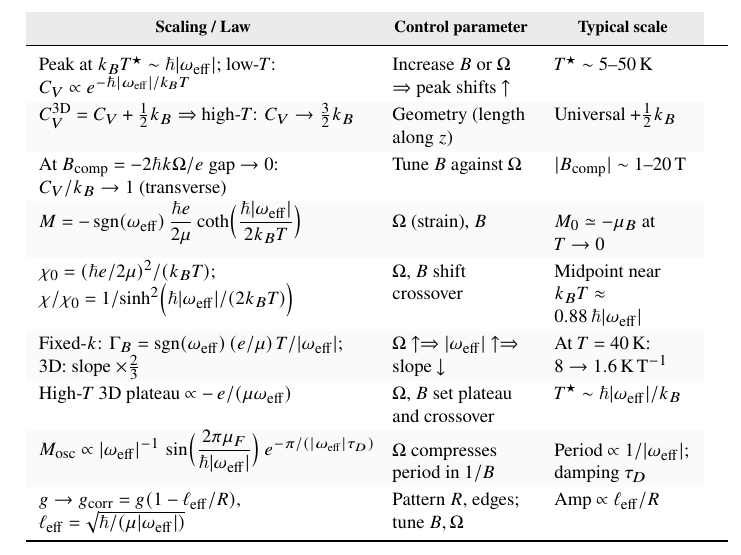}
	\caption{\footnotesize Scaling laws, control parameters, and typical scales.}
	\label{fig:tableB}
\end{figure*}

\section{Experimental relevance and testable predictions}
\label{subsec:exp_relevance}

Our results suggest several concrete, testable signatures across standard low-temperature probes. We summarize below the key observables, the predicted trends, and typical parameter scales.

\paragraph*{(i) Calorimetry: Schottky-like peak and compensated-field plateau}\cite{PRE.2018.97.052129,PRL.2022.129.147201,PRB.2016.94.125307,PRR.2025.7.033239}.
Equations~\eqref{eq:heat_capacity} and \eqref{eq:Cv3D} predict a Schottky-type maximum in $C_V(T)$ at $k_B T \sim \hbar\weffabs$, shifting to higher temperatures as either $B$ or $\Omega$ increases. At fixed $T$, sweeping $B$ reveals a pronounced enhancement of $C_V$ upon approaching the compensated point $\weff=0$ [Eq.~\eqref{eq:B_comp}], where the transverse gap collapses and the transverse equipartition plateau $C_V/k_B \to 1$ emerges (Fig.~\ref{fig:compensated_field}). For bulk samples, the longitudinal continuum raises the curves by the universal $+\tfrac{1}{2}k_B$ [Eq.~\eqref{eq:Cv3D}], driving $C_V \to \tfrac{3}{2}k_B$ at high $T$.
These features are accessible to micro/nano-calorimetry on strained heterostructures and semiconductor membranes.
\begin{figure}[tbhp]
  \centering
  \includegraphics[width=0.9\linewidth]{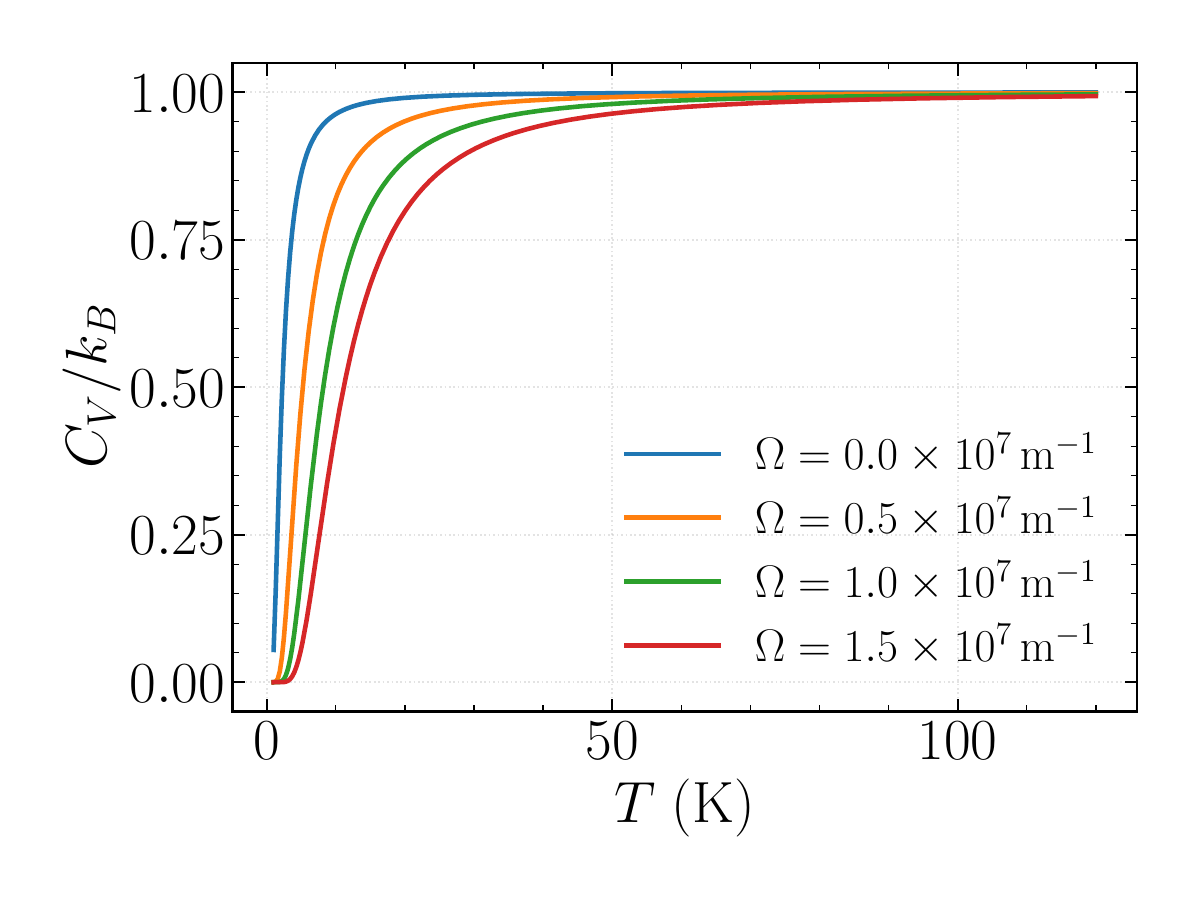}
  \caption{\footnotesize Transverse heat capacity as a function of temperature at fixed magnetic field. The heat capacity per particle in units of $k_{B}$, $C_{V}/k_{B}$, is shown as a function of temperature for a fixed magnetic field ($B=5.0$~T) and different torsion densities $\Omega$ (legend).
  The curves are computed from Eq.~\eqref{eq:heat_capacity} using the effective cyclotron frequency $\omega_{\mathrm{eff}}=\omega_{c}+\omega_{cl}$, while keeping the longitudinal momentum fixed (transverse model).
  Increasing $\Omega$ enhances $\weffabs$ and shifts the Schottky-like crossover to higher temperatures, consistent with the single-scale control $k_{B}T^{\star}\sim\hbar\weffabs$.
  In the bulk case, integrating over the free longitudinal motion simply adds the universal offset $+\,\tfrac{1}{2}k_{B}$ to all curves [Eq.~\eqref{eq:Cv3D}].}
  \label{fig:Cv_vsT_varOmega_trans}
\end{figure}
Figure~\ref{fig:Cv_vsT_varOmega_trans} illustrates the thermal fingerprint of torsion at fixed magnetic field: the transverse heat capacity per particle, $C_{V}/k_{B}$, rises through a Schottky-like crossover set by the single scale $k_{B}T^{\star}\sim\hbar\weffabs$, with $\omega_{\mathrm{eff}}=\omega_{c}+\omega_{cl}$.
As the dislocation density (torsion) increases, $\weffabs$ grows, pushing the crossover to higher $T$ and thus reducing the entropy available at a given temperature.
This behavior is the calorimetric counterpart of the magnetization/susceptibility kernel discussed below, and it directly supports our one-parameter organization of the thermodynamics.
For bulk samples, the longitudinal continuum adds the universal offset $+\tfrac{1}{2}k_{B}$ to all traces [Eq.~\eqref{eq:Cv3D}], leaving the scaling with $\weffabs$ intact.

\begin{figure}[tbhp]
  \centering
  \includegraphics[width=0.9\linewidth]{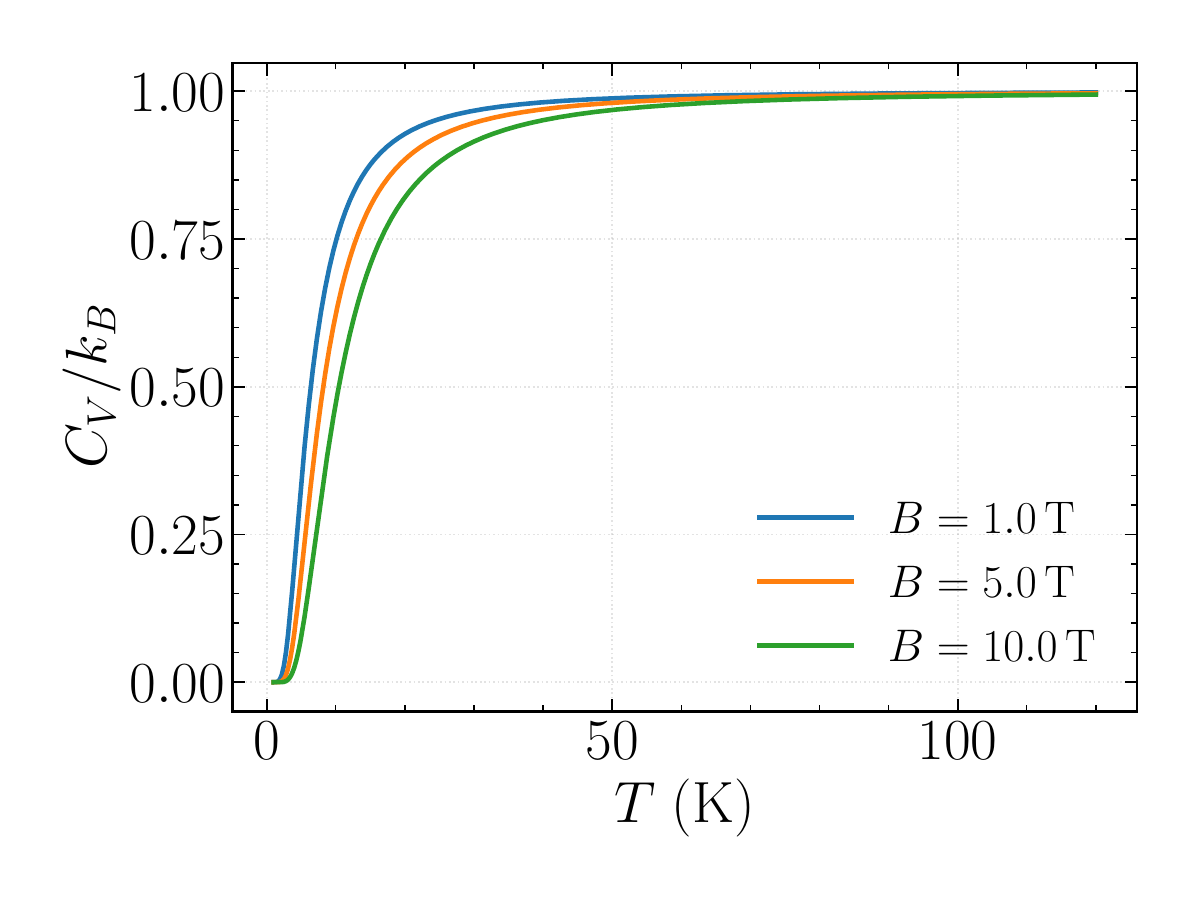}
  \caption{\footnotesize Transverse heat capacity as a function of temperature at fixed torsion. The heat capacity per particle, in units of $k_{B}$, $C_{V}/k_{B}$, as a function of temperature for a fixed torsion density ($\Omega=1.0\times 10^{7}\,\mathrm{m}^{-1}$) and several magnetic fields $B$ (see legend).
  The curves are obtained from Eq.~\eqref{eq:heat_capacity} using the effective cyclotron frequency $\omega_{\mathrm{eff}}=\omega_{c}+\omega_{cl}$ within the transverse (fixed-$k$) model.
  Increasing $B$ enhances $\weffabs$ and shifts the Schottky-like crossover to higher temperatures, consistent with the single-scale relation $k_{B}T^{\star}\sim\hbar\weffabs$.
  For bulk samples, integrating over the free longitudinal motion simply adds the universal offset $+\,\tfrac{1}{2}k_{B}$ to all curves [Eq.~\eqref{eq:Cv3D}] without affecting the scaling behavior.}
  \label{fig:Cv_vsT_varB_trans}
\end{figure}
Figure~\ref{fig:Cv_vsT_varB_trans} complements Fig.~\ref{fig:Cv_vsT_varOmega_trans} by showing the transverse heat capacity at fixed torsion while sweeping the magnetic field.
As $B$ increases, the effective cyclotron gap $\hbar\weffabs$ grows and the Schottky-like crossover in $C_{V}(T)$ shifts rigidly to higher temperatures, in quantitative agreement with the one-parameter scaling $k_{B}T^{\star}\sim\hbar\weffabs$.
This mirrors the effect of torsion and confirms that both control parameters, $B$ and $\Omega$, enter thermodynamics only through the single scale $\hbar\weffabs$.
In bulk samples, the longitudinal continuum simply offsets all curves by the universal $+\tfrac{1}{2}k_{B}$ [Eq.~\eqref{eq:Cv3D}].

\paragraph*{(ii) Torque magnetometry: diamagnetic $M(T)$ and susceptibility kernel.}
The magnetization and isothermal susceptibility,
Eqs.~\eqref{eq:M_theory}-\eqref{eq:chi_theory},
share the universal thermal kernel $1/\sinh^2(\hbar\weffabs/2k_BT)$,
leading to a torsion- (and field-) controlled crossover scale.
Figure~\ref{fig:magnetisation_vs_Omega} shows $M(T)$ fanning with $\Omega$,
while Fig.~\ref{fig:susceptibility_vs_Omega} isolates the kernel by normalizing
$\chi$ with the Curie factor $\chi_0(T)=(\hbar e/2\mu)^2/(k_B T)$.
Torque cantilevers routinely resolve these trends in 2D electron gases and vdW stacks.

\paragraph*{(iii) Magnetocaloric response and magnetic Gr\"uneisen parameter.}
Under adiabatic conditions,
\[
\Gamma_{B}(T)\equiv\Bigl(\tfrac{\partial T}{\partial B}\Bigr)_{S}
= -\,\frac{T}{C_V}\, \Bigl(\tfrac{\partial M}{\partial T}\Bigr)_{B},
\]
which yields the compact form \eqref{eq:MCE_xform}.
In the fixed-$k$ model, the kernel cancels exactly and
$\Gamma_{B}(T)=\mathrm{sgn}(\weff)\,(e/\mu)\,T/\weffabs$
[Eq.~\eqref{eq:MCE_linearT}], i.e., a strictly linear law with slope set by the inverse gap.
After integrating over the longitudinal motion (bulk 3D gas), the extra
$+\tfrac12 k_B$ in $C_V$ reduces the high-$T$ slope by a factor $2/3$
and smooths the crossover [Eq.~\eqref{eq:MCE_3D} and Eq.~\eqref{eq:MCE_3D_highT}],
as shown in Fig.~\ref{fig:gruneisen_vs_Omega_kint}.
This behavior can be probed either directly (through adiabatic field sweeps) or indirectly via
the magnetic Gr\"uneisen parameter extracted from $M(T)$ and $C_V(T)$.

\paragraph*{(iv) de Haas-van Alphen oscillations with torsion-shifted period.}
The oscillatory magnetization component,
Eq.~\eqref{eq:dHvA},
\[
M_{\mathrm{osc}}(B)=
-\frac{g\,\mu_F}{\hbar\weffabs}\,
\sin\Bigl(\tfrac{2\pi\mu_F}{\hbar\weffabs}\Bigr)\,
e^{-\pi/(\weffabs\tau_D)},
\]
displays a period in $1/B$ \emph{compressed} by the elastic contribution to $\weff=\omega_c+\omega_{cl}$
and an amplitude reduced by both the $1/\weffabs$ prefactor and Dingle damping.
Figure~\ref{fig:dHvA} shows the systematic shift with $\Omega$, which provides a
direct metrology of the screw-dislocation density via standard torque or SQUID magnetometry.

\paragraph*{(v) Finite-size corrections and mesoscopic oscillations.}
In patterned flakes and quantum rings, edge-state counting modifies the Landau degeneracy
$g\to g_{\mathrm{corr}}$ [Eq.~\eqref{eq:finite_size_correction}],
inducing oscillatory corrections in $C_V$ (Fig.~\ref{fig:finite_size})
whose period again follows $1/\weffabs$ and whose amplitude scales with $\ell_{\mathrm{eff}}/R$.
This establishes a clear mesoscopic diagnostic of elastic Landau quantization in confined geometries.

\paragraph*{Order-of-magnitude scales.}
For representative parameters ($\mu$ the electron mass, $k\sim10^9\,\mathrm{m^{-1}}$,
$\Omega\sim10^7$-$10^8\,\mathrm{m^{-1}}$), the elastic contribution
$\omega_{cl}=2\hbar k\Omega/\mu$ yields gaps
$\hbar\weffabs/k_B$ in the few-tens of kelvin range, placing the Schottky peak,
the magnetocaloric linear regime, and the Gr\"uneisen crossover within reach of
dilution to helium cryostats. The compensated field
$B_{\mathrm{comp}}=-2\hbar k\Omega/e$ [Eq.~\eqref{eq:B_comp}] then lies
in the 1-10~T decade for the same parameters.

To facilitate direct comparison, we provide:
(i) explicit $T$- and $B$-scalings for $C_V$ and $M$,
(ii) the compensated-field prediction,
(iii) a kernel-only normalization for $\chi$ and $\chi_{\Omega}$,
(iv) bulk vs fixed-$k$ distinctions for $\Gamma_B$,
and (v) finite-size benchmarks. These items allow experimentalists to
map their platform (material, strain, geometry) onto the single
control scale $\hbar\weffabs$ and to select the most sensitive observable.

\section{Experimental signatures and measurement roadmap}
\label{subsec:exp_signatures}

Figures~\ref{fig:tableA} and \ref{fig:tableB} condense the main experimental routes to test our approach, organizing signatures (first column), the associated observable and technique (second and third columns), and the predicted scalings/laws controlled by the single scale $\hbar\weffabs$ (fourth column). The last two columns highlight the control parameters ($B$, $\Omega$) and typical orders of magnitude under the working assumption of an electronic effective mass and $k\sim10^{9}\,\mathrm{m^{-1}}$.

\paragraph*{Thermal sector.}
(i) The Schottky-like peak of the transverse heat capacity $C_V(T)$ follows Eq.~\eqref{eq:heat_capacity}, peaking at $k_B T^\star\sim\hbar\weffabs$ and shifting to higher $T$ as either $B$ or $\Omega$ increases. This is the most direct thermometer of the level spacing and can be accessed with AC or pulsed micro/nano-calorimetry on chips or suspended membranes.  
(ii) For bulk, the longitudinal continuum adds the universal offset $+\tfrac{1}{2}k_B$ [Eq.~\eqref{eq:Cv3D}], which drives the high-$T$ limit to $C_V\to\tfrac{3}{2}k_B$. This offset is parameter-independent and provides a clear internal check of dimensionality (cf.\ Sec.~\ref{subsec:k_integration_details}).  
(iii) At fixed $T$, sweeping $B$ through the compensation point $B_{\mathrm{comp}}=-2\hbar k\Omega/e$ [Eq.~\eqref{eq:B_comp}] collapses the gap and produces a plateau $C_V/k_B\to1$ for the transverse sector, as shown in Fig.~\ref{fig:compensated_field}. Field-swept calorimetry thus offers an unambiguous handle on the interplay between torsion and magnetic field.

\paragraph*{Magnetization sector.}
(iv) The diamagnetic kernel is encoded by Eq.~\eqref{eq:M_theory}, with the sign tracked by $\mathrm{sgn}(\weff)$. Torque magnetometry (cantilever) or SQUID-VSM can resolve the predicted $M(T)$ trends (Fig.~\ref{fig:magnetisation_vs_Omega}).  
(v) The Curie-normalized susceptibility $\chi/\chi_0$ removes the trivial $1/T$ factor and isolates the universal kernel $1/\sinh^2 x$ [Eq.~\eqref{eq:chi_theory}], producing the family of curves in Fig.~\ref{fig:susceptibility_vs_Omega}. Increasing $\Omega$ (or $B$) pushes the onset of the strong-susceptibility regime to higher $T$, directly revealing that torsion stiffens the orbital spectrum.

\paragraph*{Magnetocaloric responses.}
(vi) The adiabatic magnetocaloric coefficient $\Gamma_B=(\partial T/\partial B)_S$ follows Eq.~\eqref{eq:MCE_xform}. In the fixed-$k$ model, the thermal kernel cancels and one obtains the exact linear law in Eq.~\eqref{eq:MCE_linearT}; for the bulk 3D gas, the longitudinal $+\tfrac{1}{2}k_B$ reduces the high-$T$ slope by $2/3$, Eq.~\eqref{eq:MCE_3D_highT}. These predictions can be extracted either from direct adiabatic sweeps or indirectly by combining $M(T)$ and $C_V(T)$ datasets.

\paragraph*{Quantum oscillations and mesoscopics.}
(vii) In the grand-canonical setting, the de Haas-van Alphen oscillations acquire the torsion-shifted frequency in $1/B$ [Eq.~\eqref{eq:dHvA}], with amplitude controlled by $1/\weffabs$ and Dingle damping $e^{-\pi/(\weffabs\tau_D)}$. Torsion ($\Omega$) thus \emph{compresses} the period in $1/B$, a clean diagnostic in torque/SQUID experiments.  
(viii) Finite-size caloric oscillations arise from the degeneracy correction $g\to g_{\mathrm{corr}}=g(1-\ell_{\mathrm{eff}}/R)$, Eq.~\eqref{eq:finite_size_correction}, with $\ell_{\mathrm{eff}}=\sqrt{\hbar/(\mu\weffabs)}$. Mesoscopic rings, flakes or patterned 2D crystals should display $\Delta C_V(B)$ oscillations whose amplitude scales with $\ell_{\mathrm{eff}}/R$ and whose period follows $1/\weffabs$ (Fig.~\ref{fig:finite_size}).

\paragraph*{How to use the table.}
The entries provide a practical roadmap to \emph{(a)} choose the observable and instrument, \emph{(b)} identify the scaling law to fit (with equation references), and \emph{(c)} select the control parameter ($B$ or $\Omega$) to move the relevant crossover. In particular, combining:  
(1) the position of the Schottky peak (or the $C_V$ plateau at $B_{\mathrm{comp}}$),  
(2) the linear-in-$T$ magnetocaloric slope, and  
(3) the $1/B$ period of $M_{\mathrm{osc}}$,  
yields three mutually consistent determinations of $\weffabs$, providing a redundant and robust experimental validation of the theory.

\section{Applications and device-level opportunities}
\label{sec:apps}

Beyond serving as a compact framework to predict caloric and magnetic responses,
the single-scale organization by $\hbar\weffabs$ enables device-level concepts. Below, we outline representative use cases that follow directly from
Eqs.~\eqref{eq:heat_capacity}, \eqref{eq:Cv3D}, \eqref{eq:M_theory},
\eqref{eq:chi_theory}, \eqref{eq:MCE_xform}-\eqref{eq:MCE_3D_highT},
\eqref{eq:Pi_Omega}-\eqref{eq:chi_Omega}, \eqref{eq:dHvA} and
the compensated-field condition \eqref{eq:B_comp}.

\paragraph*{(A) Torsion-assisted magnetocaloric microcooler.}
Use the strictly linear adiabatic coefficient in the fixed-$k$ model, $\Gamma_B=(\partial T/\partial B)_S=\mathrm{sgn}(\weff)\,(e/\mu)\,T/\weffabs$ [Eq.~\eqref{eq:MCE_linearT}], and its 3D version with the $2/3$ reduction [Eq.~\eqref{eq:MCE_3D_highT}], to design field sweeps that maximize $\Delta T$ at a
given $T$. Since $\weffabs=\omega_c+\omega_{cl}$, increasing $\Omega$ reduces the slope, so an \emph{adaptive} protocol-high $\Omega$ during precooling (minimizing entropy leakage),
low $\Omega$ during the adiabatic sweep optimizes the net temperature drop. Figure~\ref{fig:magnetocaloric_vs_Omega} provides the slope budget versus $\Omega$.

\paragraph*{(B) Dislocation-density metrology (dual-probe).}
Combine (i) the $C_V$ plateau at the compensated field
$B_{\mathrm{comp}}=-2\hbar k\Omega/e$ [Eq.~\eqref{eq:B_comp},
Fig.~\ref{fig:compensated_field}] with (ii) the torsion-compressed dHvA period in $1/B$ [Eq.~\eqref{eq:dHvA}, Fig.~\ref{fig:dHvA}]. The two independent determinations of $\weff$ (thermal and magnetic)
triangulate $\Omega$ and $k$ without fitting ambiguities. A single device can therefore act as an \emph{in situ} dislocation-density gauge.

\paragraph*{(C) Magnetoelastic “heat switch”.} Near $B_{\mathrm{comp}}$ the transverse gap collapses and $C_V/k_B \to 1$
[Eq.~\eqref{eq:heat_capacity}, Fig.~\ref{fig:compensated_field}],
whereas away from compensation, the Schottky gap suppresses heat capacity. Sweeping a small $\delta B$ around $B_{\mathrm{comp}}$ toggles the effective thermal capacity of the electron gas, implementing a field-tunable heat switch for cryogenic stabilization or periodic load matching.

\paragraph*{(D) Torsional work transducer and dilatometry readout.}
Because $\Pi_{\Omega}=-(\partial A/\partial\Omega)_{T,B}$ is the
\emph{generalized force} conjugate to $\Omega$ [Eq.~\eqref{eq:Pi_Omega}], a quasistatic stroke in $\Omega$ performs mechanical work $\delta W_{\Omega}=\Pi_{\Omega}\,d\Omega$. The differential sensitivity is set by $\chi_{\Omega}=(\partial \Pi_{\Omega}/\partial\Omega)_{T,B}$ [Eq.~\eqref{eq:chi_Omega}], which shares the universal $1/\sinh^2 x$ kernel (Fig.~\ref{fig:chiOmega_vs_Omega}). This suggests magnetoelastic transducers, where controlled dislocation injection (or strain-patterning) converts electromagnetic control ($B$) into measurable length changes via high-resolution dilatometry.

\paragraph*{(E) Mesoscopic caloric interferometer.} In patterned flakes (rings/dots), the degeneracy correction $g\to g_{\mathrm{corr}}=g(1-\ell_{\mathrm{eff}}/R)$ [Eq.~\eqref{eq:finite_size_correction}] induces oscillatory
$\Delta C_V(B)$ (Fig.~\ref{fig:finite_size}) with period set by $1/\weffabs$. Because $\ell_{\mathrm{eff}}=\sqrt{\hbar/(\mu\weffabs)}$, co-tuning $(B,\Omega)$
modulates both the period and amplitude. The device can function as a
\emph{phase-sensitive} probe of edge channels in the elastic-Landau regime.

\medskip
Each concept relies only on the single control scale $\hbar\weffabs$ and on the
sign-safe derivatives, which are already derived in the main text, so no additional microscopic assumptions are required. Prototype demonstrations can leverage the measurement roadmap of Sec.~\ref{subsec:exp_relevance}.

\subsection{Quantitative case studies and benchmarks}
\label{subsec:case_studies}

To aid experimental planning, we list concrete numbers obtained by evaluating our closed-form expressions at representative parameters
($\mu$ the electron mass, $k=10^{9}\,\mathrm{m^{-1}}$, $B=5\,\mathrm{T}$, $\mu_F=0.05\,\mathrm{eV}$). The effective cyclotron frequency is $\omega_{\mathrm{eff}}=\omega_c+ \omega_{cl}$ with
$\omega_c=eB/\mu$ and $\omega_{cl}=2\hbar k\Omega/\mu$.

\paragraph*{Thermal crossover scale $T^\star$.}
The single-scale crossover $k_BT^\star=\hbar\weffabs$ (Schottky peak position and kernel midpoint) directly sets the temperature window where all responses change rapidly.

\paragraph*{Magnetocaloric coefficient $\Gamma_B$.}
In the fixed-$k$ model, the adiabatic coefficient is strictly linear,
$\Gamma_B(T) = \mathrm{sgn}(\omega_{\mathrm{eff}})\,(e/\mu)\,T/\weffabs$
[Eq.~\eqref{eq:MCE_linearT}],
so a field sweep $\Delta B$ induces $\Delta T \approx \Gamma_B\,\Delta B$.
Including the longitudinal continuum reduces the high-$T$ slope by $2/3$
[Eq.~\eqref{eq:MCE_3D_highT}].

\paragraph*{Compensated field.}
The gap collapses when $\omega_{\mathrm{eff}}=0$, i.e.
$B_{\mathrm{comp}}=-2\hbar k\Omega/e$ [Eq.~\eqref{eq:B_comp}],
producing the transverse equipartition plateau in $C_V$.

\paragraph*{dHvA local period in $1/B$.} Using the oscillatory magnetization [Eq.~\eqref{eq:dHvA}], the phase $\varphi(B)=2\pi\mu_F/(\hbar\weffabs(B))$ gives a
\emph{local} period $\Delta(1/B)$ near a working field $B_0$ from
$\varphi(1/(1/B_0+\Delta(1/B)))=\varphi(1/B_0)+2\pi$. Because $\weffabs$ grows with $\Omega$, the period in $1/B$ \emph{decreases} (oscillations become more frequent) as the torsion density increases.

\begin{table}[t]
\caption{Concrete benchmarks at $B=5\,\mathrm{T}$ for four torsion densities.
Here, $T^\star=\hbar\weffabs/k_B$, $\Gamma_B(40\,\mathrm{K})=(e/\mu)\,T/\weffabs$ (fixed-$k$ value; multiply by $2/3$ for the 3D high-$T$ limit),
$B_{\mathrm{comp}}=-2\hbar k\Omega/e$, and $\Delta(1/B)$ denotes the local dHvA period around $B_0=5\,\mathrm{T}$.
}
\label{tab:benchmarks}
\centering
\footnotesize
\begin{tabular}{lcccc}
\toprule
$\Omega$ $[\times10^{7}\,\mathrm{m^{-1}}]$ & 0.0 & 0.5 & 1.0 & 1.5 \\
\midrule
$T^\star$ (K) & 6.71 & 10.76 & 14.82 & 18.88 \\
$\Gamma_B(40\,\mathrm{K})$ (K/T) & 8.04 & 5.03 & 3.67 & 2.96 \\
$B_{\mathrm{comp}}$ (T) & 0 & $-6.58$ & $-13.16$ & $-19.73$ \\
$\Delta(1/B)$ at 5 T (T$^{-1}$) & 0.00232 & 0.00155 & 0.00120 & 0.00096 \\
\bottomrule
\end{tabular}
\end{table}

\noindent\emph{Reading the table.}  
(1) The crossover $T^\star$ moves linearly with $\Omega$, setting the temperature where $C_V$, $\chi$ and $\Gamma_B$ change most rapidly.  
(2) At $T=40\,\mathrm{K}$, a $1\,\mathrm{T}$ sweep yields $\Delta T\simeq 8.0$, $5.0$, $3.7$, $3.0\,\mathrm{K}$ for $\Omega=0,\,0.5,\,1.0,\,1.5\times10^{7}\,\mathrm{m^{-1}}$, respectively (fixed-$k$); in bulk 3D the asymptotic high-$T$ slope is reduced by $2/3$.  
(3) The compensation field lands in the $1$-$20\,\mathrm{T}$ decade for the working $\Omega$ values, making the $C_V/k_B\to 1$ plateau experimentally accessible [Fig.~\ref{fig:compensated_field}]. (4) The dHvA period in $1/B$ \emph{decreases} with $\Omega$ (more oscillations per tesla in inverse field), while the amplitude is reduced by both the $1/\weffabs$ prefactor and Dingle damping [Eq.~\eqref{eq:dHvA}, Fig.~\ref{fig:dHvA}].

\medskip
These benchmarks turn the abstract scalings into actionable targets for torque magnetometry, micro/nano-calorimetry, and magnetocaloric measurements. They also provide cross-checks: $T^\star$ from the Schottky peak, the linear slope of $\Gamma_B(T)$, and the local dHvA period all yield consistent estimates of $\weffabs$ at the same operating point.

\section{Disorder and interactions: an effective-DOS extension}
\label{subsec:disorder_interactions}

In the presence of weak-to-moderate quenched disorder and screened interactions, the elastic-Landau ladder $E_{N'}=\hbar\weffabs\,(N'+\tfrac12)$ is accurately captured by a broadened, renormalized spectrum. Thermodynamics then follows from an effective density of states (DOS) characterized by a spectral width~$\Gamma$ and a renormalized gap scale~$\weff^\star$.

\paragraph*{Broadened DOS.}
Let $\varphi_{\Gamma}(E)$ be a normalized lineshape (Lorentzian or Gaussian), $\int dE\,\varphi_{\Gamma}=1$. The DOS reads
\begin{equation}
\rho(E)=\frac{g}{\hbar\weffabs^\star}\sum_{N'=0}^{\infty}
\varphi_{\Gamma}\Bigl(E-\hbar\weffabs^\star\bigl(N'+\tfrac12\bigr)\Bigr),
\label{eq:DOS_broadened}
\end{equation}
and the grand potential is obtained by convolution,
\begin{equation}
\Omega_G(T,\mu_F)
= -k_BT\int_{-\infty}^{\infty}\rho(E)\,
\ln\bigl[1+e^{-\beta(E-\mu_F)}\bigr]\,dE.
\label{eq:Omega_from_DOS}
\end{equation}
For canonical quantities, the same DOS enters the partition function and its derivatives.

\paragraph*{Thermal kernels and level broadening.}
Define $x^\star=\hbar\weffabs^\star/(2k_BT)$ and the Schottky kernel
$\Phi(x)=x^2\csch^2 x$ (cf.\ Eq.~\eqref{eq:heat_capacity}/$k_B$).
For a narrow lineshape ($\Gamma\ll k_BT$),
\begin{equation}
\frac{C_V(T)}{k_B}\;=\;
\Phi\bigl(x^\star\bigr)
\;+\; O\bigl((\Gamma/k_BT)^2\bigr),
\label{eq:Cv_with_broadening}
\end{equation}
i.e., the universal thermal kernel is preserved, acquiring only weak, $T$-dependent rounding. The same holds for the Curie-normalized responses $\chi/\chi_0$ and $\chi_\Omega/\chi_{\Omega0}$, which isolate $1/\sinh^2 x^\star$ (Eqs.~\eqref{eq:chi_theory}, \eqref{eq:chi_Omega}).

\paragraph*{Oscillatory sector and Dingle damping.}
Poisson resummation in the broadened ladder yields the standard Dingle factor for the $r$-th harmonic,
\begin{equation}
R_D(r,\tau_D)=\exp\Bigl[-\tfrac{\pi r}{\weffabs^\star \tau_D}\Bigr],
\qquad \Gamma=\hbar/(2\tau_D),
\label{eq:Dingle_disorder}
\end{equation}
multiplying the Lifshitz-Kosevich thermal factor
$R_T(r,T)=X_r/\sinh X_r$ with
$X_r=2\pi^2 r k_BT/(\hbar\weffabs^\star)$
[cf.\ Eqs.~\eqref{eq:RT}-\eqref{eq:RD}].
The oscillatory magnetization generalizes to
\begin{align}
M_{\mathrm{osc}}(B)
&= -\,\mathrm{sgn}(\weff^\star)\,
\frac{g\,\mu_F}{\hbar\weffabs^\star}\,
\notag\\ &\times \sum_{r\ge 1}\frac{\sin\bigl(2\pi r\,\mu_F/\hbar\weffabs^\star\bigr)}{\pi r}\,
R_T\,R_D ,
\label{eq:Mosc_with_disorder}
\end{align}
reducing to Eq.~\eqref{eq:dHvA} for $r=1$, low $T$, and finite $\tau_D$.

\paragraph*{Weak-interaction renormalizations.}
Screened interactions are incorporated via effective parameters,
\begin{align}
\mu \to \mu^\star,\quad
\omega_c^\star=\frac{eB}{\mu^\star},\quad
\omega_{cl}^\star=\frac{2\hbar k\Omega}{\mu^\star},\quad
\weff^\star=\omega_c^\star+\omega_{cl}^\star,
\label{eq:renorm_mass}
\end{align}
and, in the spin/antisymmetric channel, by a Landau factor in the susceptibility prefactor,
\begin{equation}
\chi(B,\Omega,T)\;\mapsto\;
\frac{\chi(B,\Omega,T)}{1+F_0^a},
\label{eq:FL_chi}
\end{equation}
leaving the kernel $ \csch^2 x^\star$ intact with $x^\star=\hbar\weffabs^\star/(2k_BT)$.
With the replacements $\mu\to\mu^\star$, $\weff\to\weff^\star$,
the clean-limit results
\eqref{eq:M_theory}-\eqref{eq:chi_theory},
\eqref{eq:MCE_xform}-\eqref{eq:MCE_3D_highT},
\eqref{eq:Gamma_infty}
remain valid verbatim.

\paragraph*{Slow inhomogeneity of torsion.}
For weak spatial variation $\Omega(\mathbf r)$, the gap becomes a random field $\weff^\star(\mathbf r)$. Coarse-grained observables follow from averaging over a distribution $P(\weff^\star)$:
\begin{equation}
\big\langle C_V(T)\big\rangle
= \int d\weff^\star\,P(\weff^\star)\,
k_B\left(\frac{\hbar|\weff^\star|}{2k_BT}\right)^{2}
\csch^{2}\left(\frac{\hbar|\weff^\star|}{2k_BT}\right),
\label{eq:inhom_average}
\end{equation}
which broadens crossovers while preserving their scaling with the central gap.

\paragraph*{Bulk 3D limit with longitudinal integration.}
In the $k$-integrated description (Sec.~\ref{subsec:k_integration_details}),
disorder and interactions leave the longitudinal offset unchanged:
\begin{equation}
C_{V}^{3\mathrm{D}}(T)
= k_B\,x^{\star 2}\csch^2 x^\star + \frac{1}{2}k_B 
+ O\bigl((\Gamma/k_BT)^2\bigr),
\label{eq:Cv3D_disorder_interactions}
\end{equation}
where $x^\star=\hbar\weffabs^\star/2k_BT$.
Consequently, the high-$T$ limit $C_V^{3\mathrm{D}}\to \tfrac{3}{2}k_B$ is robust, and the magnetocaloric/Gr\"{u}neisen asymptotes follow with $\weff^\star$ [Eqs.~\eqref{eq:MCE_3D_highT}, \eqref{eq:Gamma_infty}].

\paragraph*{Practical parametrization.}
For data analysis and figure generation, use
\[
\mu\to\mu^\star,\;\;\;   
\weff\to\weff^\star,\;\;\;  
M_{\mathrm{osc}}\to M_{\mathrm{osc}}\times e^{-\pi/(\weffabs^\star\tau_D)}.
\]
A finite experimental resolution can be mimicked by convolving clean expressions with $\varphi_\Gamma$; this rounds features without shifting the gap-controlled scales set by $x^\star$.

Within weak disorder and weak-to-moderate interactions, the \emph{kernel structure} of elastic-Landau thermodynamics is preserved: $C_V$, $\chi$, and $\chi_\Omega$ retain the $1/\sinh^2 x^\star$ kernels; magnetocaloric and Gr\"{u}neisen relations keep their closed forms with $\weff^\star$; and oscillatory responses acquire LK-Dingle damping with $\tau_D$. This furnishes a compact, experimentally ready parametrization in terms of $(\mu^\star,\weff^\star,\Gamma,\tau_D,F_0^a)$.

\section{Caloritronics and solid-state cooling}
\label{sec:caloritronics}

Modern caloritronics investigates how heat and entropy can be generated, routed, and converted at the nanoscale with quantum control, leveraging ultrasensitive micro/nano-calorimeters and rapidly advancing device integration. Recent chip-based platforms enable \emph{in situ} calorimetry on micron membranes at high sweep rates and sub-nanojoule resolution, opening access to low-energy excitations, phase transitions, and dissipation in mesoscopic systems~\cite{Minakov2021_applsci,PNAS.2022.119.e2205322119}. Parallel progress in spin and thermomagnetic devices has pushed heat-to-spin and spin-to-heat interconversion (spin Seebeck/Peltier) toward practical figures of merit and system-level demonstrations~\cite{APMT.2023.32.101846,Guo2024_natrevphys,NPJS.2024.236}. 

In the elastic-Landau problem, all thermodynamics is organized by the single, tunable gap $\hbar|\omega_{\mathrm{eff}}|$. This renders classic  \cite{PRB.2004.70.174429}, magnetocaloric response, and the magnetic Gr\"{u}neisen parameter - \emph{scale thermometers} directly tied to magnetoelastic control. Because $\omega_{\mathrm{eff}}=\omega_c+\omega_{cl}$ blends electromagnetic ($B$) and structural ($\Omega$) parameters, the same calorimetric protocol that tunes field can quantify strain/dislocation densities, while on-chip architectures~\cite{PNAS.2022.119.e2205322119,Minakov2021_applsci} provide the temporal and thermal resolution required to map $C_V(T,B,\Omega)$ in small samples and vdW stacks. Below, we exploit this organizing principle to propose cooling cycles and heat-switching strategies based on compensated-field operation and kernel cancellations specific to our model.

Caloritronics leverages field-tunable low-energy scales to program heat capacities, entropies, and adiabatic temperature changes in mesoscopic devices~\cite{GiazottoRMP2006,PekolaNatPhys2015}. In our system, all thermodynamics is governed by a \emph{single control scale}, the elastic-Landau gap $\hbar|\omega_{\mathrm{eff}}|$ with
\begin{equation}
\omega_{\mathrm{eff}}=\omega_c+\omega_{cl}= \frac{eB}{\mu}+\frac{2\hbar k\,\Omega}{\mu},
\label{eq:weff_calo}
\end{equation}
which unifies magnetic field $B$ and screw-dislocation density $\Omega$ into a compact operative parameter. The partition function and its derivatives (Sec.~\ref{sec2a}-\ref{sec4}) yield closed expressions for $A,U,S,C_V,M,\chi$ in terms of $x=\hbar|\omega_{\mathrm{eff}}|/(2k_B T)$. Here we exploit these results to (i) chart the \emph{operational caloritronic landscape}, (ii) specify a concrete \emph{adiabatic microcooling protocol} near the compensated line $\omega_{\mathrm{eff}}=0$, and (iii) identify limits and figures of merit relevant to on-chip calorimetry~\cite{SullivanSeidel1968,RSI.2012.83.055107}.

\subsection{Operational landscape: Schottky kernel, 3D offset, and Gr\"uneisen plateau}
\label{subsec:calo_landscape}

For the transverse ladder (fixed longitudinal $k$), the heat capacity per particle reads
\begin{equation}
\frac{C_V}{k_B}= x^{2}\,\mathrm{csch}^{2}x, 
\qquad 
x=\frac{\hbar|\omega_{\mathrm{eff}}|}{2k_B T},
\label{eq:Cv_kernel_letter}
\end{equation}
which produces a Schottky-like peak at $k_B T^\star \sim \hbar|\omega_{\mathrm{eff}}|$. Increasing either $B$ or $\Omega$ shifts $T^\star$ upward, suppresses $S(T)$ at fixed temperature, and \emph{stiffens} the caloric response (Sec.~\ref{sec3}). In bulk, integrating over the free longitudinal motion adds a universal offset,
\begin{equation}
C_V^{3\mathrm{D}}=C_V+\tfrac{1}{2}k_B,
\label{eq:calo_3Doffset}
\end{equation}
which clarifies the crossover to the classical $3\mathrm{D}$ limit $C_V\to \tfrac{3}{2}k_B$ (Sec.~\ref{subsec:k_integration_details}).

The adiabatic magnetocaloric coefficient,
\begin{equation}
\Gamma_B(T)\equiv\left(\frac{\partial T}{\partial B}\right)_S
= -\,\frac{T}{C_V}\left(\frac{\partial M}{\partial T}\right)_B,
\label{eq:Gamma_def_letter}
\end{equation}
admits an exceptionally simple law in our model: for the fixed-$k$ spectrum the thermal kernel cancels exactly, yielding the \emph{linear} relation
\begin{equation}
\Gamma^{\mathrm{fixed}\,k}_B(T)=\frac{e}{\mu}\,\frac{T}{\omega_{\mathrm{eff}}},
\label{eq:Gamma_linear_letter}
\end{equation}
while after $k$-integration one obtains the smooth 3D crossover (Sec.~\ref{subsec:magnetocaloric})
\begin{equation}
\Gamma_B^{3\mathrm{D}}(T)\xrightarrow[T\to\infty]{} -\,\frac{2}{3}\,\frac{e}{\mu\,\omega_{\mathrm{eff}}},
\label{eq:Gamma_2over3}
\end{equation}
i.e., the high-$T$ slope is reduced by a universal factor $2/3$. The same factor stabilizes device performance against sample-to-sample variations in $\omega_{\mathrm{eff}}$ at elevated temperatures.

\paragraph*{Compensated line and equipartition plateau.}
At $\omega_{\mathrm{eff}}=0$ the transverse gap collapses and the transverse heat capacity attains the equipartition plateau $C_V/k_B\to 1$ (Sec.~\ref{subsec:compensated}). In practice, sweeps across the \emph{compensated field}
\begin{equation}
B_{\mathrm{comp}}(\Omega)=-\frac{2\hbar k}{e}\,\Omega,
\label{eq:Bcomp_line_letter}
\end{equation}
reveal a pronounced $C_V(B)$ enhancement (Fig.~\ref{fig:compensated_field}), and provide a direct calibration of $\Omega$ (linear slope $-2\hbar k/e$).
\begin{figure}[tbhp]
  \centering
  \includegraphics[width=0.99\linewidth]{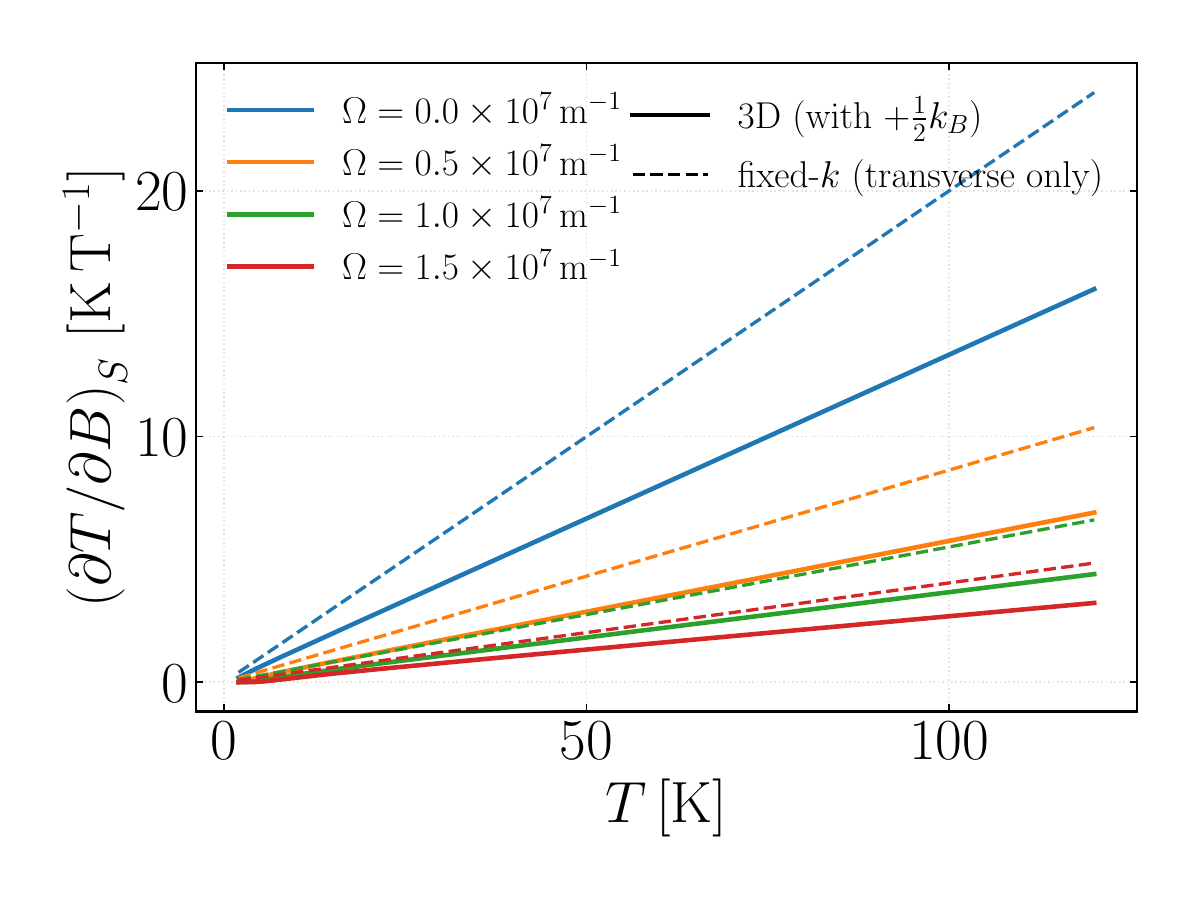}
  \caption{\footnotesize Magnetocaloric coefficient at constant entropy,
  $\Gamma_B(T)\equiv(\partial T/\partial B)_S$, as a function of temperature for representative torsion densities $\Omega$.
  Solid curves: full 3D model, which includes the additional \emph{longitudinal} contribution $\tfrac{1}{2}k_B$;
  dashed curves: fixed-$k$ (transverse-only) reference.
  Torsion suppresses $\Gamma_B$ at all temperatures, and in the high-$T$ limit the 3D slopes are reduced by a factor $2/3$ relative to the fixed-$k$ case.
  The magnetic field value used in the calculation is the same as in Sec.~\ref{sec:caloritronics} and specified in the main text.}
  \label{fig:MCE_vsT_varOmega}
\end{figure}

The magnetocaloric response at constant entropy \eqref{eq:Gamma_def_letter} ties the sign and magnitude of the temperature change under field sweeps to the thermal derivative of the magnetization and to the heat capacity $C_V$. Figure~\ref{fig:MCE_vsT_varOmega} displays $\Gamma_B(T)$ for representative torsion densities $\Omega$. All curves are positive, signaling heating upon increasing $B$ at fixed $S$ (consistent with $\partial M/\partial T<0$ in the Landau-diamagnetic regime). Torsion systematically suppresses $\Gamma_B$ over the whole temperature range, as it both reduces the magnitude of the $M(T)$ slope and mildly enhances $C_V$ via the extra longitudinal contribution. At low $T$, $\Gamma_B$ grows smoothly from zero as the lowest Landau ladder becomes thermally populated; no sharp features emerge because the torsional scale smears level spacings on the relevant window. For $k_BT\gg\hbar\omega_c$, $\Gamma_B$ is nearly linear in $T$ and its slope is controlled by the effective heat-capacity count: in the full 3D description (solid lines) $C_V\simeq \tfrac{3}{2}k_B$, whereas in the fixed-$k$ transverse-only reference (dashed lines) $C_V\simeq k_B$, yielding a universal $\sim2/3$ reduction of the high-$T$ slope—precisely as observed. Increasing $\Omega$ further lowers the slope by diminishing $|\partial M/\partial T|$, which provides a practical metrological handle: a single high-$T$ calibration of $\Gamma_B/T$ versus $\Omega$ allows one to \emph{quantify} the torsion density from standard magnetocaloric measurements.

\subsection{A compensation-line microcooler: cycle and figures of merit}
\label{subsec:microcooler}

We now outline a concrete magnetocaloric microcycle tailored to our platform. The guiding principles are: (i) exploit Eq.~\eqref{eq:Gamma_linear_letter} near (but not at) the compensated line to maximize $|\Gamma_B|$; (ii) leverage the universal offset~\eqref{eq:calo_3Doffset} to stabilize performance in bulk devices; and (iii) mitigate entropy leakage by adapting $\Omega$ during the cycle.

\paragraph*{Cycle definition.}
\emph{Step 1 (pre-cool, entropy guard).} Set a \emph{large} torsion density $\Omega_{\mathrm{hi}}$ so that $|\omega_{\mathrm{eff}}|$ is large, $C_V$ is gapped at the working $T$, and $|\Gamma_B|$ is modest. Thermalize the device and load to the base temperature $T_{\mathrm{in}}$.  

\emph{Step 2 (gain staging).} Reduce the torsion to $\Omega_{\mathrm{lo}}$ such that $|\omega_{\mathrm{eff}}|$ decreases and $|\Gamma_B|$ increases according to Eq.~\eqref{eq:Gamma_linear_letter}, but \emph{keep} $|\omega_{\mathrm{eff}}|$ finite to avoid disorder-driven level mixing.

\emph{Step 3 (adiabatic sweep).} Perform a quasi-adiabatic field sweep $\Delta B$ around $B_{\mathrm{comp}}(\Omega_{\mathrm{lo}})$; the temperature drop is $\Delta T\simeq \Gamma_B\,\Delta B$, with $\Gamma_B$ taken from Eq.~\eqref{eq:Gamma_linear_letter} (or the 3D expression for bulk).  

\emph{Step 4 (load thermalization and reset).} Thermalize the load at the reduced temperature $T_{\mathrm{out}}=T_{\mathrm{in}}+\Delta T$. Restore $\Omega\to\Omega_{\mathrm{hi}}$ for the next cycle.

\paragraph*{Performance window and optimum distance to compensation.}
At fixed $T$, $|\Gamma_B|\propto 1/|\omega_{\mathrm{eff}}|$ suggests working as close as possible to $\omega_{\mathrm{eff}}=0$. However, disorder broadening (Dingle width) and slow heat leaks penalize excessively small gaps, where $C_V$ becomes too large and the adiabaticity window narrows. There exists an \emph{optimum} distance $|\omega_{\mathrm{eff}}|_{\mathrm{opt}}$ that maximizes the net $\Delta T$ per cycle under realistic leakage rates; this optimum depends on the longitudinal contribution $+\tfrac12k_B$ and on the thermal time constant of the platform. In membrane-based nano-calorimeters~\cite{RSI.2012.83.055107}, the reduced addenda and excellent thermal isolation enlarge the practical window.

A convenient figure of merit is the differential \emph{adiabatic gain}
\begin{equation}
\mathcal{G}\equiv \frac{1}{\Delta B}\int_{\mathrm{sweep}} \Bigl|\frac{\partial T}{\partial B}\Bigr|_{S} dB,
\qquad 
\mathcal{G}\approx \Bigl|\frac{e}{\mu}\Bigr|\frac{T}{|\omega_{\mathrm{eff}}|},
\label{eq:gain_letter}
\end{equation}
evaluated at the working point; for 3D replace $e/\mu\to (2/3)\,e/\mu$ at high $T$, Eq.~\eqref{eq:Gamma_2over3}. A second metric is the \emph{stability ratio} $\mathcal{S}\equiv C_V^{3\mathrm{D}}/C_V$, which quantifies the contribution from the longitudinal continuum: $\mathcal{S}\to\infty$ at low $T$ (gapped) and $\mathcal{S}\to 3/2$ at high $T$. Operating where $\mathcal{S}$ is moderately smooths cycle-to-cycle variations without sacrificing too much gain.

\subsection{Experimental routes and readout}
\label{subsec:exp_routes}

\emph{On-chip calorimetry.} AC or pulsed calorimetry in micromachine platforms~\cite{SullivanSeidel1968,RSI.2012.83.055107} can track $C_V(T,B,\Omega)$ and $\Gamma_B(T,B,\Omega)$ with picojoule sensitivity, directly verifying: (i) the Schottky kernel~\eqref{eq:Cv_kernel_letter}; (ii) the universal $+\tfrac12k_B$ offset~\eqref{eq:calo_3Doffset}; (iii) the compensated-field plateau $C_V/k_B\to 1$; and (iv) the linear $\Gamma_B(T)$ law~\eqref{eq:Gamma_linear_letter} with the $2/3$ high-$T$ reduction~\eqref{eq:Gamma_2over3}.

\emph{Torque magnetometry and Gr\"uneisen extraction.} Combining $M(T)$ and $C_V(T)$ datasets yields $\Gamma_B$ via Eq.~\eqref{eq:Gamma_def_letter}. The same setup enables fast scouting of the optimal operating window by monitoring the collapse of the thermal kernel as a function of $B$ and $\Omega$ (Sec.~\ref{subsec:mag_results}).

\paragraph*{Robustness and limits.}
(i) \emph{Disorder.} Level broadening attenuates the sharpness of the Schottky peak but preserves the kernel form and the $2/3$ reduction; its main impact is to set the minimum practical $|\omega_{\mathrm{eff}}|$.  
(ii) \emph{Classicalization.} For $k_B T\gg \hbar|\omega_{\mathrm{eff}}|$, the Bohr-van Leeuwen limit suppresses orbital magnetism, flattening $\Gamma_B$ and favoring operation at intermediate temperatures.  
(iii) \emph{Finite size.} In mesoscopic devices, the degeneracy correction $g\to g_{\mathrm{corr}}$ (Sec.~\ref{subsec:finite_size}) produces weak field-dependent oscillations in caloric observables; these provide an internal consistency check (period $\propto 1/|\omega_{\mathrm{eff}}|$).

The caloritronic sector of elastic Landau levels is organized by a single tunable scale $\hbar|\omega_{\mathrm{eff}}|$. This yields (i) a universal Schottky kernel for $C_V$, (ii) a robust $+\tfrac12 k_B$ longitudinal offset in bulk, (iii) a strictly linear adiabatic coefficient with a universal $2/3$ high-$T$ reduction, and (iv) a compensation-line microcooler with a well-defined optimum distance to $\omega_{\mathrm{eff}}=0$. These features are straightforward to probe with state-of-the-art membrane calorimetry and torque magnetometry~\cite{GiazottoRMP2006,PekolaNatPhys2015,RSI.2012.83.055107}.

\section{Strain engineering and defect control}
\label{sec:strain_engineering}

Strain engineering has emerged as a premier route to tailor band structures, excitons, and topology in 2D and layered materials, from graphene to TMDs, including the creation of giant pseudo\-magnetic fields by nonuniform deformations~\cite{Hsu2020_sciadv,MN.2024.10.49,NML.2025.17.104,Kovalchuk2022_israel}. At the continuum scale, the geometric theory of defects provides a natural language to encode dislocations via torsion in a Riemann-Cartan manifold, establishing a direct bridge between materials microstructure and effective gauge fields~\cite{PU.2005.48.675,Yavari2012_arma,RSOS.2024.11.240711}. 

Within this geometric backdrop, a uniform density of screw dislocations contributes an elastic cyclotron term $\omega_{cl}\propto k\,\Omega$ that adds algebraically to the magnetic cyclotron frequency, thereby placing strain and field on equal footing through $\omega_{\mathrm{eff}}=\omega_c+\omega_{cl}$. This identification turns $\Omega$ into a bona fide thermodynamic field with a conjugate $\Pi_\Omega$ and susceptibility $\chi_\Omega$ governed by universal kernels. In practice, it enables (i) quantitative magnetoelastic transduction (dilatometry/nanomechanics) near the Schottky scale, (ii) a direct calibration of $\Omega$ via the compensated field $B_{\mathrm{comp}}=-2\hbar k\,\Omega/e$, and (iii) sensitive finite-size caloric fingerprints that track edge contributions through degeneracy corrections. We detail operating windows and protocols that exploit recent strain patterning and microscopy advances in 2D materials and semiconductor membranes~\cite{Hsu2020_sciadv,MN.2024.10.49,NML.2025.17.104}.

Strain engineering leverages controlled defect landscapes to tailor low-energy scales and response functions in crystalline platforms and van der Waals stacks~\cite{LevyScience2010}. In our framework, the screw-dislocation density $\Omega$ enters on the \emph{same footing} as the magnetic field through the single scale $\hbar|\omega_{\mathrm{eff}}|$ with
\begin{equation}
\omega_{\mathrm{eff}}(B,\Omega)=\frac{eB}{\mu}+\frac{2\hbar k\,\Omega}{\mu}.
\label{eq:weff_strain}
\end{equation}
Treating $\Omega$ as a \emph{thermodynamic field} makes its conjugate variable
\begin{equation}
\Pi_{\Omega}\equiv -\bigl(\partial A/\partial\Omega\bigr)_{T,B}
= -\,\mathrm{sgn}(\omega_{\mathrm{eff}})\,\frac{\hbar^{2}k}{\mu}\,
\coth\Bigl(\frac{\hbar|\omega_{\mathrm{eff}}|}{2k_{B}T}\Bigr),
\label{eq:PiOmega_letter}
\end{equation}
and the associated susceptibility
\begin{equation}
\chi_{\Omega}\equiv (\partial \Pi_{\Omega}/\partial\Omega)_{T,B}
= \Bigl(\frac{\hbar^{2}k}{\mu}\Bigr)^{2}\,
\frac{1}{k_{B}T}\,
\frac{1}{\sinh^{2}\Bigl(\frac{\hbar|\omega_{\mathrm{eff}}|}{2k_{B}T}\Bigr)},
\label{eq:chiOmega_letter}
\end{equation}
both controlled by the universal thermal kernel $1/\sinh^{2}x$ with $x=\hbar|\omega_{\mathrm{eff}}|/(2k_{B}T)$ (Sec.~\ref{subsec:torsion_response}). Equations~\eqref{eq:PiOmega_letter}-\eqref{eq:chiOmega_letter} organize the strain sector into three experimentally actionable results: (i) a \emph{gain law} for magnetoelastic transduction; (ii) a \emph{calibration protocol} for $\Omega$; and (iii) finite-size indicators via degeneracy corrections.

\subsection{Magnetoelastic transduction and operating window}
\label{subsec:strain_transduction}

Bridging thermodynamics and mechanics, \emph{magnetoelastic transduction} converts field- or defect-controlled spectral shifts into measurable strains and forces, enabling on-chip dilatometry and nanomechanical readout with sub-ångström resolution \cite{Kuechler2012,Tokiwa2011}. In straintronics, such couplings are routinely exploited to actuate or sense low-energy scales in 2D/vdW stacks and semiconductor heterostructures \cite{PRB.2020.102.235163,Manesco2020,PhysRevLett.115.177202,APL.2023.123.021702}. Within our single-scale framework, the same gap $x=\hbar|\omega_{\mathrm{eff}}|/(2k_BT)$ that organizes $C_V$, $M$, and $\chi$ also controls the \emph{torsional work} $\Pi_\Omega$ and its susceptibility $\chi_\Omega$, providing a direct, calibration-friendly figure of merit for device operation. The dimensionless efficiency $\eta_\Omega$ below makes this explicit and identifies an optimum window slightly \emph{above} the Schottky scale $T^\star\sim \hbar|\omega_{\mathrm{eff}}|/k_B$, where the kernel $1/\sinh^2 x$ boosts response while dissipation remains modest.

Defining a dimensionless ``magnetoelastic efficiency''
\begin{equation}
\eta_{\Omega}(T,B)\equiv \frac{|\Pi_{\Omega}|}{k_{B}T}
=\frac{\hbar^{2}|k|}{\mu k_{B}T}\,
\coth\Bigl(\frac{\hbar|\omega_{\mathrm{eff}}|}{2k_{B}T}\Bigr),
\label{eq:etaOmega}
\end{equation}
we see that (i) $\eta_{\Omega}\to\text{const}$ as $T\to0$ (frozen ladder), and (ii) $\eta_{\Omega}\propto 1/|\omega_{\mathrm{eff}}|$ for $k_{B}T\gtrsim\hbar|\omega_{\mathrm{eff}}|$. Hence, for \emph{transduction} (dilatometry/nanomechanics) the optimal window is slightly \emph{above} the Schottky scale $T^{\star}\sim\hbar|\omega_{\mathrm{eff}}|/k_{B}$, where $\chi_{\Omega}$ peaks through the kernel $1/\sinh^{2}x$ while dissipation remains modest.

High-resolution capacitive dilatometry~\cite{KuechlerRSI2012} and piezoelectric nanomechanical platforms can read out $\Pi_{\Omega}$ via length changes induced by controlled dislocation injection (or patterned torsion). Mapping $\Pi_{\Omega}(T,B)$ along iso-$|\omega_{\mathrm{eff}}|$ lines isolates the kernel and collapses geometrical prefactors, providing a compact comparison with Eqs.~\eqref{eq:PiOmega_letter}-\eqref{eq:chiOmega_letter}.

\subsection{Direct calibration of \texorpdfstring{$\Omega$}{Omega}}
\label{subsec:strain_calibration}

A two-observable protocol removes the longitudinal wave number $k$ as a fit parameter.  
(i) From the \emph{compensated field}
\begin{equation}
B_{\mathrm{comp}}(\Omega)= -\frac{2\hbar k}{e}\,\Omega,
\label{eq:Bcomp_strain}
\end{equation}
swept at fixed $T$, obtain the slope $\partial B_{\mathrm{comp}}/\partial\Omega=-2\hbar k/e$.  
(ii) At a single temperature $T_{0}$, measure $\Pi_{\Omega}(T_{0})$. The pair $\{B_{\mathrm{comp}}'(\Omega),\Pi_{\Omega}(T_{0})\}$ uniquely fixes $\{k,\Omega\}$ through Eqs.~\eqref{eq:PiOmega_letter}-\eqref{eq:Bcomp_strain} without additional assumptions. Cross-checks with $C_V(T)$ (offset $+\tfrac12 k_{B}$ in bulk) and with $M(T)$ reinforce consistency.

\subsection{Finite-size caloric fingerprints}
\label{subsec:strain_finitesize}

In mesoscopic geometries the Landau degeneracy acquires edge corrections, $g\to g_{\mathrm{corr}}=g\bigl(1-\ell_{\mathrm{eff}}/R\bigr)$ with $\ell_{\mathrm{eff}}=\sqrt{\hbar/(\mu|\omega_{\mathrm{eff}}|)}$ (Sec.~\ref{subsec:finite_size}). Because $\ell_{\mathrm{eff}}$ depends on $\Omega$ through $\omega_{\mathrm{eff}}$, \emph{field and torsion} co-modulate the degeneracy and thus the free energy. The ensuing oscillatory caloric terms $\Delta C_V(B,\Omega)$ are the strain counterpart of dHvA oscillations and scale with $\ell_{\mathrm{eff}}/R$, providing a sensitive internal diagnostic of edge dominance in patterned flakes and rings.

The strain sector reduces to a kernel-controlled pair $\{\Pi_{\Omega},\chi_{\Omega}\}$ with a clean operating window and a direct calibration route for $\Omega$. Finite-size caloric oscillations furnish complementary evidence of magnetoelastic quantization and boundary participation.

\section{Metrology and quantum oscillations}
\label{sec:metrology}

\emph{Quantum oscillations as precision probes.}
de Haas-van Alphen (dHvA) and Shubnikov-de Haas (SdH) oscillations remain gold-standard probes of Fermi surfaces, effective masses, and scattering in quantum materials~\cite{ShoenbergBook,ComprehensiveZrSiS2024}. Emerging experiments combine torque magnetometry on microcantilevers with field/angle/temperature sweeps to resolve subtle band features and topological responses in metals and semimetals~\cite{ComprehensiveZrSiS2024}. 

In the elastic-Landau regime, the elastic contribution shifts the oscillation period by a constant \emph{offset field} $B_{\mathrm{tor}}=2\hbar k\,\Omega/e$, making the signal exactly periodic in $1/B_{\mathrm{eff}}$ with $B_{\mathrm{eff}}=B+B_{\mathrm{tor}}$. We build on the Lifshitz-Kosevich framework to formulate a robust \emph{phase-unwarping} protocol that extracts $B_{\mathrm{tor}}$ from a single field sweep by maximizing the FFT sharpness after reparametrization in $1/(B+\delta B)$. 
Strictly, the oscillations are exactly periodic in $1/B_{\mathrm{eff}}$ with $B_{\mathrm{eff}}=B+B_{\mathrm{tor}}$ (torsion offset).
When traces are analyzed as a function of $1/B$ without reparametrization, the period is only approximately uniform over finite windows; our phase-unwarping protocol recovers exact periodicity by scanning $\delta B$ in $1/(B+\delta B)$ and maximizing FFT sharpness.
Consistency is checked against (i) the IQHE fan-diagram shift dictated by the St\v{r}eda relation, which translates all plateau transitions rigidly by $B_{\mathrm{tor}}$, and (ii) the collapse of $M(T)$ and $\chi(T)$ onto universal kernels along iso-$|\omega_{\mathrm{eff}}|$ lines~\cite{ShoenbergBook,LK1956,Dingle1952}. Recent case studies on nodal-line and correlated oxides underscore the maturity of these techniques for quantitative materials metrology~\cite{ComprehensiveZrSiS2024}.
\begin{figure}[tbhp]
  \centering
  \includegraphics[width=0.99\linewidth]{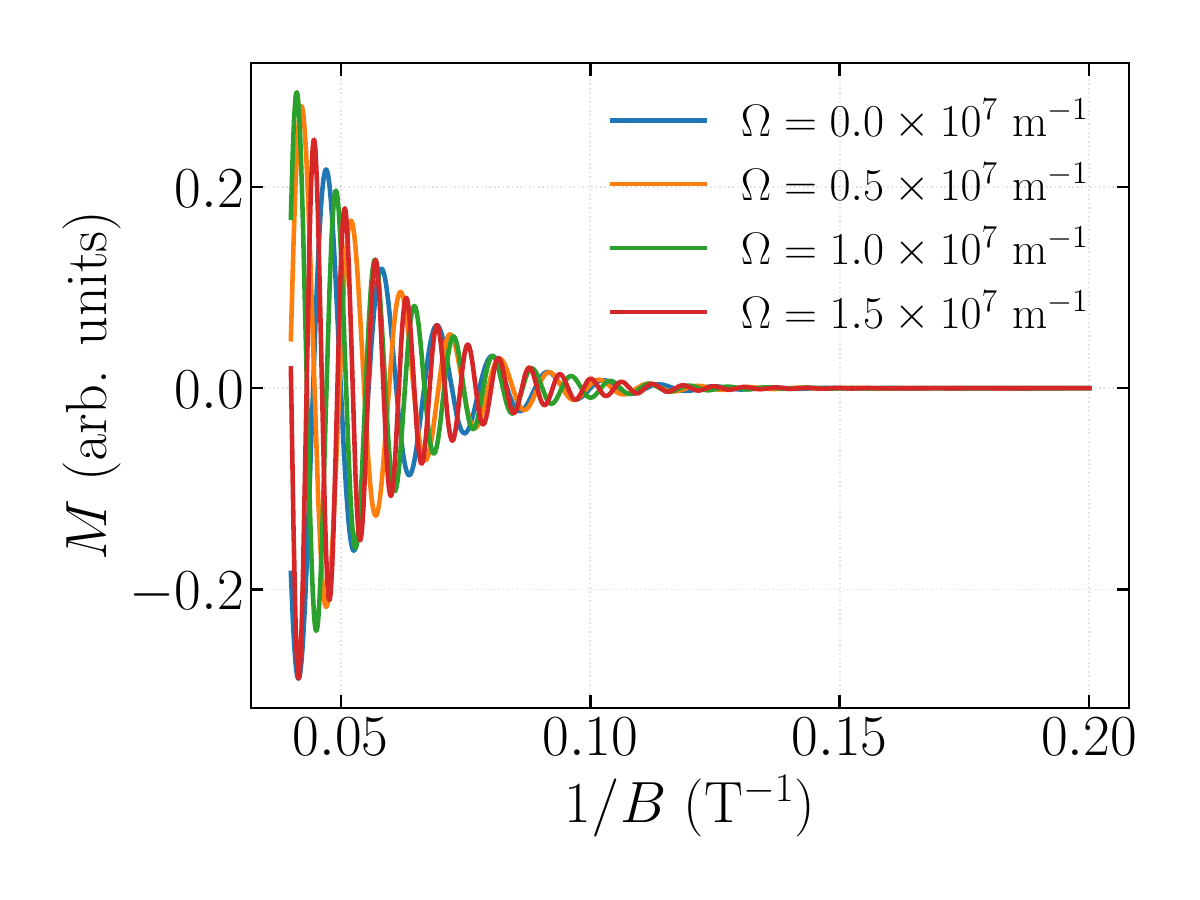}
  \caption{\footnotesize Magnetization oscillations as a function of inverse magnetic field at fixed temperature, $M(1/B)$, for representative torsion densities $\Omega$ (see legend). Increasing $\Omega$ shifts the oscillation period in $1/B$ to smaller values (higher frequency), i.e., compresses the $1/B$ spacing, consistent with the torsion-enhanced effective frequency predicted by our framework. The solid curves are computed at fixed emperature using the same parameters as in Sec.~\ref{sec:metrology}, and the overall amplitude scale
  is arbitrary.} \label{fig:met_M_invB_varOmega}
\end{figure}

The dHvA signal $M(1/B)$ provides two metrological handles to extract $\Omega$ [Fig.~\ref{fig:met_M_invB_varOmega}]: (i) a systematic increase of the oscillation frequency in $1/B$ (or Onsager frequency in an FFT analysis) as $\Omega$ grows, reflecting the torsion-enhanced Landau spacing; and (ii) an $\Omega$-dependent damping of the amplitude, stemming from the reduced number of occupied levels and the stronger thermal/Dingle attenuation at the effective cyclotron scale.
\begin{figure}[tbhp]
\centering
\includegraphics[width=\linewidth]{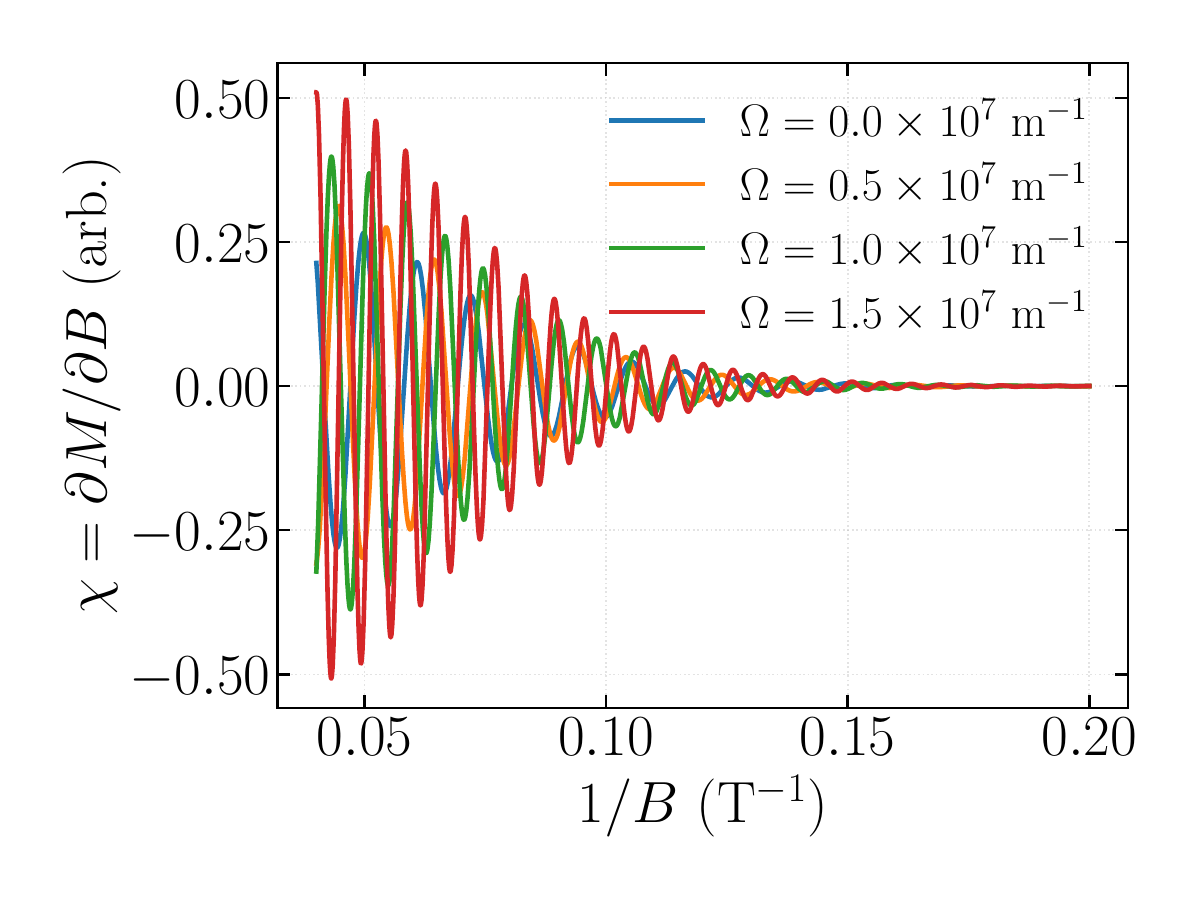} 
\caption{\footnotesize Differential magnetic susceptibility $\chi=\partial M/\partial B$ as a function of inverse magnetic field $1/B$ at fixed temperature for several torsion densities $\Omega$ (see legend). The oscillations of $\chi(1/B)$ share the same $1/B$-period as $M(1/B)$; thus, the dHvA frequency increases systematically with $\Omega$ (compressed period). Differentiation with respect to $B$ enhances the high-field (small $1/B$) signal and modifies the envelope, producing larger initial amplitudes for higher $\Omega$ and a faster decay with $1/B$. The vertical scale is arbitrary.}
    \label{fig:met_chi_invB_varOmega}
\end{figure}

Torque magnetometry and SQUID readout often access the differential response $\chi=\partial M/\partial B$ [Fig.~\ref{fig:met_chi_invB_varOmega}].
Because $\chi$ is the field derivative of the dHvA signal, it preserves the oscillation frequency in $1/B$ (thus the torsion-induced period compression), while amplifying the high-field oscillations and reshaping the amplitude envelope.
Both the frequency shift and the $\Omega$-dependent damping provide complementary handles to quantify the torsion density from standard susceptibility traces.
\begin{figure}[tbhp]
\centering
\includegraphics[width=\columnwidth]{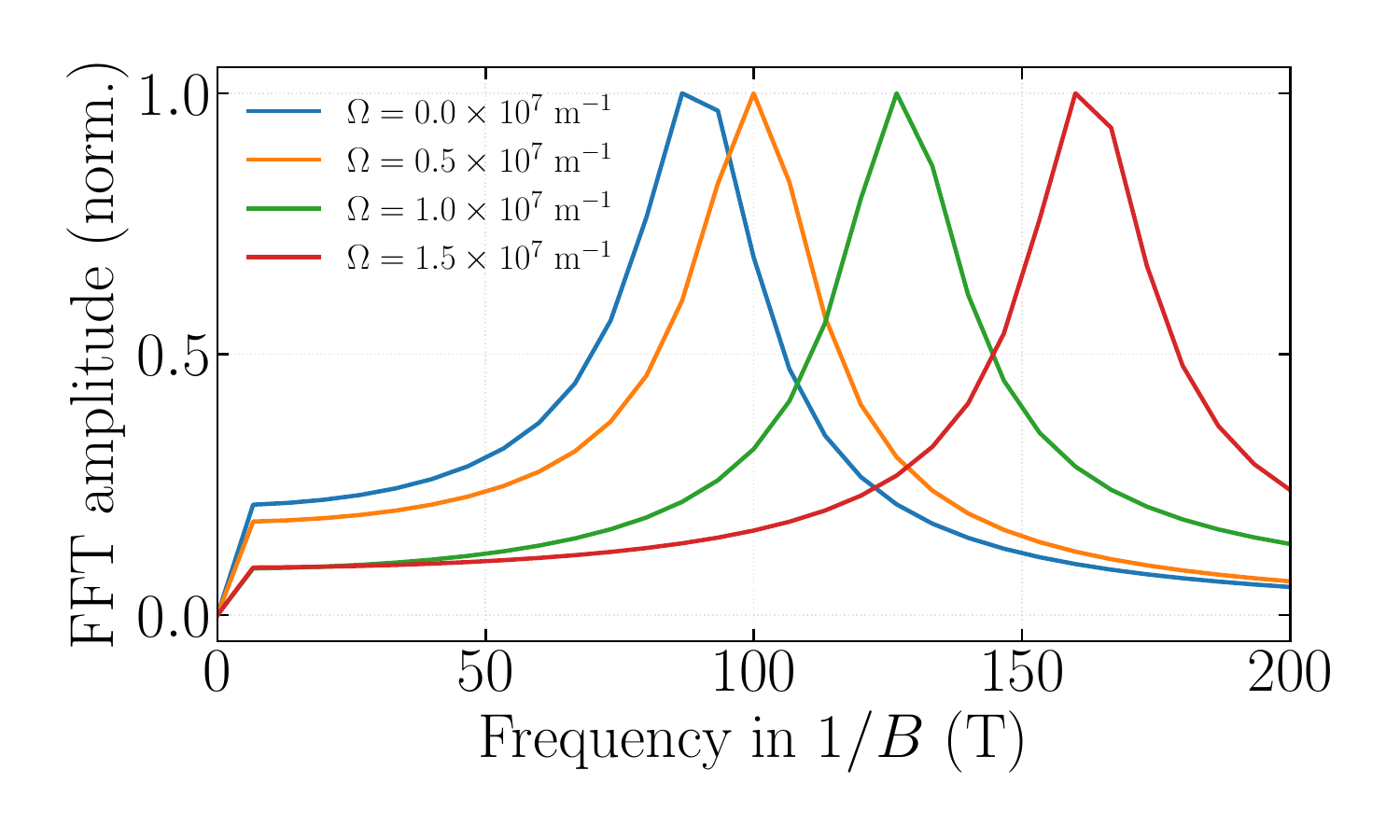}
\caption{\footnotesize 
Fast Fourier transform (FFT) of the magnetization oscillations in inverse magnetic field, $M(1/B)$, for representative torsion densities $\Omega$ (see legend). The amplitude is normalized to unity at the global maximum for each trace. Increasing $\Omega$ shifts the dominant FFT peak to \emph{higher} frequencies in $1/B$, revealing the compression of the dHvA period predicted by the elastic-torsion coupling (\,$\omega_{\mathrm{eff}}$ renormalization\,). Peak broadening reflects the finite field window and damping effects.}
\label{fig:met_fft_M_invB_varOmega}
\end{figure}

Quantum oscillations probe the Fermi surface and low-energy scales with exquisite precision. In our setting, the elastic contribution shifts the effective cyclotron scale and compresses oscillation periods through
\begin{equation}
B_{\mathrm{eff}}= B + B_{\mathrm{tor}},
\qquad
B_{\mathrm{tor}}=\frac{2\hbar k\,\Omega}{e},
\qquad
\omega_{\mathrm{eff}}=\frac{eB_{\mathrm{eff}}}{\mu}.
\label{eq:Beff_metrology}
\end{equation}
We formulate a practical metrology suite with three components: (i) dHvA/SdH phase-unwarping to extract $B_{\mathrm{tor}}$; (ii) St\v{r}eda analysis of IQHE plateau shifts; and (iii) cross-validation with thermodynamic kernels $M(T)$ and $\chi(T)$.

\subsection{dHvA/SdH with torsion shift: phase-unwarping protocol}
\label{subsec:unwarping}

For a broadened Landau ladder, the oscillatory magnetization (or conductivity) admits the Lifshitz-Kosevich (LK) structure~\cite{ShoenbergBook,LK1956}
\begin{equation}
\mathcal{O}_{\mathrm{osc}}(B)\propto
\sum_{r\ge 1}\frac{1}{r^{\alpha}}
R_{T}(r)\,R_{D}(r)\,
\cos\Bigl[\tfrac{2\pi r}{\hbar|\omega_{\mathrm{eff}}|}\,\mu_{F}+\phi_{r}\Bigr],
\label{eq:LK_osc}
\end{equation}
with $R_{T}(r)=X_{r}/\sinh X_{r}$, $X_{r}=2\pi^{2} r k_{B}T/(\hbar|\omega_{\mathrm{eff}}|)$, and Dingle factor $R_{D}(r)=\exp\{-\pi r/(|\omega_{\mathrm{eff}}|\tau_{D})\}$~\cite{Dingle1952}. Because $|\omega_{\mathrm{eff}}| \propto |B+B_{\mathrm{tor}}|$, oscillations are \emph{strictly periodic} in $1/B_{\mathrm{eff}}$.

We propose a robust \emph{phase-unwarping} algorithm to extract $B_{\mathrm{tor}}$ directly from field sweeps $B\in[B_{\min},B_{\max}]$:

\begin{enumerate}
\item For a trial offset $\delta B$, reparametrize the trace by $x(\delta B)=1/(B+\delta B)$ and resample uniformly in $x$.
\item Compute the FFT amplitude $A(\delta B,f)$ versus frequency $f$ (conjugate to $x$) after subtracting a low-order polynomial background.
\item Define a sharpness score (e.g., peak height over integrated sideband power). The optimal $\delta B^{\star}$ maximizes the score and estimates $B_{\mathrm{tor}}$ with uncertainty from the curvature of the score near $\delta B^{\star}$ (bootstrap for error bars).
\end{enumerate}

The method is insensitive to the unknown $\mu_{F}$ and to smooth backgrounds, and remains accurate when only the fundamental harmonic $r=1$ survives thermal/Dingle damping.

\subsection{IQHE fan diagram and St\v{r}eda consistency}
\label{subsec:streda}

In the integer QHE, plateau transitions occur when $\mu_{F}$ crosses the center of a broadened Landau band, $\mu_{F}=E_{n}(B_{\mathrm{eff}})$. Solving for $B$ yields a \emph{linear} torsion shift of critical fields:
\begin{equation}
B_{c}^{(n)}(\Omega)= B_{c}^{(n)}(0) - \frac{2\hbar k}{e}\,\Omega,
\qquad
B_{c}^{(n)}(0)=\frac{\mu}{e}\,\frac{2\mu_{F}}{\hbar(2n+1)}.
\label{eq:Bc_shift}
\end{equation}
The St\v{r}eda relation $\sigma_{xy}=e\,(\partial n/\partial B)_{\mu,T}$~\cite{Streda1982} implies that the entire Hall fan is rigidly translated by $B_{\mathrm{tor}}$ without changing its $1/B_{\mathrm{eff}}$ spacing. A joint fit of (i) the fan offsets and (ii) the dHvA/SdH unwarping returns a consistent $B_{\mathrm{tor}}$ (hence $\Omega$).

\subsection{Cross-validation with thermodynamic kernels}
\label{subsec:crossval}

The temperature kernels derived in Sec.~\ref{subsec:mag_results} provide a purely thermodynamic cross-check: plotting $M(T)$ and $\chi(T)$ along iso-$|\omega_{\mathrm{eff}}|$ lines collapses curves onto the universal $\coth x$ and $1/\sinh^{2}x$ forms. The extracted $|\omega_{\mathrm{eff}}|$ sets the $1/B_{\mathrm{eff}}$ period of quantum oscillations \emph{and} the position of IQHE plateau transitions, closing the metrological loop.

\paragraph*{Noise and systematic effects.}
Finite windows, detrending, and apodization introduce weak low-frequency backgrounds in FFTs; these can be quantified and bounded by comparing multiple windows and by subtracting mean trends before unwarping. Dingle broadening reduces amplitudes but does not bias the unwarping maximum; classicalization as $k_{B}T\gg\hbar|\omega_{\mathrm{eff}}|$ suppresses the overall signal, setting the upper temperature limit of the protocol.

Quantum-oscillation metrology in the elastic-Landau regime reduces to extracting an \emph{offset field} $B_{\mathrm{tor}}=2\hbar k\Omega/e$ by phase-unwarping, validating it with IQHE fan shifts (St\v{r}eda), and cross-checking against thermodynamic kernels. The three routes are mutually consistent and return $\Omega$ with controlled uncertainty, independent of smooth backgrounds, and with minimal modeling assumptions.

\section{Conclusions and Outlook}\label{sec5}

The results presented here establish a single-scale framework for magnetoelastic Landau quantization in media with a uniform areal density of screw dislocations under a uniform magnetic field. Starting from the elastic-Landau ladder, the canonical partition function at fixed longitudinal momentum $k$ yields closed expressions for the Helmholtz free energy, internal energy, entropy, heat capacity, magnetization, susceptibility, and magnetocaloric responses, all controlled by the gap variable $x=\hbar|\omega_{\mathrm{eff}}|/(2k_{B}T)$ with $\omega_{\mathrm{eff}}=\omega_{c}+\omega_{cl}$. This organization exposes universal hyperbolic kernels that govern temperature crossovers and link thermodynamic and magnetic derivatives in a transparent way. Integrating the free longitudinal motion adds a universal $+\tfrac{1}{2}k_{B}$ to the heat capacity per particle, restoring the expected classical limits and clarifying the approach to Bohr-van Leeuwen at high temperature.

A central physical consequence is the compensated-field regime, in which the elastic and electromagnetic cyclotron contributions cancel and the transverse gap collapses. Exactly at compensation, the heat capacity reaches an equipartition plateau while magnetic responses vanish, providing a crisp and tunable signature of magnetoelastic interference. The same single scale controls quantum oscillations: within the Lifshitz-Kosevich analysis, torsion is encoded by the substitution $B\mapsto B_{\mathrm{eff}}$, which compresses the period in $1/B$ and modifies amplitudes through $1/|\omega_{\mathrm{eff}}|$ and Dingle damping. This observation motivates a practical phase-unwarping protocol that determines the torsional offset directly from a single sweep and dovetails with the St\v{r}eda relation that translates the integer Hall fan without altering level spacings in $1/B_{\mathrm{eff}}$.

Mesoscopic samples inherit finite-size corrections to the Landau degeneracy set by the effective magnetic length $\ell_{\mathrm{eff}}\propto 1/\sqrt{|\omega_{\mathrm{eff}}|}$. These corrections generate oscillatory contributions to caloric observables as the field is swept, offering an additional diagnostic channel in patterned flakes, rings, and quantum dots where boundaries are well controlled. Because both field and torsion flow through $|\omega_{\mathrm{eff}}|$, the magnitude and period of these oscillations can be tuned \emph{in situ}.

From an experimental perspective, the framework suggests a compact metrology suite. On-chip calorimetry accesses Schottky-like features and the magnetic Gr\"uneisen parameter; torque magnetometry probes the low-$T$ saturation of $M$ and its thermal decay; transport quantifies the $B_{\mathrm{eff}}$ shift of the SdH/dHvA frequency and the IQHE plateau fan. Cross-correlating these measurements closes a loop around the single scale $\hbar|\omega_{\mathrm{eff}}|$ and enables direct extraction of the dislocation areal density. In this setting, moderate disorder and weak interactions smooth amplitudes but preserve the kernel structure, so that the main scaling relations remain robust.

The analysis has a clear range of validity. The fixed-$k$ thermodynamics captures the discrete transverse spectrum and is most accurate when $k_{B}T\lesssim\hbar|\omega_{\mathrm{eff}}|$ or for quasi-2D/mesoscopic geometries dominated by a single longitudinal mode; bulk comparisons should include the longitudinal integration, which yields the universal $+\tfrac{1}{2}k_{B}$ offset in the heat capacity and the $2/3$ reduction of the high-$T$ magnetocaloric slope. Close to compensation, level broadening from disorder regularizes the formal singularities and sets the minimal field window where universal trends survive. Spin, Zeeman splitting, and strong-interaction effects fall outside our minimal model and constitute natural extensions when materials and temperatures make them relevant.

Looking ahead, three directions appear particularly promising. First, quantitative phase-unwarping of oscillation data in a controlled field window should benchmark the extraction of $B_{\mathrm{tor}}$ and validate the $B_{\mathrm{eff}}$ substitution across materials classes. Second, nanoscale dilatometry and alternating-field thermometry in devices with engineered dislocation densities can map the predicted kernel collapse in $C_{V}$, $(\partial M/\partial T)_{B}$ and $\Gamma_{B}$. Third, device-level opportunities emerge in caloritronics and straintronics: magnetocaloric microcoolers operating near the Schottky scale $T^{\star}\sim \hbar|\omega_{\mathrm{eff}}|/k_{B}$, magnetoelastic heat switches exploiting the compensated-field plateau, and dilatometric transducers tuned for maximal torsional susceptibility. In all cases, the single scale $\hbar|\omega_{\mathrm{eff}}|$ furnishes a compact design rule that connects materials, geometry, and performance.

\medskip
\noindent\emph{Perspectives.}
Several extensions appear natural. On the theory side, incorporating Zeeman splitting, anisotropic masses, and Dirac/Weyl (or moiré) dispersions will test how the single-scale structure survives in multiband and relativistic settings; spatially nonuniform and time-dependent torsion (dislocation textures and phonon-torsion coupling) can be treated semiclassically to address Nernst/Ettingshausen and AC magnetocaloric response; controlled disorder and interactions (self-energy broadening, Fermi-liquid parameters, emergent criticality) will quantify how kernel universality deforms near instabilities. On the materials side, first-principles/atomistic estimates of $\Omega$ and device-level dislocation engineering (TEM characterization, nanoindentation, vdW stacking) can be tied to our phase-unwarping protocol and to combined torque-calorimetry-transport campaigns, enabling metrology of $B_{\mathrm{tor}}$ from a single sweep. These directions should clarify robustness, uncover regimes where $\Gamma_{B}$ changes sign, and translate the single-scale design rule into practical caloritronic and straintronic components.

\section*{Acknowledgments}
F.A.\ acknowledges the Inter University Centre for Astronomy and Astrophysics (IUCAA), Pune, India, for granting a visiting associateship. E.\ O.\ Silva acknowledges the support from grants CNPq/306308/2022-3, FAPEMA/UNIVERSAL-06395/22, FAPEMA/APP-12256/22, and (CAPES) - Brazil (Code 001).

\begin{appendices}

\section{Elastic Landau spectrum: notation and thermodynamic formulas}
\label{app:map-SF}

We follow Ref.~\cite{JPCM.2008.20.125209} and only fix notation and the steps leading to the thermodynamic observables used in the main text. The Riemann-Cartan background of a uniform density of screw dislocations is described by
\begin{equation}
    ds^2 = (dz+\Omega\,\rho^2 d\phi)^2 + d\rho^2 + \rho^2 d\phi^2,
\end{equation}
where $\Omega$ encodes torsion (set by the Burgers-vector density) as 
\begin{equation}
    \Omega = \frac{b\,\sigma}{2},
\end{equation}
with $b$ being the Burgers vector (length) and $\sigma$ the areal density of screw dislocations (m$^{-2}$); thus $b\,\sigma$ has units of inverse length.

\paragraph*{Spectrum.}
Including a longitudinal wavevector $k$ and a uniform field $B\hat{\mathbf z}$, the Landau-like spectrum is
\begin{equation}
    E_{nmk}=\hbar(\omega_{cl}+\omega_c)\left(n+\frac{|m|}{2}+\frac{m}{2}+\frac{1}{2}\right)+\frac{\hbar^2 k^2}{2\mu},
    \label{eq:SF37}
\end{equation}
with $\omega_c=eB/\mu$ and the ``elastic cyclotron'' frequency
\begin{equation}
    \omega_{cl} \equiv \frac{2\hbar k\,\Omega}{\mu}.
\end{equation}
Equation~\eqref{eq:SF37} is Eq.~(37) of Ref.~\cite{JPCM.2008.20.125209} in our symbols. It is convenient to define the single control scale
\begin{equation}
    \omega_{\rm eff}\equiv \omega_c+\omega_{cl}=\frac{eB}{\mu}+\frac{2\hbar k\,\Omega}{\mu},
\end{equation}
which governs level spacing and the $1/B$-period of quantum oscillations used in Secs.~\ref{sec:caloritronics}-\ref{sec:metrology}.

\paragraph*{Fixed-$k$ vs.\ 3D averaging.}
In the ``transverse/fixed-$k$'' model adopted in our plots, $k$ is a parameter set by longitudinal confinement/lead geometry; thermodynamics then follow from the 2D ladder in \eqref{eq:SF37}. For a bulk 3D gas, one averages over $k$ (equivalently, the longitudinal quadratic mode adds the classical $\tfrac{1}{2}k_B$ at high $T$); we indicate which option is used in each figure/caption.

\paragraph*{Degeneracy and effective magnetic length.}
The per-area degeneracy of each transverse ladder is $g=A/(2\pi \ell_{\rm eff}^2)$ with
\begin{equation}
    \ell_{\rm eff}^{\,2}=\frac{\hbar}{\mu\,\omega_{\rm eff}}=\frac{\hbar}{eB+2\hbar k\,\Omega},
\end{equation}
making explicit the additivity of $B$ and $\Omega$ emphasized in Ref.~\cite{JPCM.2008.20.125209}.

\paragraph*{Partition function and thermodynamics (fixed-$k$).}
For a single transverse channel (per spin),
\begin{equation}
    Z = \sum_{n=0}^{\infty} e^{-\beta\hbar\omega_{\rm eff}(n+1/2)}
      = \frac{1}{2\sinh\left(\tfrac{1}{2}\beta\hbar\omega_{\rm eff}\right)}.
\end{equation}
The internal energy $U$, heat capacity at fixed $B$ ($C_B$), and magnetization $M$ are
\begin{align}
    U(T,B;\Omega,k) &= -\frac{\partial}{\partial\beta}\ln Z
    = \frac{\hbar\omega_{\rm eff}}{2}\coth\left(\frac{\beta\hbar\omega_{\rm eff}}{2}\right),\\[2pt]
    C_B(T,B;\Omega,k) &= \frac{\partial U}{\partial T}
    = k_B\left(\frac{\beta\hbar\omega_{\rm eff}}{2}\right)^{2}\mathrm{csch}^{2}\left(\frac{\beta\hbar\omega_{\rm eff}}{2}\right),\\[2pt]
    M(T,B;\Omega,k) &= -\left.\frac{\partial F}{\partial B}\right|_{T}
    = -\frac{\partial}{\partial B}\left(-k_B T \ln Z\right)\notag\\
    &= -\frac{\hbar}{2}\,\frac{\partial\omega_{\rm eff}}{\partial B}\,
      \coth\left(\frac{\beta\hbar\omega_{\rm eff}}{2}\right),
\end{align}
with $\partial\omega_{\rm eff}/\partial B=e/\mu$. The susceptibility $\chi=\partial M/\partial B$ follows by differentiation. For a 3D gas, integrate over the longitudinal $k$-distribution (or include the longitudinal $\tfrac{1}{2}k_B$ contribution in the high-$T$ limit).

\paragraph*{Limiting cases.}
(i) $\Omega\to0$ reproduces standard Landau thermodynamics.
(ii) $B\to0$ yields the \emph{elastic} Landau ladder with spacing $\hbar\omega_{cl}\propto k\,\Omega$, used as the ``torsion-only'' reference in the MCE figure.
(iii) At high $T$, $C_B \to k_B$ per transverse channel (plus the longitudinal $\dfrac{1}{2}\,k_B$ in 3D).

\section{Grand-canonical derivation of torsion-shifted de Haas-van Alphen oscillations}
\label{app:dhva}

Here we derive the oscillatory part of the grand potential for the elastic-Landau ladder and the corresponding de Haas-van Alphen (dHvA) magnetization, making explicit how torsion enters only through the single scale $\omega_{\rm eff}=\omega_c+\omega_{cl}$.

\paragraph*{Setup.}
For fixed longitudinal wave number $k$, the transverse spectrum is
\begin{equation}
E_{N'}=\hbar\omega_{\rm eff}\Bigl(N'+\tfrac12\Bigr),\qquad 
\omega_{\rm eff}\equiv \frac{eB}{\mu}+\frac{2\hbar k\,\Omega}{\mu},
\end{equation}
with a macroscopic degeneracy per level $g\propto A/(2\pi\ell_{\rm eff}^2)$,
$\ell_{\rm eff}^{\,2}=\hbar/(\mu\,\omega_{\rm eff})$.
In the grand canonical ensemble at temperature $T$ and Fermi level $\mu_F$,
\begin{equation}
\Omega_G(T,\mu_F;B,\Omega)
= -k_BT\,g \sum_{N'=0}^{\infty}
\ln\Bigl[1+e^{-\beta\bigl(E_{N'}-\mu_F\bigr)}\Bigr].
\label{eq:appOmega}
\end{equation}

\paragraph*{Poisson summation.}
Introduce a discrete density of states (DoS)
$\rho(E)=g\sum_{N'}\delta\bigl(E-\hbar\omega_{\rm eff}(N'+\tfrac12)\bigr)$
and rewrite
$\Omega_G=-k_BT\int dE\,\rho(E)\ln\bigl[1+e^{-\beta(E-\mu_F)}\bigr]$.
Applying the Poisson summation formula to the series in $N'$ with spacing $\hbar\omega_{\rm eff}$ and Maslov phase $+1/2$ yields
\begin{equation}
\rho(E)=\bar\rho
+2\,\bar\rho\sum_{r=1}^{\infty}
\cos\Bigl(\tfrac{2\pi r\,E}{\hbar\omega_{\rm eff}}\Bigr),
\qquad
\bar\rho=\frac{g}{\hbar\omega_{\rm eff}}.
\end{equation}
Thus $\Omega_G=\Omega_{\rm sm}+\Omega_{\rm osc}$, with the oscillatory piece
\begin{align}
\Omega_{\rm osc} &=
-2\,\bar\rho \sum_{r=1}^{\infty}
\int_{-\infty}^{\infty} dE\,
\ln\bigl[1+e^{-\beta(E-\mu_F)}\bigr]\,
\cos\Bigl(\tfrac{2\pi r\,E}{\hbar\omega_{\rm eff}}\Bigr).
\label{eq:appOmegaosc}
\end{align}
Differentiating w.r.t.\ $\mu_F$ and integrating back, one obtains the standard Lifshitz-Kosevich (LK) thermal damping factor $R_T$ and a Dingle factor $R_D$ for disorder broadening. The result can be written compactly as
\begin{align}
\Omega_{\rm osc}
= -\frac{g\,\mu_F^2}{\hbar\omega_{\rm eff}}
\sum_{r=1}^{\infty}\frac{1}{\pi^2 r^2}\,
&\cos\Bigl(\frac{2\pi r\,\mu_F}{\hbar\omega_{\rm eff}}\Bigr)\notag\\ &\times
R_T(r,T)\,R_D(r,\tau_D),
\label{eq:appOmegaosc_final}
\end{align}
with
\begin{align}
R_T(r,T)&=\frac{X_r}{\sinh X_r},\\
X_r&=\frac{2\pi^2 r k_B T}{\hbar\,|\omega_{\rm eff}|},\\
R_D(r,\tau_D)&=\exp\Bigl(-\frac{\pi r}{|\omega_{\rm eff}|\,\tau_D}\Bigr).
\label{eq:appRT_RD}
\end{align}

\paragraph*{Oscillatory magnetization.}
The orbital magnetization follows from
$M_{\rm osc}=-(\partial\Omega_{\rm osc}/\partial B)_{T,\mu_F}$.
Using $\partial|\omega_{\rm eff}|/\partial B=(e/\mu)\,\mathrm{sgn}(\omega_{\rm eff})$,
one finds
\begin{align}
M_{\rm osc}(B)
= -\,\mathrm{sgn}(\omega_{\rm eff})\,&
\frac{g\,\mu_F}{\hbar\,|\omega_{\rm eff}|}\,
\sum_{r=1}^{\infty}\frac{1}{\pi r}
\sin\Bigl(\frac{2\pi r\,\mu_F}{\hbar\,|\omega_{\rm eff}|}\Bigr)\notag \\ &\times
R_T(r,T)\,R_D(r,\tau_D).
\label{eq:appMosc}
\end{align}
In the common case where higher harmonics are suppressed by $T$ and/or disorder,
retaining only $r=1$ leads to
\begin{equation}
M_{\rm osc}(B)\simeq
-\,\mathrm{sgn}(\omega_{\rm eff})\,
\frac{g\,\mu_F}{\hbar\,|\omega_{\rm eff}|}\,
\sin\Bigl(\frac{2\pi \mu_F}{\hbar\,|\omega_{\rm eff}|}\Bigr)\,
e^{-\pi/(|\omega_{\rm eff}|\tau_D)}.
\label{eq:appMosc_simple}
\end{equation}

\paragraph*{Period and torsion shift.}
Because $\omega_{\rm eff}=\tfrac{e}{\mu}\,B+\tfrac{2\hbar k}{\mu}\,\Omega$,
the phase $\varphi=2\pi\mu_F/(\hbar|\omega_{\rm eff}|)$ is periodic in 
$1/B_{\rm eff}$ with
\begin{equation}
B_{\rm eff}=B+B_{\Omega},\qquad
B_{\Omega}\equiv \frac{2\hbar k\,\Omega}{e}.
\end{equation}
Hence, torsion $\Omega$ \emph{compresses} the dHvA period in $1/B$ by increasing $|\omega_{\rm eff}|$, while simultaneously reducing the amplitude through both the prefactor $1/|\omega_{\rm eff}|$ and the Dingle factor in Eq.~\eqref{eq:appRT_RD}.

\paragraph*{Remarks on signs and robustness.}
The kernels $R_T$ and $R_D$ depend only on the gap scale $|\omega_{\rm eff}|$. The explicit sign appears as an overall $\mathrm{sgn}(\omega_{\rm eff})$ in Eq.~\eqref{eq:appMosc} through the chain rule for $\partial|\omega_{\rm eff}|/\partial B$. Fixing the field orientation so that $\omega_{\rm eff}>0$ removes this global sign without changing the periodicity or amplitudes.

Equations \eqref{eq:appOmegaosc_final}-\eqref{eq:appMosc_simple} are the torsion-shifted LK formulas used in Sec.~\ref{sec:metrology} to connect the oscillation period and envelope to the defect parameter $\Omega$. They reduce to the standard Landau-case results when $\Omega\to 0$.

\section{Longitudinal-$k$ integration: partition function, \texorpdfstring{$+\tfrac{1}{2}k_B$}{+1/2 kB} in \texorpdfstring{$C_V$}{CV}, and consequences for \texorpdfstring{$\Gamma_B$}{Gamma\_B}}
\label{app:k-integration}

This appendix derives, step by step, how the integration over the free longitudinal motion adds a universal $+\tfrac{1}{2}k_B$ to the heat capacity per particle and how this modifies the magnetocaloric and Gr\"{u}neisen responses used in the main text.

\subsection*{A. Factorization of the spectrum and fixed-$k$ partition function}

For a given longitudinal wave number $k$, the elastic-Landau spectrum factorizes as
\begin{equation}
E_{N',k}=\hbar\omega_{\rm eff}\Bigl(N'+\tfrac{1}{2}\Bigr)+\frac{\hbar^{2}k^{2}}{2\mu},
\qquad
\omega_{\rm eff}=\frac{eB}{\mu}+\frac{2\hbar k\,\Omega}{\mu},
\end{equation}
so that the canonical partition function per transverse channel (and per spin) at fixed $k$ reads
\begin{align}
Z(T;k)
&= \sum_{N'=0}^{\infty}
\exp\Bigl[-\beta\hbar\omega_{\rm eff}(N'+\tfrac12)\Bigr]\;
\exp\Bigl[-\beta \hbar^{2}k^{2}/(2\mu)\Bigr]
\notag\\
&= \frac{1}{2\sinh\bigl(\tfrac{1}{2}\beta\hbar\omega_{\rm eff}\bigr)}\;
\exp\Bigl[-\beta \hbar^{2}k^{2}/(2\mu)\Bigr].
\label{eq:appZk}
\end{align}
Equations for the fixed-$k$ internal energy, heat capacity, and magnetization used in the main text follow immediately from $Z(T;k)$.

\subsection*{B. Passing from sum to integral over \texorpdfstring{$k$}{k}}

In a box of length $L$ along $z$ (periodic boundary conditions), the allowed longitudinal wave numbers are $k_n=2\pi n/L$ $(n\in\mathbb{Z})$. In the thermodynamic limit,
\begin{equation}
\sum_{n} f(k_n)\;\longrightarrow\; \frac{L}{2\pi}\int_{-\infty}^{\infty} f(k)\,dk.
\end{equation}
The 3D partition function per transverse channel is then
\begin{align}
Z_{3\rm D}(T)
&= \frac{L}{2\pi}\int_{-\infty}^{\infty} dk\; Z(T;k)\notag \\&
= \frac{L}{2\pi}\,\frac{1}{2\sinh(\tfrac{1}{2}\beta\hbar\omega_{\rm eff})}
\int_{-\infty}^{\infty} dk\, e^{-\beta \hbar^{2}k^{2}/(2\mu)}.
\end{align}
The remaining Gaussian integral yields $\sqrt{2\pi\mu/(\beta\hbar^{2})}$. Introducing the thermal de Broglie wavelength
\begin{equation}
\lambda_T \equiv \sqrt{\frac{2\pi\hbar^{2}\beta}{\mu}}=\sqrt{\frac{2\pi\hbar^{2}}{\mu k_B T}},
\end{equation}
we arrive at the compact result
\begin{equation}
Z_{3\rm D}(T)=
\frac{L}{\lambda_T}\,\frac{1}{2\sinh\bigl(\tfrac{1}{2}\beta\hbar\omega_{\rm eff}\bigr)}.
\label{eq:appZ3D}
\end{equation}

\subsection*{C. Internal energy and heat capacity: universal \texorpdfstring{$+\tfrac{1}{2}k_B$}{+1/2 k\_B}}

From the $k$-integrated partition function \eqref{eq:appZ3D},
the Helmholtz free energy is $F=-k_B T\ln Z_{3\rm D}$.
Differentiating with respect to $\beta=1/k_BT$ gives
\begin{align}
U_{3\rm D}(T)
&= -\frac{\partial}{\partial\beta}\ln Z_{3\rm D}\notag\\
&= \underbrace{\frac{\hbar\omega_{\rm eff}}{2}\,
\coth\Bigl(\frac{\beta\hbar\omega_{\rm eff}}{2}\Bigr)}_{\text{transverse (fixed-$k$) part}}
\;+\;
\underbrace{\frac{1}{2}k_B T}_{\text{longitudinal (free mode)}},
\label{eq:appU3D}
\\[4pt]
C_{V}^{3\rm D}(T)
&= \Bigl(\frac{\partial U_{3\rm D}}{\partial T}\Bigr)_{B,\Omega}\notag\\
&= k_B\Bigl(\frac{\beta\hbar\omega_{\rm eff}}{2}\Bigr)^{2}
   \csch^{2}\Bigl(\frac{\beta\hbar\omega_{\rm eff}}{2}\Bigr)
\;+\;\frac{1}{2}k_B.
\label{eq:appCv3D}
\end{align}
Thus, relative to the purely transverse/fixed-$k$ ladder, the longitudinal
free quadratic mode adds a universal $+\tfrac{1}{2}k_B$ to the heat capacity per
particle. At high temperature, when the transverse ladder is thermally smeared,
the total tends to the classical 3D ideal-gas value, $C_V\to \tfrac{3}{2}k_B$.

\subsection*{D. Magnetocaloric coefficient and Gr\"{u}neisen parameter}

At constant entropy, the adiabatic magnetocaloric response and the magnetic Gr\"{u}neisen parameter are related by
\begin{align}
&\Bigl(\frac{\partial T}{\partial B}\Bigr)_{S}
= -\,\frac{T}{C_V}\Bigl(\frac{\partial M}{\partial T}\Bigr)_{B},\\
&
\Gamma_B\equiv -\frac{1}{T}\Bigl(\frac{\partial T}{\partial B}\Bigr)_{S}
=\frac{1}{C_V}\Bigl(\frac{\partial M}{\partial T}\Bigr)_{B}.
\end{align}
From $F=-k_B T\ln Z$ with $Z$ of \eqref{eq:appZ3D}, and using
$\partial\omega_{\rm eff}/\partial B=e/\mu$, the orbital magnetization is
\begin{equation}
M(T,B;\Omega)= -\Bigl(\frac{\partial F}{\partial B}\Bigr)_{T}
= -\,\frac{\hbar e}{2\mu}\,\coth\Bigl(\frac{\beta\hbar\omega_{\rm eff}}{2}\Bigr),
\end{equation}
so that
\begin{equation}
\Bigl(\frac{\partial M}{\partial T}\Bigr)_{B}
= -\,\frac{\hbar e}{2\mu}\,\frac{\hbar\omega_{\rm eff}}{2k_B T^{2}}\,
\csch^{2}\Bigl(\frac{\beta\hbar\omega_{\rm eff}}{2}\Bigr).
\end{equation}
Inserting this and $C_V^{3\rm D}$ from \eqref{eq:appCv3D} yields
\begin{align}
\Gamma_B^{3\rm D}(T)
&= \frac{1}{C_V^{3\rm D}}\Bigl(\frac{\partial M}{\partial T}\Bigr)_{B}
= -\,\frac{\hbar e}{2\mu k_B}\,
\frac{\beta\,\csch^{2}x}{x^{2}\csch^{2}x+\tfrac{1}{2}},
\label{eq:appGamma3D}
\end{align}
where $x\equiv \beta\hbar\omega_{\rm eff}/2$. Its limits are transparent:

\emph{Low $T$ (large $x$)}: $x^{2}\csch^{2}x\to 0\Rightarrow \Gamma_B^{3\rm D}\to 0$ exponentially.

\emph{High $T$ (small $x$)}: using $x^{2}\csch^{2}x=1-\tfrac{x^{2}}{3}+\dots$,
\begin{equation}
\Gamma_B^{3\rm D}(T\to\infty)\;\longrightarrow\;
-\frac{2}{3}\,\frac{e}{\mu\,\omega_{\rm eff}},
\end{equation}
i.e., the longitudinal $+\tfrac12 k_B$ reduces by $2/3$ the fixed-$k$ plateau $\Gamma_B^{\rm fixed\,k}=-e/(\mu\,\omega_{\rm eff})$.

Similarly, for the adiabatic magnetocaloric coefficient,
\begin{align}
\Bigl(\frac{\partial T}{\partial B}\Bigr)_{S}^{3\rm D}
&= -\,\frac{T}{C_V^{3\rm D}}\Bigl(\frac{\partial M}{\partial T}\Bigr)_{B}
= \frac{\hbar e}{2\mu k_B}\,
\frac{x\,\csch^{2}x}{x^{2}\csch^{2}x+\tfrac{1}{2}},
\end{align}
with the same low-$T$ suppression and a high-$T$ linear asymptote reduced by a factor $2/3$:
\begin{equation}
\Bigl(\frac{\partial T}{\partial B}\Bigr)_{S}^{3\rm D}
\xrightarrow[T\to\infty]{} 
\frac{2}{3}\,\frac{e}{\mu}\,\frac{T}{\omega_{\rm eff}}.
\end{equation}

\subsection*{E. Practical note}

Equations~\eqref{eq:appCv3D}--\eqref{eq:appGamma3D} justify, in bulk samples, the use of the three-dimensional heat capacity $C_V^{3\rm D}$ (offset by $+\tfrac{1}{2}k_B$) when interpreting magnetocaloric and Gr\"{u}neisen measurements, whereas mesoscopic/quantum-wire geometries that effectively fix $k$ are well described by the purely transverse formulas.

\section{St\v{r}eda relation and Hall quantization with \texorpdfstring{$\omega_{\rm eff}$}{omega\_eff}}
\label{app:streda}

For completeness, we record the form of the St\v{r}eda relation in our setting. The Hall conductivity is
\begin{equation}
\sigma_{xy}=e\Bigl(\frac{\partial n}{\partial B}\Bigr)_{\mu_F,T}
= -\,e\,\frac{\partial^{2}\Omega_G}{\partial\mu_F\,\partial B}\Bigg|_{T},
\end{equation}
where $\Omega_G$ is the grand potential and $n=-\partial\Omega_G/\partial \mu_F$ the areal density. In the limit $T\to 0$, for broadened Landau bands with Chern number $C_n=1$ each, one recovers
\begin{align}
&\sigma_{xy}(0,\mu_F;B,\Omega)
= \frac{e^{2}}{h}\,\nu,\\ 
&\nu =\sum_{n=0}^{\infty}\Theta\big(\mu_F-E_n(B_{\rm eff})\big),
\end{align}
with $E_n=\hbar|\omega_{\rm eff}|(n+\tfrac12)$ and $B_{\rm eff}\equiv B+B_{\Omega}$, $B_{\Omega}=2\hbar k\,\Omega/e$. Plateau transitions occur at $\mu_F=E_n(B_{\rm eff})$, i.e.
\begin{equation}
B_c^{(n)}(\Omega)=\frac{\mu}{e}\,\frac{2\mu_F}{\hbar(2n+1)}-\frac{2\hbar k\,\Omega}{e},
\end{equation}
an exactly \emph{linear} torsional shift of the critical fields. Finite temperature enters through the Fermi kernel and disorder via a Dingle-type broadening of the DoS; neither alters the rigid $B\mapsto B_{\rm eff}$ shift induced by $\Omega$.

\section{Note on the Effective Quantum Number \texorpdfstring{$N'$}{N'}}
\label{app:Nprime}

It is convenient to simplify the energy spectrum by introducing an effective quantum number $N'$ that combines the principal quantum number $n$ and the magnetic quantum number $m$. This maps the spectrum onto the levels of a quantum harmonic oscillator and greatly simplifies the partition function.

We define
\begin{equation}
    N' = n + \frac{|m| + m}{2}.
    \label{eq:N_prime_definition}
\end{equation}
With this definition, the spectrum of Eq.~\eqref{eq:SF37} rewrites as
\begin{equation}
    E_{N'} = \hbar|\omega_{\rm eff}|\left(N' + \frac{1}{2}\right) + E_{\rm trans},
\end{equation}
where $E_{\rm trans} = \hbar^2 k^2 / (2\mu)$.

This procedure is robust and accounts for all states, including those with $m<0$. Indeed:

\begin{itemize}
    \item \textbf{Case $m \ge 0$.} $|m|=m$ and
    $N' = n + (m+m)/2 = n+m$, hence $N'\in\{0,1,2,\dots\}$.

    \item \textbf{Case $m < 0$.} $|m|=-m$ and
    $N' = n + (-m+m)/2 = n$, again $N'\in\{0,1,2,\dots\}$.
\end{itemize}

Therefore, for any allowed pair $(n,m)$, $N'$ is always a nonnegative integer. The sum over $N'\ge 0$ in the partition function covers all states; the possible multiplicity of $(n,m)$ pairs yielding the same $N'$ is absorbed into the degeneracy factor $g$ used in the calculations.

\section{Numerical cookbook: from \texorpdfstring{$(B,\Omega,k,\mu)$}{(B, Omega, k, mu)} to thermodynamic and magnetic observables}
\label{app:cookbook}

This appendix is written as a reproducible narrative rather than a compact formula sheet. Each subsection explains what is being computed, why it matters physically, and how to implement it robustly in double precision. All symbols and units follow the conventions used in the main text.

\subsection*{A. Constants, inputs, and units}

To ensure unit safety and numerical reproducibility, all intermediate quantities should be kept in SI. The independent inputs are the magnetic field $B$ (tesla), the torsion density $\Omega$ (m$^{-1}$), the longitudinal wavenumber $k$ (m$^{-1}$), the effective mass $\mu$ (kg), and the temperature $T$ (kelvin). The fundamental constants used throughout are the exact SI values
\begin{align}
e&=1.602176634\times 10^{-19}\,{\rm C},\\
\hbar &=1.054571817\times 10^{-34}\,{\rm J\,s},\\
k_B &=1.380649\times 10^{-23}\,{\rm J/K}.
\end{align}
Working in SI avoids hidden conversion factors. Conversions to eV or other laboratory units should be performed only at the end for plotting axes or tables.

\subsection*{B. Single control scale and compensation line}

All observables in this work descend from a single effective cyclotron frequency that blends the magnetic and elastic contributions. It is convenient to compute first
\begin{align}
\omega_c &= \frac{eB}{\mu},\qquad
\omega_{cl}=\frac{2\hbar k\,\Omega}{\mu},\\[2pt]
\omega_{\rm eff} &= \omega_c+\omega_{cl}
= \frac{eB+2\hbar k\,\Omega}{\mu},
\end{align}
and then the dimensionless gap variable
\begin{equation}
x=\frac{\hbar\,|\omega_{\rm eff}|}{2k_B T}.
\end{equation}
The compensation line is the locus where the transverse gap collapses, $\omega_{\rm eff}=0$. At fixed $(k,\Omega)$ this occurs at
\begin{equation}
B_{\rm comp}=-\,\frac{2\hbar k\,\Omega}{e}.
\end{equation}
Close to compensation ($x\to 0$), one must switch to series expansions for the thermal kernels to avoid loss of significance, as described next.

\subsection*{C. Thermal kernels and stable asymptotics}

Two special functions appear repeatedly,
\begin{equation}
\coth x=\frac{\cosh x}{\sinh x},\qquad
\csch x=\frac{1}{\sinh x},
\end{equation}
always evaluated at $x>0$. When $x$ is very small, direct evaluation of hyperbolic functions suffers from cancellation. In that regime, it is numerically stable to use the series
\begin{align}
\coth x &\simeq \frac{1}{x}+\frac{x}{3}-\frac{x^{3}}{45}
\,+\,\mathcal{O}(x^{5}),
\\
\csch^{2}x &\simeq \frac{1}{x^{2}}-\frac{1}{3}+\frac{x^{2}}{15}
\,+\,\mathcal{O}(x^{4}),
\end{align}
which are accurate for $x\lesssim 10^{-3}$. In the opposite extreme, $x\gtrsim 10$, the large-$x$ tails $\coth x\simeq 1+2e^{-2x}$ and $\csch^{2}x\simeq 4e^{-2x}$ are both safe and fast.

\subsection*{D. Thermodynamics per transverse channel (fixed $k$)}

When the longitudinal motion is held fixed, the transverse sector behaves as a gapped harmonic mode controlled solely by $x$. The Helmholtz free energy, internal energy, and heat capacity per channel take the compact forms
\begin{align}
A(T) &= k_B T\,\ln\bigl[2\sinh x\bigr]
\,+\,\text{const},\\[2pt]
U(T) &= \frac{\hbar\,|\omega_{\rm eff}|}{2}\,\coth x,\\[2pt]
C_V(T) &= k_B\,x^{2}\,\csch^{2}x.
\end{align}
These expressions make the low- and high-temperature limits transparent. For $x\gg 1$, the system is frozen into the ground level and $C_V$ decays exponentially; for $x\ll 1$, the transverse equipartition value $C_V\to k_B$ is recovered. The ``const'' in $A$ is irrelevant to derivatives with respect to $T$ and fields.

\subsection*{E. Including the free longitudinal motion (3D limit)}

In bulk samples, the $z$-direction is free and contributes additively. Integrating over the continuum of longitudinal states adds a universal offset to energy and heat capacity. One finds
\begin{align}
U_{3\rm D}(T)&= U(T)+\frac{1}{2}\,k_B T,\\[2pt]
C_{V}^{3\rm D}(T)&= C_V(T)+\frac{1}{2}\,k_B,
\end{align}
so that the classical three-dimensional limit $C_{V}^{3\rm D}\to \tfrac{3}{2}k_B$ emerges when both transverse and longitudinal sectors are thermally populated.

\subsection*{F. Magnetic responses per channel}

The orbital magnetization and the isothermal susceptibility follow from field derivatives of the free energy. With the sign handled only by the chain rule, the working expressions are
\begin{align}
M(B,\Omega,T) &= -\,\mathrm{sgn}(\omega_{\rm eff})\,
\frac{\hbar e}{2\mu}\,\coth x,
\\[2pt]
\chi(B,\Omega,T) &= \Bigl(\frac{\hbar e}{2\mu}\Bigr)^{2}\,
\beta\,\csch^{2}x,
\qquad \beta=\frac{1}{k_B T}.
\end{align}
At low temperature, $M$ saturates to a Landau-like value and $\chi$ vanishes exponentially; at high temperature, both decay in agreement with the Bohr-van Leeuwen theorem.

\subsection*{G. Magnetocaloric slope and magnetic Gr\"{u}neisen parameter}

The adiabatic temperature response to a field change can be expressed either as $(\partial T/\partial B)_S$ or through the magnetic Gr\"{u}neisen parameter. Using the Maxwell identity and the fixed-$k$ heat capacity, kernels cancel exactly, and one obtains
\begin{align}
\Bigl(\frac{\partial T}{\partial B}\Bigr)_{S}
&= \frac{e}{\mu}\,\frac{T}{\omega_{\rm eff}},
\\[2pt]
\Gamma_B &= -\,\frac{1}{T}
\Bigl(\frac{\partial T}{\partial B}\Bigr)_{S}
= -\,\frac{e}{\mu\,\omega_{\rm eff}}.
\end{align}
After including the longitudinal continuum, the same formulas hold with $C_V\to C_{V}^{3\rm D}$, which reduces the high-temperature slopes by a factor $2/3$ and smooths the crossover controlled by $x$.

\subsection*{H. Effective length, degeneracy, and sample totals}

To connect per-channel formulas to macroscopic totals, one restores the Landau-like degeneracy per level, which depends on the effective magnetic length. The two key quantities are
\begin{align}
\ell_{\rm eff}&=\sqrt{\frac{\hbar}{\mu\,|\omega_{\rm eff}|}}
=\sqrt{\frac{\hbar}{|eB+2\hbar k\Omega|}},
\\[2pt]
g_{\rm bulk} &= \frac{A}{2\pi\,\ell_{\rm eff}^{2}}
\propto |\omega_{\rm eff}|,
\end{align}
with $A$ the sample area. In mesoscopic geometries, boundary corrections deplete the bulk degeneracy. A minimal disk model of radius $R$ gives
\begin{equation}
g_{\rm corr}\simeq g_{\rm bulk}\Bigl(1-\frac{\ell_{\rm eff}}{R}\Bigr),
\end{equation}
adequate to estimate finite-size trends in caloric observables. Sample totals follow by multiplying any per-channel result by $g(B,\Omega)$ or $g_{\rm corr}$.

\subsection*{I. Grand-canonical quantum oscillations (dHvA) in practice}

The oscillatory part of the magnetization in the grand-canonical ensemble can be written as a harmonic series modulated by thermal and Dingle damping. Keeping the sign only in the chain rule and restoring the degeneracy factor, one may use
\begin{align}
M_{\rm osc}(B) &=
-\,\mathrm{sgn}(\omega_{\rm eff})\,
\frac{g\,\mu_F}{\hbar\,|\omega_{\rm eff}|}
\sum_{r=1}^{\infty}\frac{1}{\pi r}\,
\sin\Bigl(\frac{2\pi r\,\mu_F}{\hbar\,|\omega_{\rm eff}|}\Bigr)
\notag\\[-2pt]
&\qquad\times
R_T(r,T)\,R_D(r),
\end{align}
where the damping factors are
\begin{align}
R_T(r,T)&=\frac{X_r}{\sinh X_r},\qquad
X_r=\frac{2\pi^{2} r\,k_B T}{\hbar\,|\omega_{\rm eff}|},
\\[2pt]
R_D(r)&=\exp\Bigl[-\frac{\pi r}{|\omega_{\rm eff}|\,\tau_D}\Bigr].
\end{align}
The oscillation period is strictly uniform in $1/B_{\rm eff}$, with $B_{\rm eff}=B+2\hbar k\Omega/e$. In typical parameter ranges, the first harmonic ($r=1$) suffices.

\subsection*{J. A practical workflow from inputs to plots}

A robust computational path proceeds as follows in prose form. One first evaluates $\omega_c$, $\omega_{cl}$, and $\omega_{\rm eff}$ from the inputs, and then forms $x=\hbar|\omega_{\rm eff}|/(2k_B T)$. The thermal kernels $\coth x$ and $\csch^{2}x$ are computed either directly or through the small-$x$ series if $x$ is tiny. With these at hand, the per-channel thermodynamics $A$, $U$, and $C_V$ follow immediately, and the 3D versions are obtained by adding the universal longitudinal offsets. Magnetization, susceptibility, and magnetocaloric quantities are then read from the formulas above. Whenever sample totals are desired, one computes $\ell_{\rm eff}$ and multiplies by $g(B,\Omega)$ or by $g_{\rm corr}$ if finite-size effects are relevant. Finally, for dHvA traces, the $r=1$ harmonic with $R_T$ and $R_D$ already captures the main oscillatory content.

\subsection*{K. Sanity checks and limiting behaviors}

Performing several quick checks helps catch mistakes before plotting. In the low-temperature regime ($x\gg 1$), the heat capacity must vanish exponentially, the magnetization must approach $-(\hbar e/2\mu)\,\mathrm{sgn}(\omega_{\rm eff})$, and the susceptibility must be strongly suppressed. In the classical regime ($x\ll 1$), one should recover $C_V\to k_B$ for the transverse model and $C_{V}^{3\rm D}\to \tfrac{3}{2}k_B$ after integrating the longitudinal motion; both $M$ and $\chi$ decay with temperature, consistent with Bohr-van Leeuwen. Exactly at compensation, $|\omega_{\rm eff}|\to 0$ at fixed temperature, the transverse heat capacity tends to $k_B$ while the diamagnetic response vanishes.

\subsection*{L. Common pitfalls and how to avoid them}

The most frequent numerical issue is the accidental omission of absolute values: the thermal kernels always see $|\omega_{\rm eff}|$, whereas overall signs enter only from chain-rule derivatives through $\mathrm{sgn}(\omega_{\rm eff})$. A second source of trouble is the evaluation of $\coth x$ and $\csch^{2}x$ at extremely small $x$ using standard library calls; switching to the series expansions eliminates cancellation errors. A third pitfall is mixing per-channel results with totals without explicitly multiplying by $g(B,\Omega)$, which can lead to apparent mismatches with experimental magnitudes.

\subsection*{M. Worked numerical example}

To anchor scales, consider electrons with $\mu=m_e=9.109\times 10^{-31}\,$kg, $k=10^{9}\,$m$^{-1}$, $\Omega=10^{7}\,$m$^{-1}$, $B=5\,$T, and $T=20\,$K. The cyclotron pieces are
\begin{align}
\omega_c&=8.79\times10^{11}\,{\rm s^{-1}},\qquad
\omega_{cl}=2.32\times10^{12}\,{\rm s^{-1}},\\
\omega_{\rm eff}&=3.20\times10^{12}\,{\rm s^{-1}},
\qquad
\frac{\hbar\omega_{\rm eff}}{k_B}=24.4\,{\rm K},
\end{align}
so that $x=24.4/(2\times 20)=0.61$. Using the fixed-$k$ formulas, one finds $U=\tfrac12\hbar|\omega_{\rm eff}|\coth x=1.73\times10^{-21}\,{\rm J}$, $C_V=k_B x^2\csch^2 x=1.17\times10^{-23}\,{\rm J/K}$, $M=-(\hbar e/2\mu)\coth x=-1.18\times10^{-23}\,{\rm J/T}$, and $\chi=((\hbar e)/(2\mu))^{2}\beta\csch^{2}x=7.06\times10^{-48}\,{\rm J/T^2}$. The three-dimensional values follow by adding $+\tfrac12 k_B$ to $C_V$ and $+\tfrac12 k_B T$ to $U$.

\subsection*{N. Final remarks on plotting and FFTs}

When analyzing Shubnikov-de Haas or de Haas-van Alphen oscillations, the true periodicity is in $1/B_{\rm eff}$ rather than in $1/B$. Reparametrizing the trace with $B_{\rm eff}=B+2\hbar k\Omega/e$ sharpens the dominant FFT peak and centers the frequency at a value independent of torsion. If one prefers to keep the horizontal axis as $1/B$, the period appears compressed by a smooth Jacobian factor, which is a plotting artifact rather than a change of the underlying physics.

\end{appendices}

\bibliographystyle{apsrev4-2}

\end{document}